\documentclass[11pt,fleqn,a4paper]{article}
\usepackage{graphicx}

\begin{document}
\thispagestyle{empty}
\renewcommand{\baselinestretch}{1.2}  
\small\normalsize
\frenchspacing
\noindent  
{\Large \textbf{Quantum mechanics in terms of realism}}\hspace{1.5pt}\footnote{The present version is another thorough revision of the 1996 version, taking experimental and theoretical progress into account.}
\\
\\
{\bf{Arthur Jabs}}
\renewcommand{\baselinestretch}{1}
\small\normalsize
\\
\\
\\
Alumnus, Technical University Berlin.

\noindent
arthur.jabs@alumni.tu-berlin.de 
\\
\\
(14 July 2019)\\
\newcommand{\rmi}{\mathrm{i}}
\newcommand{\bfitr}{\emph{\boldmath $r$}}
\newcommand{\spsi}[1]{\psi_{\textrm{\footnotesize{#1}}}}
\vspace{5pt}

\noindent
{\bf{Abstract.}}
We expound an alternative to the Copenhagen interpretation of  the  formalism
of nonrelativistic quantum mechanics. The basic difference is  that  the  new
interpretation is formulated in the language of epistemological  realism. The $\psi$  function  is  no
longer interpreted as a probability amplitude of the observed behaviour of
elementary particles but as  an  objective  physical  field  representing  the
particles themselves. The particles are  thus  extended  objects  whose  extension
varies in time according to the  variation  of $\psi$.  They  are  considered  as
fundamental regions of space with some kind of nonlocality. There is no wave-particle duality. The point-particle-like behaviour  is explained by spatial contraction in a deterministic reduction process.
\vspace{10pt}

\noindent
 {\bf Key words:} foundations of quantum mechanics, interpretation,    
realism, wavepackets, 
measurement, reduction,  collapse, entanglement, nonlocality,  Einstein-Podolsky-\linebreak[4]Rosen problem, Bell inequality
\vspace{10pt}
\vspace{10pt}

\newcommand{\hsp}{\hspace{60pt}}
\newcommand{\hspp}{\hspace{80pt}}
\newcommand{\hsppp}{\hspace{90pt}}

\newpage

\noindent
{\bf CONTENTS}
\bigskip

\noindent
 {\bf 1 Realism and Wave Aspect}  3

1.1~~Difficulties in Present-Day Quantum Theory  3                       

1.2~~Realism  5

 1.3~~Classical Wavepackets  8                           

1.4~~The Heisenberg Relations  10
                                                        
\smallskip
\noindent {\bf 2 Elimination of the Particle Aspect}  13

2.1~~Reduction and Measurement  13
                                        
2.2~~The Stern-Gerlach Experiment   15     
                                        
2.3~~Micro-Determinism and Macro-Indeterminism   17

\smallskip
\noindent {\bf 3 Nonlocality}  20

3.1~~One-Particle Nonlocality   20
 
3.2~~Entanglement and Multi-Particle Nonlocality   23                   
 
3.3~~Similar (Identical) and Condensed Wavepackets   27

\smallskip
\noindent {\bf 4 Nonlocality and Superluminal Signaling}  30
 
4.1~~The EPR Problem and Nonlocality  30                                    
 
4.2~~Superluminal Signaling  33                                          
 
4.3~~Bell's Inequality  35

4.4~~Experiments  38

\smallskip
\noindent  {\bf 5 Quantum Statistics with Wavepackets}  40

5.1~~Field Quantization   40                                             

5.2~~The Many Aspects of the Condensed Wavepackets 41                    

5.3~~The Balance Relation   44                                             

5.4~~The Bose and Fermi Distributions   48                                 

5.5~~Quantum Count Fluctuations   51                                       

\smallskip \noindent
{\bf Acknowledgement}  57

\smallskip
 \noindent
{\bf Appendix A:} 
Wavepacket Spreading Formulas   57
 
 \vspace{5pt}
 \noindent
{\bf Appendix B:} 
Derivation of the Bell Inequality  65

 \vspace{5pt}             
 \noindent
{\bf Appendix C:}
EPR Joint Probability Formulas  67
  
\vspace{5pt}                                   
\noindent
{\bf Appendix D:}
EPR Probabilities in Different Systems of Eigenfunctions  70

\vspace{5pt}                                   
\noindent
{\bf Notes and References}  73-88

\newpage
\begin{flushright}
\emph{The environment as we perceive it is our invention.}

Heinz von Foerster
\end{flushright}

\bigskip

\noindent {\bf 1. REALISM AND WAVE ASPECT}

\bigskip
\noindent {\bf 1.1~~Difficulties in Present-Day Quantum Theory}
\medskip

\noindent
There is no doubt that quantum  theory  is  one  of  the  most  successful
physical \mbox{theories}. Yet there is  also  no  doubt  that  it  contains  serious
difficulties. These difficulties are nowadays felt more and more strongly  by
those concerned with the unification of quantum theory and relativity and the
future basis of physics. The difficulties may  be  divided  into  two  kinds:
conceptual and mathematical.

The conceptual  difficulties  are  related  to  the  so-called  Copenhagen
interpretation. Any physical theory consists  of  a  mathematical  formalism,
that is, a set of mathematical symbols and the  rules  for  connecting  these
among themselves, and a set of interpretation rules, connecting  the  symbols
of the mathematical formalism with the concepts of our  sensory  experience.
The Copenhagen interpretation represents that  set  of  interpretation  rules
that is presented more or less explicitly  in
current textbooks on quantum mechanics. Actually, it is difficult to  say
who exactly constitutes the ``Copenhagen  school'', supporting  the  Copenhagen
interpretation; certainly Bohr and Heisenberg, but also Dirac, Pauli and  von
Neumann \cite{Jam}, \cite{Bohr63}. Also, many versions of  ``the  Copenhagen  interpretation'', from orthodox to  liberal, can be found when  different  authors  or  textbooks  are consulted. 

The difficulties of the Copenhagen interpretation may be characterized  in
the following way:

1) The wave function $\psi ({\bf x},t)$ is not taken as an  objective  physical  field
like the electromagnetic field, but  as  a  probability  amplitude.  And  the
probabilities to which it refers are not the probabilities that something  is
true or something will happen, whether it is observed or not (as in statistical mechanics) but  the probabilities of specified outcomes  of  measurements or observations. Most importantly, the observer is not just
another physical object but the linguistic ego, something  which  appears  nowhere
 as a mathematical symbol in the formalism. An electron  does  not  have  an  exact
location as long as we do not observe it,  but  it  does  have  one  when  we
observe it. ``The `trajectory' arises only by our observing it'' \cite{Hei}.  In  this  way  the
Copenhagen interpretation  speaks  of  position  and  of  momentum,  angular
momentum components and others only as  ``observables'',  not  as  real  properties which 
objects have regardless of whether  we observe them  or not.  The
observer and the measurement  are  therefore  indispensable  elements  in  defining  the
theory. In all other physical theories the observer's  only  function  is  to
test and apply the theory, not to define it.  The  Copenhagen  interpretation
thus rejects the language of epistemological realism. In my opinion, this  is
the most serious difficulty with that interpretation.

2) According to the Copenhagen interpretation  it  is  impossible,  in
principle, to explain the probabilistic behaviour in quantum physics  as the result of  some
underlying deterministic processes.  This
means that the Copenhagen interpretation rejects determinism. This point will
be examined further in  Chap.~2.

3) In the Copenhagen view concepts that refer to pointlike particles  (with single,
sharp position) are applied to  micro-objects, whereas  the $\psi $
function is the solution of a field equation, namely a  partial  differential
equation, like Schr\"odinger's, and may show up wave-like  behaviour.  This  has
been called the ``wave-particle duality'', and it  has  been  asserted  that  a
unifying picture of micro-objects cannot exist.

As is well known Einstein, von Laue, Schr\"odinger, Planck, and  de  Broglie
never  accepted  the  Copenhagen  interpretation.  In  fact, any really thoughtful scholar finds it difficult  to  digest.  Inquiring
students are usually silenced by authoritarian statements such as  that  they
do not yet know enough and will understand later, or that their questions are
not relevant to physics.

In particular, the defenders of the  Copenhagen  interpretation  
assert that a unifying picture is really unnecessary.  They  say  that  the
formalism plus some working rules for its application  give  us  the  correct
prescriptions for calculating  the  probabilities  of  the  outcomes  of  any
experiment, and that that is all we want. But as if some  tectonic  tensions  were
felt, discussions concerning the  foundations  of  quantum
mechanics continue. Moreover, it is noticeable that  in  the  course  of  historical
development as well as in the mind of  any  particular  scientist  the  final
mathematical formalism describing a set of physical phenomena emerges from  a
more or less pictorial view, conception or model.  A  good  picture  is  very
helpful since it has the same logical structure  as  the  region  of  reality
which it aims to represent, and it leads to a correct  mathematical  formulation  of
this reality. An example is Faraday's intuitive picture of lines of force and
their subsequent mathematical formulation by Maxwell. ``It is mainly with  the
hope of making these [Faraday's] ideas the basis  of  a  mathematical  method
that I have undertaken this treatise'', Maxwell writes \cite{Max}.  A bad picture leads
to no or to an only partially correct formalism. In this latter  case  it  may
happen that the emerging formalism  describes  the  known
phenomena correctly in its initial stage, but when it is developed further  to  include  more  and
more experimental facts it sooner or later comes off the track. This is  what
I think has happened to quantum theory. I think  that  the  lack  of  a  good
picture is responsible for the  mathematical  difficulties.

The present-day mathematical difficulties arise with the
attempt to extend nonrelativistic quantum  mechanics  into  the  relativistic
domain, that is, into quantum electrodynamics and relativistic quantum  field
theory. Here, marvelous successes in nuclear and particle physics and relativistic quantum field theory have been obtained. Nevertheless, divergent integrals have shown up in the perturbation  expansions  as
solutions of the basic equations. Even if these integrals are made finite  by
means of renormalization procedures or are avoided by means of Epstein-Glaser methods \cite{Eps}, \cite{Scha} nobody knows whether  the  expansions
converge, and nobody has found an exact solution of the  equations  including
interactions in the real world of 3+1  dimensions,  although  enormous  efforts
have been undertaken \cite{Jau},  \cite[Sec.~11.1]{Bjo}. Thus Dirac \cite{Dir72}, \cite{Dir81} writes:
\begin{quote}
I feel pretty sure that the changes which will be needed  to  get  over
the present difficulties facing  quantum  theory  and  appearing  as  a
resistance between the quantum  theory  and  relativity  will  be  very
drastic just as drastic as the change from Bohr orbits to  the  quantum
mechanics of Heisenberg and Schr\"odinger and therefore  one  should  not
become  too  much  attached  to  the  present  quantum  mechanics.  One
shouldn't build up ones whole philosophy as though this present quantum
mechanics were the last  word.  If  one  does  that,  one  is  on  very
uncertain ground and one will in some future time have to change  one's
standpoint entirely.
\end{quote} 
In the present paper we are, however, only concerned with nonrelativistic quantum mechanics.

\vspace{10pt}
\noindent {\bf 1.2~~Realism}
\medskip

\noindent
About the  nature  of  the  expected  changes  Einstein writes \cite{Ein48} - \cite[p. 667]{Schi} - \cite{Ein53}.  In particular in \cite[p. 6]{Geo}:
\begin{quote}
But in any case my conception starts from a thesis  which  is  strongly
rejected by most present-day theoreticians: \emph{There is something like the
``real state''} of a physical system, which independent of any observation
or measurement exists objectively and which can in  principle  be  described by means of  physical  terms  [Which  adequate  terms  or  basic
concepts have to be employed for this is in my opinion unknown  at  the
present moment (material points? field? concepts that have still to  be
invented?)]. Because of  its  ``metaphysical''  nature,  this  thesis  of
reality does not have the purpose of providing a statement of fact:  it
has really only a \emph{programmatic} character. However, everybody, including
the quantum  theoreticians,  sticks  consistently  to  this  thesis  of
reality so long as he does  not  discuss  the  foundations  of  quantum
theory. Nobody doubts, for example, that there has been  at  a  certain
time a certain position of the moon's center of gravity even if no real
or potential observer existed.
\end{quote}
This thesis of Einstein's is what we mean by epistemological  realism.  We
do not attempt to give a fool-proof  definition  of  realism.  We  emphasize,
however, that the type of realism adopted here does not insist that  physical objects with their properties exist independently  of  whether or not we  observe
them (this would be naive realism); it only means  that  the  laws  of
nature can be formulated {\it{as if}} that were the case. 

Realism is not  a matter to be proved or disproved, it is a way of
speaking, a language.
Compare the language of realism with English, and the Copenhagen language with German. Nobody will deny that everything that can be expressed in German can also be expressed in English, although some things can be expressed in a much shorter and simpler way in the one language than in the other. With realism quantum mechanics is formulated in the same language as classical mechanics and every other physical theory, and the power that lies in a realist language is available for quantum mechanics, too. To quote Wittgenstein \cite{Wit}:
\begin{quote}
For \emph{this} is what disputes between Idealists,  Solipsists  and  Realists
look like. The one party attack the normal form  of  expression  as  if
they were attacking a statement, the others defend it, as if they  were
stating facts recognized by every reasonable human being.
\end{quote}
If we were to distinguish our type of realism from naive realism we would
call it epistemological or linguistic realism.

Von Laue \cite{Lau},  Schr\"odinger \cite{Schr35a}, \cite{Schr35b}, \cite{Schr53},  and Planck \cite{Pla}, \cite{Jan} have always  shared  a
realist attitude with Einstein. And in the course of the years  the  number  of
physicists who openly advocate realism  in  quantum  theory  has  continually
increased. Jammer's book \cite{Jam74} already quotes Bohm, Bunge, de Broglie,  Jaynes,
Ludwig, Popper and Renninger. And we want to add the papers by Janossy \cite{Jan},  de
 Broglie \cite{Bro},   Bunge and Kalnay \cite{Bun75} - \cite{Bun67b},    Bell \cite{Bel73d} - \cite{Bel90a},   
Rayski \cite{Ray73},    L\'evy-Leblond \cite{Lev},    Stapp \cite{Sta77}, \cite{Sta85},    Roberts \cite{Rob78},   
Maxwell \cite{Max82},    Burgos \cite{Bur},    Popper \cite{Pop},  \cite{Pop85},  Pearle \cite{Pea},    
Bohm,  Hiley   and   Kaloyerou \cite{Bohm87},    Rohrlich \cite{Roh},    
 Dorling \cite{Dor},    and  Dieks \cite{Die88}. 
Actually, it is  difficult  to  do  justice  to
everybody because there are several types of  realism,  because statements in favour of realism range from very outspoken to rather casual.  Scientists I found particularly outspoken in favor of realism are Popper,  Bunge,  and
Bell.  Bell \cite[p. 40]{Bel87} in  particular   postulated   ``\emph {be}ables''   to   replace   the
``\emph {observ}ables''  (Sec.~2.1), and his work will concern us in Chap.~4. 

Thus in the present work we show how one can overcome the  conceptual  difficulties of quantum mechanics by interpreting the formalism in terms of  realism. The formalism is: (1) that presented in the usual textbooks \cite{Mes}, \cite{Coh}, plus: (2) a mathematical description of the reduction (collapse) process, introduced in Sec.~2.1 \cite{Jab12}.  Actually, since 1996 a number of other "ontological" or realist interpretations have appeared, but according to them they have nothing to do with my work.

The essential point is that the wave function $\psi ({\bf x},t)$ is taken as  an  objective  physical  field,
comparable in this respect to the function $F_{\mu \nu }({\bf x},t)$  as  the  source-free electromagnetic
field. This implies 
that $\psi $ is not merely a device for calculating the probabilities
of specified outcomes of observations; it does not  merely  describe
``knowledge'' \cite{Hei58}. \vspace{5pt}

It is mainly for convenience  of  presentation  that  the  nonrelativistic
formalism is chosen, with the familiar Schr\"odinger equation and wave function
as the basis of the reinterpretation. We think that  the  concepts  developed
will prove fruitful in the relativistic domain as well; at least  we  do  not
know of any argument that would point to the contrary. Everything that can be
described  by  the  Schr\"odinger  equation  can  also  be  described  by   the
Klein-Gordon, Dirac etc. equation. Thus  we  might  as  well  have  used a Lorentz scalar, spinor, vector,
etc. instead of the Schr\"odinger scalar $\psi ({\bf x},t)$. Moreover, we include photons in our considerations, that is, we treat classical electromagnetic radiation pulses on the same footing as pulses of Schr\"odinger or similar waves.

\medskip

The realist conception of the wave function already resolves the problems of Wigner's friend and Schr\"odinger's cat. Usually the whole measurement apparatus consists of a long chain of sub-apparatuses (amplifier, channel analyzer, transmitter ...). In the most orthodox version  of  the  Copenhagen  interpretation  it is my becoming conscious of the result that completes the measurement. Now, a friend of mine may form a sub-apparatus in that chain  in
that he, for example, reads off the pointer position on a  display  and  then
telephones it to me. The difficulty arises as  soon  as  I  credit  my friend with the  same
capabilities  as I have because this implies  that  the  result
has already obtained in the apparatus due to his  being conscious of it.  This  is
essentially the conflict between the Copenhagen description where the  observer is the linguistic ego and any realist description, where the observer is
just another physical object and the  result appears in a  physical  process that
occurs whether it is noticed or not.
\vspace{5pt}

Another difficulty with the Copenhagen view is described in the  example
of Schr\"odinger's cat \cite[p.~812]{Schr35a}. Consider a closed box containing a cat, a  certain
amount of radioactive nuclei, a Geiger counter and a cat-killing device,  all
protected against the cat. Circumstances are arranged so that the probability
that the Geiger counter discharges at the decay of at least one nucleus  within  one
hour is just 1/2. If the counter discharges it triggers the
cat-killing device, which consists of a hammer and a flask of  prussic  acid.
The flask is smashed, the acid is released, and  the  cat  is  poisoned.  The
probability that after one hour the cat is dead is  1/2.  Since  the  box  is
closed we cannot know after an hour whether the cat is dead or alive, unless
we cautiously open the box and  look into it.

In orthodox quantum mechanics, where the wave function represents our knowledge, there is one wave function $\psi_{\rm L}$ that represents our knowledge that there is a living cat in the box and another function $\psi_{\rm D}$ that there is a dead cat, but the situation is not described by the sum of the probabilities but by the superposition of the probability amplitudes
\begin{displaymath}(1.1)\hsp
c_{\rm L}\psi_{\rm L}+c_{\rm D}\psi_{\rm D}
\end {displaymath}
in the two-dimensional Hilbert space of dead and living cats. (1.1) is then interpreted as the wave function of neither a dead nor a living cat but a superposition of both.
Only when we look into the box  a reduction occurs and the cat's wave function becomes either $\psi_{\rm L}$ or $\psi_{\rm D}$. 

The cat is a macro-object, and in the  realm  of  macro-objects  the
language of realism is spoken: the cat is either alive or  dead  even  if  we
do not observe it. The radioactive nucleus is a micro-object, and in the
realm of micro-objects the language of realism is forbidden by the  verdict  of
Copenhagen. In the example there is a chain of  reactions  beginning  in  the
microworld with the decay of the unstable nucleus and ending  in  the  macroworld with the death of the cat. If both the micro- and the macro-object  are
described by a $\psi $ function, the character of the $\psi $ function  must  change  when
the chain of reactions crosses the borderline between the two realms. This is
another difficulty.

In our interpretation the language of realism is spoken in the  microworld
as well as in the macroworld, and the character of the $\psi $ function  is  always
that of a real physical field representing real physical objects, micro- or macroscopic.
The point is however that these objects, when their wavepackets are superposed,  must exist at the same time \cite{Cat}. Probabilities may refer to different times, but  the wavepackets may not. The wavepackets $\psi_{\rm L}$ and $\psi_{\rm D}$, on the contrary, exist at different times, $\psi_{\rm L}$ before and $\psi_{\rm D}$ after the decay of the radioactive nucleus. No such superposition is met elsewhere in the standard formalism of quantum mechanics, not even for micro-objects. 

In fact I do think that Schr\"odinger considered the cat example in order to point out to what incredible features the superposition (1.1) would lead.

Thus, although the particular superposition (1.1) for the cat is is not allowed in realism, this does not mean that there is no superposition at all of wavefunctions representing macro-objects. The restriction is that these wavefunctionss must represent something that really exists \emph{at the same time}. Examples are the recent experiments with large molecules and clusters \cite{Eibenberger}, \cite{Arndt}, if  one  accepts these to be already macro-objects.

\vspace{20pt}
\noindent {\bf 1.3~~Classical Wavepackets}
\medskip

\noindent
The identification of an elementary  particle  with  a  field  means  that
quantum mechanics becomes a field theory, albeit a special one. The Schr\"odinger equation, or any of
the quantum equations of motion, in any case is a field equation, that is,  a
\emph {partial} differential equation, with the solution $\psi $  depending  on  the  four
independent variables $x, y, z$ and $t$. The equations of  motion  of  the  point
particles of classical mechanics, on the contrary, are \emph{ordinary}  differential
equations for the three functions $x(t), y(t)$ and $z(t)$. On this point Einstein \cite{Ein34} 
writes:

\begin{quote}
\noindent
The most difficult point for such a field theory at present is how  to
include the atomic structure of matter and energy. For  the  theory  in
its basic principles is not an atomic one in  so  far  as  it  operates
exclusively with continuous functions of space, in contrast to classical mechanics whose most important feature, the material point,  squares
with the atomistic structure of matter ...  And yet a theory may perfectly well  exist,  which  is  in  a
genuine sense an atomistic one (and  not  merely  on  the  basis  of  a
particular interpretation), in which there  is  no  localizing  of  the
particles in a mathematical model. For example, in order to include the
atomistic character of electricity, the field equations  only  need  to
involve that a three-dimensional volume of space on whose boundary  the
electrical density vanishes everywhere,  contains  a  total  electrical
charge of an integral amount. Thus in a continuum theory, the atomistic
character could be satisfactorily expressed  by  integral  propositions
without localizing the particles which constitute the atomistic system.
Only if this sort of representation of the  atomistic  structure  be
obtained could I regard the quantum problem within the framework  of  a
continuum theory as solved.
\end{quote}
\noindent
The idea that elementary particles are  not pointlike but are extended  objects  has  repeatedly
appeared in the literature. However, the size has always been  considered  to
be fixed, for example equal to the Compton length of the electron $\phantom{i}^{-}\llap{$\lambda$}_{\rm C} = \hbar /mc$,  the  classical electron radius $r_{\rm cl} = e^{2}/mc^{2}$, or the Planck length $l_{\rm P} = (\hbar G/c^{3})^{1/2}${\emph .}
In the interpretation presented here the size of any individual  particle  is
variable, namely equal to the size of the $\psi $ function traditionally associated
with it. Mathematically, the $\psi $ function need not have a  sharp  boundary  but
for our purposes it may be considered to have  the  extension  given  by  the
usual standard  deviation $\Delta x :=\langle(x - \langle x\rangle)^{2}\rangle^{1/2}$, which varies in  time
according to the variation of $\psi ({\bf x},t)$.
\vspace{5pt}

\begin{figure}[h]
\begin{center}
\includegraphics[width=0.9\textwidth]{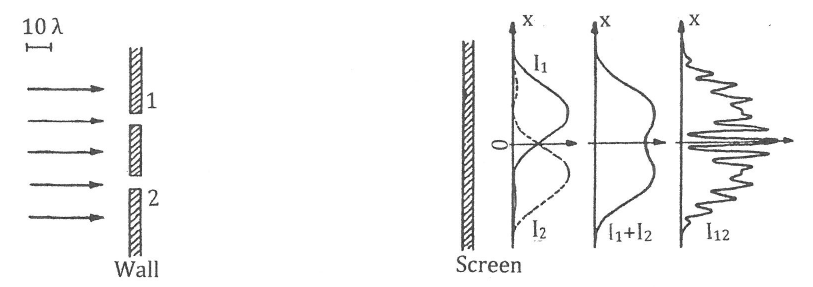}
\caption{The double-slit experiment.}
\end{center}
\end{figure}
The most convincing experiments showing the wave aspect of the elementary particles is the double-slit experiment.
Consider a beam of electrons with average momentum $p$  and  little  spread
about this value. The beam is directed towards a wall,  as  shown  in  Fig.
1. The wall contains two parallel slits, which  can  be  opened  and  closed.
Behind the wall there is a detecting screen  which  registers  the  intensity
{\emph I}({\emph x}) of the beam (number of electron counts per second) as a function of  the
distance $x$ from the center O. If only slit 1 is open the  intensity  function
{\emph I}({\emph x}) will look like $I_{1}$; if only slit 2 is open it will look like $I_{2}$. If  both
slits are open the intensity function is not, however, the sum $I_{1}+I_{2}$, but will
look like $I_{12}$. The shape of $I_{12}$ is obtained simply by regarding  the  beam  of
electrons as a plane wave with a wavelength $\lambda  = h/p$  and  calculating  the
interference of waves originating from slit 1 and slit 2. These  interference
effects constitute the difference between $I_{12}$ and $I_{1}+I_{2}$.

The important point is that  no  matter  how  low  the  intensity  of  the
incoming beam, the intensity function on the screen, when both  slits
are open, is always given  by $I_{12}$,  provided  we  compensate  for  the  lower
incoming intensity by a longer exposure time in order to  have  the  same  total
amount of energy (or total number of electrons) deposited on the  screen.  We
may adjust the incident intensity until it is so low that it  corresponds  to
one incoming electron per hour. Thus, even a single electron  must  correspond
to a number of wave trains capable of interference with one another.

The electron covers  both  slits,  so  that  in  any  single
passage both slits are involved and determine the final interference pattern.
Only if the wave function has a transverse width (normal to the direction of its
motion) that is smaller than the distance between the slits will no  double-slit interference effects be observed.

In order to emphasize that we consider the  elementary  particles  not  as
pointlike but as extended objects we call them   \emph{wavepackets} (one word). The function $\psi ({\bf x},t)$ then is the mathematical
representation of a  wavepacket.  Sometimes  we  shall  neglect  the
difference between the wavepacket and its mathematical representation  and  just
call $\psi$ a wavepacket. The term `wavepacket' does not, however, in any way mean a
restriction to a linear superposition of plane waves, and  even  when  it  is
mathematically expressed as such it  does  not  mean  that  plane  waves  are
physical constituents of the  particles.  It  just  means  the  region(s)  of
non-vanishing $\psi$.  Compare it in this respect with a water drop or a soap bubble.  A particle may be regarded as a ``matter pulse'',  on  an
equal footing with an electromagnetic radiation pulse. Indeed, in our  interpretation the radiation pulse, under certain conditions, \emph {is} also a  particle,
namely a photon. For each kind of particle the ``matter''  field is  specified by  additional  parameters like mass or charge.

\vspace{10pt}
\noindent {\bf 1.4~~The Heisenberg  Relations}
\medskip

\noindent
The Heisenberg relations
 \begin{displaymath}(1.2)\hsp
 \Delta x \Delta p_{x} \ge  \hbar /2
 \end{displaymath}
\noindent
play a central role in the continuing discussions on the physical meaning  of
quantum theory. Many different interpretations have been advanced \cite{Jam74}, and it
is not our intention to review them here. Rather, we shall  pick  out  a  few
points that serve to clarify the present interpretation.

The derivations of the Heisenberg relations presented in most textbooks consider operators for position and momentum and their Hermiticity and noncommutativity, and the Schwarz inequality for the state vectors. This might suggest that the Heisenberg relations are basic structural features of the quantum mechanical formalism. We do not think so. The relation can also be derived by using only classical wavepackets, as is amply demonstrated by Rapp \cite{Rapp}.

This equation
as it stands
is still a classical  field  equation \cite{Bog}. The  appearance  of  the  typical
quantum constant $h$ in it only indicates the  kind  of  fields  to  which  the
equation refers, namely to Schr\"odinger, de Broglie  or  matter  fields  whose
characteristic wave quantities $\omega $ and $\lambda $ are related to the particle quantities
$E$ and $p$ by the relations $E =  \hbar \omega $ and $p = h/\lambda $.  Even  the  special  boundary
conditions introduced in order to get ``quantized''  solutions,  which  can  be
distinguished from one another by parameters that take on  discrete  (eigen)values, are not enough to provide a genuine  quantum  character.  The  energy
eigenvalues in the hydrogen  atom,  for  example,  follow  from  the
normalizability postulate.  Eigenvalues already appear in classical macro-physics, and it is significant
that the title of Schr\"odinger's famous papers was ``Quantization as a  problem
of eigenvalues''.

\hspace{2pt}

\noindent
 In 1928  Schr\"odinger  said  \cite{Schr28}:
\begin{quote}
One may believe either (1) that matter has {\emph{really}} a wave structure. Then the uncertainty principle is an immediate consequence. Or (2) one may think that the uncertainty principle is the more fundamental.
\end{quote}
Of course, ours is the first option.
Moreover, Dirac in 1972 said \cite{Dirac72}:
\begin{quote}
So if one asks what is the main feature of quantum mechanics, I feel inclined now to say that it is not noncommutative algebra. It is the existence of probability amplitudes which underly all atomic processes.
\end{quote}

\noindent
The importance of the Heisenberg relations stems from  the  importance  of
the wave\-packet nature of the elementary particles.  In  the macroworld  only  the  first  moment  of  the  function
$| \psi ({\bf x},t)| ^{2}$, that is, the centre value $\langle {\bf x\rangle }$  plays  a  role  (Ehrenfest  theorem),
whereas in the microworld  the  higher  moments  come  into  play.
Already the first step, the inclusion of only  the  second  moments $\Delta x$,  marks the essential differences from the macroworld. This is  why the wavepacket concept captures more characteristics of the elementary particle than does the point particle concept.

The function $\psi ({\bf x},t)$ represents all of the wavepacket (particle),  whether we know it or not. It need not be an eigenfunction of some operator. Only when the wavepacket passes a specific physical situation (e.g. a magnetic field) it assumes the form of  a superposition of eigenfunctions of the associated observable. And in a subsequent collapse (Sec.2.1) it assumes the form of a single eigenfunction. In the case of a Heisenberg relation, different physical situations are necessary for the wavepacket to assume eigenfunctions of different observables. This concords with Bohr's statement.
\cite[p. 233]{Schi}:
\begin{quote}
As  repeatedly  stressed,  the  principal  point  is  here  that   such
measurements demand mutually exclusive experimental arrangements.
\end{quote}
Once we are given the shape of  the  wavepacket  (which
includes its ranges), no uncertainty is left concerning this packet. We therefore
do not speak of uncertainty relations; rather we would speak of  complementarity, or just Heisenberg  or Born-Heisenberg relations. As $\Delta p_x$ is the width of the wavepacket $\tilde{\psi}(p_x)$ in momentum space, where  $\tilde{\psi}(p_x)$   is the Fourier transform of $\psi(x)$ (cf. Appendix A), 'Fourier reciprocity relation' would appear the most fitting denomination. 

A classical analogy may help: Consider a certain amount of dough of any form. Then shape a tower ("experimental arrangement"). Let its height ($"\Delta x"$) be large. Then its base ($"\Delta p"$) is small. Alternatively shape a pancake. Its height is small, but its base is large. And you cannot shape a tower and a pancake with the same procedure.
\vspace{3pt}

The relations (1.2) may be generalized to any two  quantities  which  are
represented by noncommuting Hermitean operators {\emph A} and B \cite{Schr30a}, \cite{Rob29}
\begin{displaymath}\hspp
\Delta A \Delta B \ge \mbox{$\frac{1}{2}$} | \langle [A,B]\rangle |  >0,
 \end{displaymath}
\noindent and our realist interpretation  is  essentially the same as that of the original Heisenberg relations. This interpretation is independent  of
whether the operators have discrete or continuous eigenvalues.  For  example,
the relation between the components of angular momentum
\begin{displaymath}\hspp
\Delta l_{x} \Delta l_{y} \ge  \mbox{$\frac{1}{2}$} \hbar  | \langle l_{z}\rangle |  >0,
 \end{displaymath}
\noindent means that the particle which is an eigenpacket of $l_{z} (\neq 0)$ cannot at the same
time be an eigenpacket of $l_{x}$ and $l_{y}$, rather it has the finite ranges $\Delta l_{x}$  and
$\Delta l_{y}$. When the same packet in an appropriate physical situation  changes  into
an eigenpacket of $l_{x}$, say, then  its  range $\Delta x$  shrinks  to  zero  while  it
acquires the finite ranges $\Delta l_{y}$ and $\Delta l_{z}$.
\newpage

\vspace{20pt}
\noindent
{\bf{2~~ELIMINATION OF THE PARTICLE ASPECT}}

\vspace{10pt}
\noindent
{\bf{2.1~~Reduction and Measurement}}
\medskip

\noindent
So far we have only considered the wave aspect, which does not present any genuine quantum feature. The quantum features come in with the particle aspect. This aspect will now be discussed -- and eliminated.

To see why the Copenhagen interpretation holds the concept of classical particles let us return to Fig.~1 in Sec. 1.3, and let us observe the detecting screen behind the wall with the two slits in it. What is observed is that the interference picture $I_{12}$ is built up gradually as time proceeds. It is made of a distribution of pointlike spots whose spatial density finally exhibits the interference picture. This is  impressively shown in Fig.~2, from an experiment  corresponding to the two-slit experiment, done with electrons \cite{Tonomura}.
Moreover, from other experiments (some cited in \cite{Jab14}, \cite{Jab17}) one knows that a one-particle wavepacket never produces more than one spot at a time.

These observations are indeed suggestive of being produced by point-particles. Now, since the double-slit interferences make it impossible for the particle to have a  (sharp)
position at all times, that position is ascribed to them only at  the  moment
of measurement. Thus, ``position'' is no longer a permanent objective  property
of a particle but an ``observable'', which comes into existence only in the act
of observation. If there is no observation, a position must not be ascribed to
the particle.  As with  position,  other  quantities  like  the  components  of
momentum and angular momentum must not be ascribed to the particle, except at
the  moment  of  the  respective  measurement.  All  these   quantities   are mere 
``observables''. Thus, the Copenhagen interpretation has both a wave and a particle inside the wave. This is the pernicious wave-particle duality.

We do away with that  duality. There are no classical point-particles in the microworld. Here we replace the concept of a classical point-particle by the concept of a quantum wavepacket. A quantum wavepacket suffers a reduction or collapse, and thereby it can contract
 to a pointlike extension in space (more on this below). Those effects that formerly were ascribed to the action of a classical point-particle, now are ascribed to the collapse of the quantum wavepacket. And when we speak of a particle in our theory we always mean a one-quantum wavepacket. 
 \begin{figure}[h]
\begin{center}
\includegraphics[width=0.78\textwidth]{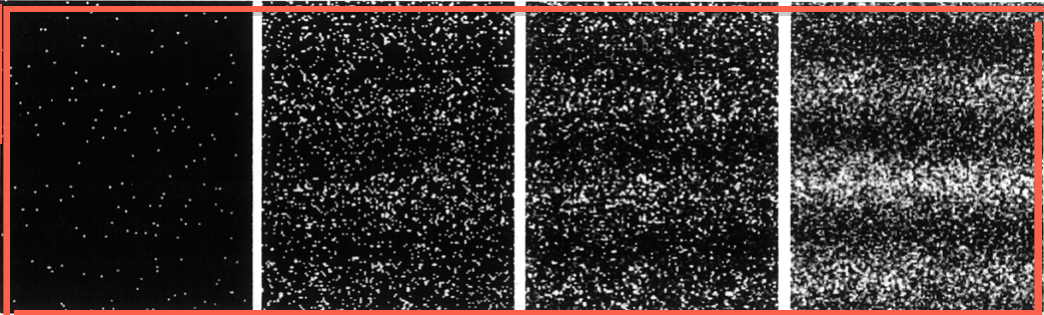}
\caption{Experiment on the buildup of an electron interference pattern \cite{Tonomura}}.
\end{center}
\end{figure}
Notice that the spots in the detection screen are not points but have some extension. A spot is the result of a cascade  of  processes  which  is  initiated  by  the
ionization of an atom in the detecting screen  by the incoming wavepacket. The localization is thus limited in practice by the extension of the black spot and  in
principle by the extension of the initiating atom or of the smallest 
wavepacket in the screen with which the incoming packet interacts.

The  concept of reduction in quantum mechanics was first mentioned shortly by Heisenberg \cite[p. 186]{Hei} and was worked out very  clearly
by von Neumann \cite{Neu}. It is therefore  often called  von  Neumann's  axiom.  It
assumes that immediately after the measurement the  wavepacket  is  a  superposition of eigenfunctions of the respective operator  which  belong  to  the
interval of eigenvalues specified by  the  measurement apparatus \cite[p. 298]{Mes}, \cite[p. 221]{Coh}.  This covers the case of continuous as well as discrete eigenvalues. In the special  case  of  a  discrete  spectrum,  a  non-degenerate eigenvalue and sufficient measurement accuracy,  the  wavepacket
immediately after the measurement  will  be  a  completely  specified  eigenfunction of the chosen observable.

In our conception, reduction is different. It is conceived to be a real physical process quite independent of any measurement, though measurement needs it (see below). It occurs when a certain physical criterion is satisfied.

Actually, a quantitative physical criterion for the occurrence of the reduction process has been formulated in detail in \cite{Jab12}. The reduction formulas (4.1), (2.1) below are on an equal footing with the Schr\"odinger equation in describing the temporal evolution of the wavepackets.  We here only want to sketch those features which are used in the present paper. Reduction, in our picture,  occurs when two wavepackets, $\psi_1$ and $\psi_2$, in ordinary space for the first time overlap and satisfy a certain criterion. Both then contract suddenly to the place of overlap. The criterion depends on the parameters of both wavepackets in a symmetric way:

\begin{displaymath}(2.1)\hsp
|\alpha_1-\alpha_2| \leq \mbox{{\normalsize{$\frac{1}{2}$}}}\,\alpha_{\textrm s}
\end{displaymath}
\begin{displaymath} (2.2)
\hspace{62pt} 
\bigg[\int_{\rm R^3} |\spsi{1}(\bfitr,t)| \, |\spsi{2}(\bfitr,t)| \, {\textrm d}^3r\bigg]^2 \ge \alpha/{2\pi}.
\end{displaymath}

\noindent $\alpha_{\textrm s}$ is \mbox{Sommerfeld's fine-structure~constant}.
$\alpha_1$ and $\alpha_2$ are the absolute phase constants of wavepacket 1 and 2, respectively. They are nonlocal `hidden' variables. and in our reduction they do play an important role. They are physical because there are situations  where they can be determined. In the superposition the absolute phases become relative and determine the position of the interference fringes. 
The phase constant $\alpha$ in (2.2) is the smaller one of $\alpha_1$ and $\alpha_2$.
In an ensemble of wavepackets, the phase constants are taken to be pseudorandom numbers. That is, they  only seem to be random, but in reality they are determined by certain initial conditions. These phase constants determine when and where a reduction occurs. We call this micro-determinism.

\vspace{5pt}
Now comes quantum measurement. By this we mean one
where the wavepacket nature of the elementary particles cannot  be
neglected, that is, where a location is  measured  with  an
error interval that is smaller than, or of the order  of,  the  corresponding spatial 
width of the wavepacket.

As every physicist can verify, in any quantum mechanical measurement the measurement apparatus fans out the incoming wavepacket in ordinary space into spatially separated eigenpackets of the chosen observable (self-adjoint operator). When one of these eigenpackets and some wavepacket in the sensitive region of the apparatus satisfy the criterion, the reduction associates the place of contraction with an eigenvalue of the observable. There must be special wavepackets in the sensitive region which under contraction initiate an avalanche of effects that result in an observable spot. These wavepackets in the screen are very small, so the region of overlap of such a wavepacket with the incoming wavepacket is also very small, and the whole of the incoming wavepacket has contracted to this very small region. This mimics the action of a pointlike particle.

In order to predict the outcome of a particular measurement, the observer would have to know the shape of the incoming wavepacket and the phase constant of every special grain in the screen of the measuring apparatus. This is virtually impossible. 
Moreover, the observer cannot know effects coming from events outside his past light-cone \cite{Jay}.

We thus have  determinism in the microworld (micro-determinism), but  unpredictability (macro-indeterminism) in the macroworld, just as in throwing dice.

\vspace{20pt}
\noindent {\bf 2.2~~The Stern-Gerlach Experiment}
\medskip

\noindent
The Stern-Gerlach experiment is well suited for a demonstration of  the new concept of measurement outlined above. In quantum mechanics it has traditionally been regarded  as  the prototype of a measurement.

First let us briefly recall the facts. We consider hydrogen atoms  in  the
ground state which move in the $y$ direction with velocity $\nu $  through  an  inhomogeneous magnetic field ${\bf B}$ produced by  a  Stern-Gerlach  magnet \cite{Ger22} - \cite{Wre}, \cite[Sec.~III.10]{Mes}. The
magnet is positioned so that along the path of the atoms both ${\bf B}$  and  grad{\emph B}$_{z}$
point in the $z$ direction. This is then the ``spin-reference axis'',  or  simply
the ``axis'' of the apparatus. The  hydrogen  atom  has  a  permanent  magnetic
moment $\vec{\mu }$ which comes from the magnetic dipole moment  of  the  electron,  the
contribution of the proton being  negligible.  Therefore  it  is  mainly  the
electron that interacts with the magnetic field, and it is the  spin  of  the
electron
\begin{displaymath}(2.3)\hsp
{\bf s} = - (m/e)\vec{\mu }
 \end{displaymath}
\noindent that determines the precession in the magnetic field. Accordingly, in  (2.3)
$m$ is the electron mass. We might thus just speak of electrons moving  through
the Stern-Gerlach magnet. We note,  however,
that the Stern-Gerlach magnet does not work for free electrons. This  is  due
to   the   Lorentz   force   and   to   the   spreading   of   the   electron
wavepacket \cite{Pauli29} - \cite[p.~214]{Mot}. If we want to perform the Stern-Gerlach  experiment  for
free electrons we may scatter  electrons  by  atoms \cite[Chap.~IX]{Mot}, \cite{Lam}. We  call  such
devices Stern-Gerlach-\emph{type} apparatuses.

The experimental arrangement is chosen so that the laws of classical mechanics  and
electrodynamics predict that the atom when it has spent the time $\Delta t$  in  the
Stern-Gerlach magnet will be deflected along the $z$ direction by the angle
\begin{displaymath}(2.4)\hsp
\alpha _{z} = p_{z}/p_{y} = \mu _{z}(\partial B_{z}/\partial z) \Delta t/p_{y} ,
 \end{displaymath}
\noindent where $p_{y}$ and $p_{z}$ are the momentum components of the atom,  and $\mu _{z}$  is  the $z$
component of $\vec{\mu }$. When a beam of atoms goes through  the  Stern-Gerlach  magnet
with the spins of the atomic electrons initially oriented at  random, $\mu _{z}$  can
take on all values between $+\mu $ and $- \mu $, and the deflection angles  can  take  on
all values between the corresponding extreme values $\pm \mu (\partial B_{z}/\partial z)(\Delta t/p_{y})$. On the
screen behind  the  magnet, the sensitive region,  one  would  therefore  observe  one  single  spot
elongated along the $z$ direction. What is actually observed, however, is  two
separate spots corresponding to the above two extreme values of $\alpha _{z}$, with
\begin{displaymath}\hspp
\mu  = \frac{e\hbar}{2m} =\frac{ec}{4\pi} \lambda_{\rm C}
 \end{displaymath}
\noindent corresponding to the electron spin value $s = \hbar /2$ in formula (2.3). The upper
spot on the screen thus corresponds to spin-up electrons and the  lower  spot
to spin-down electrons with respect to the axis of the apparatus.

So far our  considerations  of  the  Stern-Gerlach  experiment  have  been
independent of our interpretation since they have been formulated with  beams
consisting of many atomic electrons. Difficulties arise when the behaviour
of the individual electron wavepackets of the beam is considered. Any single
wavepacket is fanned out into two coherent parts, one corresponding to spin up and
the other corresponding to spin down, and it covers both the  upper  and  the
lower path in portions that can  be  calculated  by  the  standard  formulas (cf. formula (2.5) below).
According to the orthodox version of the Copenhagen interpretation it is only in
a subsequent measurement, for example, when a black spot at  the  proper  ``up''
position on the screen is observed, that the packet contracts and is  reduced
to a pure spin-up eigenpacket. When we choose not to look at the  measurement
device, no reduction can occur \cite{Wig}.

In the Copenhagen interpretation the Stern-Gerlach experiment is called  a
meas\-urement  of  the  initial $z$  component  of  the  spin  of   the   atomic
electron \cite{Mer}, \cite[p.~593]{Bohm51} In the present interpretation  this  is  different.  Consider  an
atomic electron which is initially described by a wave  function  that  is  a
product of a spatial and a spin function.  Such  a  wavepacket  never  has  a
definite pointlike position, but it always has a definite spin component, in the  sense
that the wave function can always be written as a  spin-up  eigenfunction  of
the spin-component operator $s_{z^\prime }$, with \emph{some} axis $z^\prime $, which of course need  not
coincide with the axis $z$ of the Stern-Gerlach apparatus.  This  is  connected
with the fact that the group SU(2) is locally isomorphic to O(3). We may call
the direction of the axis $z^\prime $ the spin direction of  the  electron  before  it
entered the apparatus. 

Now the inhomogeneous magnetic field accomplishes that the eigenfunctions of the spin component of the incoming electron become spatially separated: the spin-up component eigenfunction goes upward (say) and the spin-down eigenfunction downward. Both functions then enter the sensitive screen in which contraction can occur. It is only when the function contracts at a cluster in the upper region of the screen, say, that the electron has become a pure spin-up electron with respect  to  the  apparatus axis $z$, whatever the electron's initial spin direction $z^\prime $. Because of
total angular momentum conservation, the angular momentum of the apparatus is
thereby also changed. This has been verified experimentally in  the  case  of
photon polarization apparatuses, which  in  principle  function  like  Stern-Gerlach apparatuses \cite{Bet}.

In our interpretation, the operation of  the
Stern-Gerlach apparatus on an individual incoming electron and  the  observation of its respective final position on the  screen  is not a  measurement  of  the
electron's initial spin  component.  We  may  indeed  use  the  Stern-Gerlach
apparatus for such a measurement, but not  without  further  steps:  a  large
number of equal electrons must be given. Let the electrons enter  the  magnet
one after the other. Then the direction of the axis of the apparatus must be
varied, and from the abundance ratios of  the  up  and  down  spots  for  the
various chosen directions the original spin direction can be  derived,  using
the standard formulas of quantum mechanics. For example, let the spin of  the
incoming electrons be restricted to lie in the $x$-$z$ plane perpendicular to the
direction of motion $(y$ axis). The probability of inducing a spot in the  ``up''
or ``down'' position, respectively, is then given by
\begin{displaymath}(2.5)\hsp
P_{\mbox{up}/\mbox{down}} =\mbox{$\frac{1}{2}$}(1 \pm  \cos \vartheta )
 \end{displaymath}
\noindent where $\vartheta $ denotes the angle between the spin axis of the incoming electron  and
the axis of the apparatus. The ratio of  the  corresponding  abundances  will
give $(P_{\mbox{up}})/(P_{\mbox{down}}) = \tan ^{-2}(\vartheta /2)$, hence $\vartheta $  and  the  spin  direction  of  the
incoming electrons becomes known. Alternatively, we may rotate the  apparatus
round the $y$ axis until a position is obtained where  only  spots  at  the   ``up''
position are observed. This signifies $\vartheta =0$ in (2.5), and the apparatus axis
coincides with the spin direction of the incoming electrons.

If  the  electron  were  a
classical gyroscope, the measurement of the place where it  hits  the  screen
behind the magnet could be considered as a measurement of  its  initial  spin
component  (formula  (2.4)).  However,  the  electron  is  not  a  classical
gyroscope, and to describe the quantum situation  in  the  same  way  as  the
classical situation is misleading. In our interpretation, expressing  a  wave
function as a superposition of certain eigenfunctions in general is  no  more
than a mathematical procedure. Only in special physical situations  like  the
one within the Stern-Gerlach magnet are the spatially separated eigenfunctions physical parts of a wavepacket.

\vspace{20pt}
\noindent {\bf 2.3~~Micro-Determinism and Macro-Indeterminism}
\medskip

\noindent
As explained in Sec.~2.1 we have micro-determinism but macro-indeterminism (unpredictability). Of course macro-determinism  is only possible with micro-determinism. 

Both determinism  and  indeterminism  are  compatible  with  realism.  Determinism means a programme. It means  the  expectation that as science advances we will be able to make more and more phenomena
predictable, by means of laws of nature, and that this process is infinite.

It is well known that Einstein favoured determinism \cite{Ein24} - \cite{Spe}: ``At any  rate
I am convinced that \emph{he} is not playing dice'' \cite{Ein26}. And  this  has  often  been
considered as Einstein's  main  criticism  of  quantum  theory.  However,  as
indicated by Einstein \cite{Ein50} and emphasized by Pauli in a letter to Born \cite{Pauli54}:
\begin{quote}
Einstein (as he explicitly  repeated  to  me)  does  not  consider  the
concept of ``determinism'' to be as fundamental as it is frequently  held
to be $\ldots 
$ Einstein's point of  departure  is  ``realistic''  rather  than
``deterministic.''
\end{quote}
What Einstein had in mind is obviously micro- as well as macro-determinism.

We comment on this topic here because the Copenhagen interpretation takes the stand of strict indeterminism, micro and thus also macroscopic, and apodictically decrees
a definite limit to the described  process of the deterministic  programme.  It
maintains that the probabilities in quantum mechanics  are  unlike  those  in
statistical mechanics and can never be explained by  underlying determining processes that would specify the physical situation in more detail.  The
probability statements in quantum mechanics, according  to  this  interpretation, are the last word. Even if we knew all the laws and all the wavepackets
in the world, we would not be able,  in  principle,  to  calculate  the  exact
future result of an individual measurement. Only in some degenerate cases can
we obtain probabilities that reach the value 1 and thus give certainty.  In
general, identical initial conditions do not lead to identical results.

Why does the Copenhagen interpretation assume such  an  extreme  position?
Admittedly,  so  long as no  theory  is available that  can  specify  the
hypothetical underlying processes postulated in the  deterministic  attitude, it might seem reasonable, from the viewpoint of economy of concepts,  to
eliminate the concept of these processes  altogether.  This  would  give  the
additional bonus that the indeterminacy no longer points at  an  inability
of the quantum theorists to build a complete theory,  but  is  a  property  of
nature.

It seems that the attitude of the Copenhagen school received additional support
from von Neumann's demonstration \cite[Secs.~IV.1 and IV.2]{Neu} that some  basic  features  of  quantum
mechanical states are incompatible with the introduction of additional hidden
variables besides $\psi $ in order to further specify the physical  situations  and
to restore micro-determinism. This statement seems to have been taken to mean
that no deterministic theory at all  is  possible.  When  Bell  examined  the
case \cite{Bel64}, \cite{Bel66} he found  that those basic features of the quantum  mechanical states which von Neumann postulated also for the  states  in  a  hidden-variable theory, are actually more than can reasonably be postulated in  such
a theory. Thus, von Neumann's proof, although  mathematically  correct,  leaves
the real question  untouched  and  does  not  exclude  deterministic  hidden-variable theories. The same conclusion had been reached by Grete Hermann in 1935 \cite{Her35}.

On the other hand, Bell's investigations revealed that any
hidden-variable theory  which  after  averaging  over  the  hidden  variables
reproduces the formulas of quantum mechanics must  have  a  grossly  nonlocal
structure. A more detailed account of this specific aspect will be  given  in
Chap.~4. Here, the essential lesson is that von Neumann's  proof  does  not
exclude micro-deterministic theories, and  the
apodictic exclusion  of  micro-determinism  in  the  Copenhagen  interpretation  is
unjustified. 

Let us, therefore, consider what a deterministic programme might look like  in
a realist interpretation. In fact, such a programme is presented
 in  \cite{Jab12}. The point is that in an ensemble of quantum wavepackets their overall phase constants $\alpha$ in the reduction criterion (2.1), (2.2) are taken as \emph{pseudo}random numbers, determined by certain initial conditions, in the spirit of the theory of deterministic chaos. The criterion (2.1), (2.2) then reproduces the Born probability rules in measurements.

\newpage

\noindent {\bf  3~~NONLOCALITY}

\vspace{15pt}
\noindent {\bf 3.1~~One-Particle Nonlocality}
\medskip
\vspace{5pt}

\noindent
In addition to the property of reduction ascribed to the quantum wavepackets in the preceding sections  in order to account for the experimental results, there is the property of nonlocality, which will be described in this and the subsequent sections. 

These quantum properties  may
appear rather strange. This is the price one has  to  pay. It is not to be expected that the difficulties  that  have
beset quantum theory for more than 90 years can be overcome by  some  cheap
trick. On the other hand, the experimental observations now will fit into a coherent picture.
\vspace{1pt}

We will now consider more closely how the black spot in an emulsion screen
is brought about. Consider a beam  of  electromagnetic
radiation falling onto a screen. Let us first treat the beam classically as a
continuous field with the time-averaged energy flux  density  $\bar S  = \epsilon _{o} c  \overline {E^2}$  in
${\rm W}/{\rm m}^{2}$, say. The registering screen, on the other hand, is conceived to consist
of atoms in the sense of quantum mechanics. The incoming radiation will  then
cause black spots at certain places on the screen, and their density classically is proportional to the energy flux density of  the  radiation  at  these
places. The resulting pattern will exhibit a granular structure but this in itself does
not demonstrate that the incoming radiation has particle or  quantum  properties; it only shows that the screen has, which we  have  presupposed  anyway.
The wind is not made of particles only because it causes an  integral  number
of trees to fall in the forest, as Marshall and Santos  put  it \cite{Mars}.

To  see
what may be called the quantum or particle aspect of the  incoming  radiation recall that
 according to Sec.~2.1 the incoming radiation in the limit of low intensity (not energy) does not excite more than one atom at a time, even if the energy $h\nu$ would have sufficed for this. The surplus energy goes into the kinetic energy of the atomic electron \cite{Ein05}. The black spot is the result of a cascade of  processes  which
is initiated by the ionization of an atom. This  atom,  in  order  to  become
ionized,  needs  some  threshold  energy $E_{\rm{thr}}$.  Of  course,  knowing  quantum
mechanics we assume that $h\nu  \ge  E_{\rm {thr}}$, but this is not sufficient for ionization
in the classical picture. Imagine that the atom gets this energy by absorbing
and accumulating all the energy of the  incoming  classical  radiation  which
arrives on its area $\sigma $, like a dust collector that  is  put  in  a  stream  of
polluted air. With the energy flux density of  the  radiation $\bar S$  the  energy
accumulated by the atom during the time $t$ is $E$  = $\bar S  \sigma ${\emph t}.  When  the  atom  has
accumulated the energy $E_{\rm {thr}}$ it becomes ionized and initiates the formation of
the black spot. The accumulation time needed for this is
\begin{displaymath}\hspp
t_{\rm {acc}} = E_{\rm {thr}}/(\bar S \sigma ).
\end{displaymath}
\noindent Now, the value for $t_{\rm{acc}}$ according to this formula turns out to be of the
order of hours or years in situations where  effects  are  actually  observed
immediately after the arrival of the radiation. 

As an example consider the  interference experiments of Reynolds et al. \cite{Rey69}, \cite{Rey66}. Light of wavelength $\lambda=4.358\times {10}^{-7}$ m  and $\Delta\lambda=10^{-12}$ m passing a Fabry-Perot interferometer produced an interference pattern on the multi alkali cathode of an image intensifier tube. The entire interference pattern had an area of $1.5\times3.1$ mm$^2 $= $47\times10^{-6}$ m$^2$. From the size of the coherence length of the light (cf. Appendix A, (A25) f.) and from the density of the excited atoms in the source it could be concluded that there was only one photon in the apparatus at a time. The wavelength of $4.358\times {10}^{-7}$ m means photons of energy 2.85 eV. So, a light energy of $15\times2.85$ eV = 43 eV/s $(\bar S = 1.5\times10^{-13}$ W/m$^2$) passed over the area of the pattern.

Each photon wavepacket covers the whole interference pattern. With a minimum linear size of the photon-absorbing molecule of $8\times10^{-10}$ m, the energy passing per second over the area of a molecule located in an interference maximum ($\approx2\times$ average energy) may be approximately $2\times15\times(64\times10^{-20}/47\times10^{-6})\times2.85$ eV/s. With a threshold energy of 1.36 eV ($\lambda=0.9\times10^{-6}$ m) necessary for the emission of an electron \cite{Fin}, the accumulation time is $1.36/(11.6\times10^{-13})$ s = $1.17\times10^{12}$ s = 37500 years!

The interference pattern would thus only  appear
after centuries, but then it would  appear  fully  in  one  flash.  Actually  the
interference patterns were obtained in 15 seconds, and in  various  runs  with
reduced exposure time the authors verified  that  the  pattern  is  built  up
gradually as time proceeds. The first black  spots  are  induced  immediately
after the arrival of the radiation. In fact, upper limits of the time lags between the arrival of the radiation and the ionization of the molecule as short as $3\times10^{-9}$ s \cite{Law} and $10^{-10}$ s \cite{For} have been reported. Of  course
the total energy absorbed by all atoms during the whole accumulation time  is
the same in both cases. The difference is that classical theory would have it
absorbed all in the last moment (at $t_{\rm{acc}})$, whereas experiment shows  that  it
is absorbed in many small portions distributed over the accumulation time.

Even with somewhat different assumptions one arrives at the
same conclusion, as already shown by Campbell \cite{Cam}, Planck \cite{Pla21}, Mandel \cite{Man76}, Paul \cite{Pau04}, and others \cite{Whe}. It is thus safe to conclude that the energy for ionizing the atom  is  not
the energy contained in the cylinder that the atom has cut out of  the  field
up to the moment of its ionization.

Now, the field quanta are spatially extended wavepackets of mean energy $\langle E\rangle =h\langle\nu\rangle$, and in the act of measurement they \emph{contract} to an effectively pointlike region (Sec.~2.1).  And what we have here concluded for the photon wavepackets we conceive to hold for any wavepackets.

This contraction is different from the shrinking (or spreading) of the wavepacket governed by the Schr\"odinger equation (Appendix~A). In order to account for the situations described above, the contraction must occur with superluminal velocity in the reference system of the measurement apparatus:

With the dimension of the interference pattern ($\le$ lateral dimension of the wavepacket) in \cite{Rey69}, \cite{Rey66} of 3.1 cm and a time lag between the arrival of the radiation and the ionization of the atom of $10^{-10}$ s \cite{For} the contraction velocity would be 3.1$\times 10^8$ m/s. This is only slightly larger than light velocity, but wavepackets with larger dimensions are easily met: 

In fact, the one-particle wavepacket may consist of several non-overlapping spatially well separated parts. In \cite{Kwi95}, for example, each photon of a low intensity radiation beam was split by a beam splitter into two separate parts, and either part was directed into a different detector. No coincidence counts between the two detectors could be observed. That is, if the photon is detected in detector 1 it immediately contracts to a small spot in that detector, so that there is no longer any part of the wavepacket at detector 2. The detectors were separated by 20 m. With a time lag between arrival and detection of $10^{-10}$ s \cite{Pan} ($0.074\times 10^{-10}$ s \cite{Hal}) the contraction had to occur at a velocity of 20  $c$ (270 $c$). 

In other words, the count in one detector effects that there is no count in the other detector. In view of the multi-particle nonlocality to be considered in Sec.~3.2 we regard this as one-particle nonlocality \cite{Note6}. And we ascribe this one-particle nonlocality also to massive wavepackets, for example to the atoms in the Stern-Gerlach apparatus (Sec.~2.2) and the neutrons in the single-crystal neutron interferometer \cite{Bon}.

There are no parts of an electron wavepacket, say,  which could dynamically interact with each other. We want to call this \emph{internal structurelessness} of the wavepackets. 
In fact, we may regard the success of  the  Schr\"odinger  equation  as a 
confirmation  of  the  absence   of   dynamic self-interactions (i.e. those that are explicitly represented by interaction terms in the Schr\"odinger equation) \cite{Schr27}.  Consider   the
Schr\"odinger equation for an electron in an electrical potential $V\!({\bf x},t)$
\begin{displaymath}(3.1)\hsp
i\hbar  \frac{\partial \psi ({\bf x},t)}{\partial t} = - {\hbar ^{2}\over 2m}\Delta \psi ({\bf x},t)  - eV\!({\bf x},t)\,\psi ({\bf x},t) .
 \end{displaymath}
\noindent Let us for the moment regard the  quantity $\rho ({\bf x},t) := - ~e| \psi ({\bf x},t)| ^{2}$  as  the
charge density of the electron, and let us write the potential $V\!({\bf x},t)$ as  the
sum of two terms
\begin{displaymath}(3.2)\hsp
V\!({\bf x},t) = V_{\rm o}({\bf x}) + V_{\rm e}({\bf x},t) ,
 \end{displaymath}
\noindent where $V_{\rm o}({\bf x})$ is the potential produced by the  atomic  nucleus  (proton)  plus
some outside charges, and $V_{\rm e}({\bf x},t)$ is the potential  produced  by  the  charge
distribution of the electron itself. This can be written as
\begin{displaymath}(3.3)\hsp
V_{\rm e}({\bf x},t) =\frac {1}{4\pi \epsilon _{\rm o}} \int \frac{\rho ({\bf x^\prime },t)}{ | {\bf x} -  {\bf x^\prime | }} d^{3}x^\prime  .
 \end{displaymath}
\noindent $V_{\rm e}({\bf x},t)$ represents some action of the electron on itself. Inserting (3.2) and
(3.3) into (3.1) leads us to the nonlinear integro-differential equation
\begin{displaymath}\hsp
i\hbar  \frac{\partial \psi ({\bf x},t)}{\partial t}
 = - \frac{\hbar ^{2}}{2m}\Delta \psi ({\bf x},t) - eV_{\rm o}({\bf x}) \psi ({\bf x},t)
\end{displaymath}
\begin{displaymath}(3.4)\hspace{90pt}
+ \quad  \psi ({\bf x},t) \frac{e^2}{4\pi \epsilon _{\rm o}}  \int \frac{| \psi ({\bf x^\prime },t)| ^2}{| {\bf x} - {\bf x^\prime | }}d^{3}x^\prime , 
\end{displaymath}
\noindent which differs from the familiar  Schr\"odinger  equation  by  the  last,  self-interaction term. On the other hand, it is the familiar Schr\"odinger equation (3.1),
and not Eq. (3.4), that gives the correct results, for example  for  the
hydrogen eigenfunctions.

Of course, the absence of dynamic interactions between  spatial  parts
of a wave\-packet does not exclude the  existence  of  recoil  effects  of  the
emitted radiation on the emitting wavepacket as a whole, as it is  considered
by Barut and collaborators  in  their  approach  to  quantum  electrodynamics
without canonical quantization \cite{Bar}.

In this context we may quote Lorentz \cite{Lor}:

\begin{quote}
In speculating on the structure of these minute particles we  must  not
forget that there may be many possibilities not dreamt of  at  present;
it may very well be that other internal  forces  serve  to  ensure  the
stability of the system, and perhaps, after all, we are wholly  on  the
wrong track when we apply to the parts  of  an  electron  our  ordinary
notion of force.
\end{quote}

and also Dirac \cite{Dir38}:

\begin{quote}
it is possible for a signal to be transmitted faster than light through
the interior of an electron.  The  finite  size  of  the  electron  now
reappears in a new sense, the interior of the electron being  a  region
of failure, not of the field equations of electromagnetic  theory,  but
of some of the elementary properties of space-time.
\end{quote}

The contraction in any case shows that the quantum wavepacket  must be a special object of its own kind. We have already stated that a one-quantum wavepackets can never induce more than one effect at a time. This is just another aspect of the here described contraction to one single place.
Does the contraction occur even with infinite velocity? In which reference system? \cite{Simul}. In any case a velocity inside  the  wavepacket has no direct physical meaning: there cannot be observers sitting inside the wavepacket at determined positions and reading off synchronized clocks. Might the failure of the elementary properties of spacetime mentioned by Dirac go so far that there is no space at all (i.e. no distance to travel) inside the wavepacket, as speculated in the 1996 version of this article? Note also Bell's remark in \cite{Bel86}:
\begin{quote}
Behind the apparent Lorentz invariance of the phenomena, there is a deeper level which is not Lorentz invariant.
\end{quote}

\bigskip
\noindent {\bf 3.2~~Entanglement and Multi-Particle Nonlocality}
\medskip

\noindent
In the preceding  sections  single elementary particles or single field quanta were considered. Now we  shall
extend our considerations to include systems of several particles. 

Multi-particle systems in quantum mechanics are described by a configuration 
space wave function 
\begin{displaymath}
\hspace{130pt} \Psi ({\bf x}_{1},{\bf x}_{2},\ldots 
,{\bf x}_{N},t). 
\end{displaymath}
This does not prevent us  from  maintaining a realist interpretation. Essentially, what the configuration-space formalism effects is to introduce correlations between wavepackets, as we shall see below (and in Chap.~4).

\vspace{5pt}
It  is  of  particular  importance  to  distinguish  between  multi-particle wave functions that can be written (perhaps after some transformation to a different system of eigenfunctions) as a product of one-particle functions and those that cannot.

In those that can, the particles are independent of each other.
In those that cannot, the particles, viz. the wavepackets representing them, in a sense are dependent on each other and are called \emph{entangled}, a term coined by Schr\"odinger \cite{Schr35c}, \cite{Schr35d}, and they form a \emph{system of entangled wavepackets}.

We will consider  these entangled wavepackets  more closely. A two-particle entangled wavepacket, for example, may be written in the form
\begin{displaymath}(3.5)\hspace{42pt}
\Psi = a_1 \,\varphi_1(x_1; u_1, m_1)\,\varphi_2(x_2; u_2, m_1)
+a_2 \,\varphi_1(x_2; u_2, m_2)\,\varphi_2(x_1; u_1, m_2).\hspace{20pt}
\end{displaymath}
The function $\varphi_1$ (in both  parts) represents one particle, and $\varphi_2$ the other. The parameters $u_1,\,u_2$ determine the spatial shapes of the wavepackets, which includes their centre position and their width. The parameters $m_1,\,m_2$ are additional properties, for example spin or polarization components \cite{Kwi92} or energy and arrival time \cite{Kwi93}. The time variable $t$ is the same for all, and is omitted. For simplicity only one spatial dimension $x$ is considered.

It may happen that the wavepackets $\varphi_1$ and $\varphi_2$ develop so as to occupy disconnected regions $R, L$ of space, that is, the distance between their centres being large compared with their widths. Let us write this as
\begin{displaymath}(3.6)\hsp
\Psi = a_1 \varphi_1(R, m_1)\,\varphi_2(L, m_2)
+a_2 \varphi_1(R,m_2)\,\varphi_2(L,m_1).
\end{displaymath}
The two wavepackets are still entangled because both components $m_1$ and $m_2$ appear in each paket. In the realist interpretation Eq. (3.6) means that particle 1 neither has the property $m_1$ nor the property $m_2$, and the same holds for particle 2. Considering the experiments where such processes occur \cite{Fre} - \cite{Hag} it seems that a necessary condition for independent particles getting entangled is that the wavepackets must overlap to some degree
\begin{displaymath}(3.7)\hsp
\varphi_1(x,t)\; \varphi_2(x,t)\neq0
\end{displaymath}
at some instant of time. It does not seem that dynamic interactions are sufficient for establishing entanglement. An electron and a proton, for example, in principle interact with each other via the Coulomb force even if the proton is in the Andromeda galaxy and the electron on Earth. I do not think that there is any physicist who would assume that the two are entangled, with the properties of entanglement described below, if they never satisfied condition (3.7). An exact mathematical specification of the entangling condition is not the concern of the present article. Its concern is only to emphasize that it is a real physical process, occurring with real physical wavepackets, not just a loss of `information' or `knowledge'.

How do entangled wavepackets become disentangled? This also happens  in the reduction  [49], sketched in Sec.~2.1. We here extend that sketch and mention that when one of the overlapping wavepackets on his part is entangled with another one, these two become disentangled.

\vspace{5pt}
We now come to the most remarkable feature of entangled wavepackets. As early as 1932   \cite{Ehr}   Ehrenfest  had emphasized that the mere  use  of  a  non-product configuration space  wave
function  implies  some  kind  of  sinister
action at a distance. Indeed, the conditional  probability  that  particle  1
acts (induces an effect) in $d^{3}x_{1}$ about ${\bf x}_{1}$, if particle 2 acts in $d^{3}x_{2}$ about ${\bf x}_{2}$, is
\begin{displaymath}(3.8)\hsp
P({\bf x}_{1}| {\bf x}_{2},t)d^{3}x_{1} = {| \Psi ({\bf x}_{1},{\bf x}_{2},t)| ^{2}d^{3}x_{1} d^{3}x_{2}\over d^{3}x_{2} \int | \Psi ({\bf x}_{1},{\bf x}_{2},t)| ^{2}d^{3}x_{1}} ,
 \end{displaymath}
\noindent and this depends on ${\bf x}_{2}$, that is, on the result of a  simultaneous  action  of
 the second particle.  On  the  other  hand,  the  distance
between the particles in ordinary space $| {\bf x}_{1} - {\bf x}_{2}| $ may be many kilometers  long, and the correlations are independent of whether or not there are dynamic interactions between the two particles. With
a product form of $\Psi(1,2)$ the probability (3.8) would be independent of ${\bf x}_{2}$, and  there
would be no  correlations.

Indeed, the two  events may occur at spacelike intervals of spacetime, that is, if a signal from one event to the other were to connect the two, this signal would have to proceed with superluminal speed. There are many experimental confirmations of this. In  \cite{Yin} a lower bound of that speed was found to be four orders of magnitude larger than the speed of light. As in the case of one-particle nonlocality an interesting question is: can there be a reference system where  the speed is infinite \cite{Simul}? 

What has been said here concerning the case of two wavepackets holds  also for $N$ entangled one-particle wavepackets \cite{Note5}. 

A macro-body, which in the classical Newton-Euler description is a system of point particles, in the present description is a system of wavepackets, where in addition to the dynamic interactions between them there are entanglement and disentangling contractions between them. 

Why are the correlations between spacelike separated events (spacelike correlations, for short) such a remarkable feature? Could it not be that the spacelike distance is caused by so far unknown common causes in the overlap of the past light cones of the events, like the consecutive illumination of a series of places on the Moon by searchlight pulses emitted from a place on the earth? In Chap.~4 we shall return to this question and show that  the observed spacelike correlations predicted by quantum mechanics can lead to to a violation of the Bell inequality \cite{Gis}, which excludes common causes in the past.

We will call such special spacelike correlation, ascribable to entanglement, {\emph multi-particle nonlocality}. Therefore the range of the nonlocality in our theory is  limited to the extension of the system of entangled wavepackets. And one-particle and multi-particle nonlocality are conceived to be basically of the same nature, and we will   just speak of \emph{nonlocality} in both cases. 

As we also shall see in Sec.~4.2  this nonlocality does not lead to a superluminal connection between cause and effect and does not allow superluminal signaling. 
\newpage

\noindent {\bf 3.3~~Similar (Identical) and Condensed Wavepackets} 
\medskip

\noindent
The case of identical particles deserves special consideration. Following Dirac's book \cite{Dir67} we call the particles `similar' rather than `identical'. One
reason for this is the ambiguity  in  the
meaning of the word identical. We may say ``Lord Kelvin and  William  Thompson
are identical'', which means that the two are one and the same person; but  we
may also speak of identical twins, which means two different persons. 

In quantum mechanics the effects of similarity go beyond those met in classical mechanics \cite{Jab10}. In classical  mechanics  similarity  or  indistinguishability always  means  essentially  the  indistinguishability  of  equal
billiard balls. Imagine one billiard ball in your right hand and the other in
your left hand. You are blindfolded and somebody else takes the balls out  of
your hands, then puts them back. If you look at them again  you  cannot  tell
whether or not they have been interchanged. However, if you were  not  blindfolded, you could follow their paths and decide which ball  was  initially  in
your right hand and which was in your left  hand.  Or,  imagine  a  situation
where the balls are in rapid movement around  each  other  so  that  you  can see
nothing but a fuzzy cloud of  whirling  balls.  Nevertheless,  when  you  are
allowed to use more refined methods of observation, you will always be able to
follow the paths of the balls individually.

Clearly,  in  classical  physics
with its mass points representing the centres of mass of impenetrable bodies,
there is no indistinguishability that could not be resolved in principle.  In
quantum physics this is no longer true. There are situations, associated with
wave function overlap, where the observer is unable, in principle, to distinguish the particles, in the sense that he is unable to follow the path  of  a
given particle unmistakably through all processes. In other  words,  for  him
the particles lose their individuality. In the realist  programme this is conceived not as  any  incapability  on  the  part  of  the
observer but as an objective  physical  fact.

The wave function representing a system of similar particles must be symmetric or antisymmetric under the exchange of function parameters \cite{Jab10}. Consider the product function
\begin{displaymath}(3.9)\hsp
\Psi=\varphi_1(R\uparrow)\,\varphi_2(L\downarrow) \hspace{10pt}
\end{displaymath}
representing one particle with spin up ($\uparrow$) in the spatial region $R$ and another particle with spin down ($\downarrow$) in the separate region $L$. Considering the case where there has been the possibility of spin flip in a previous overlap of the two wavepackets so that they can no longer be identified by their spin components, function (3.9) can be superposed with the exchange function
\begin{displaymath}(3.10)\hsp
\Psi=\varphi_1(L\downarrow)\,\varphi_2(R\uparrow) \hspace{10pt},
\end{displaymath}
and we obtain the (anti)symmetric function 
\begin{displaymath}(3.11)\hsp
\Psi=\varphi_1(R\uparrow)\,\varphi_2(L\downarrow) \pm
\varphi_1(L\downarrow)\,\varphi_2(R\uparrow) \hspace{10pt}.
\end{displaymath}
However, as emphasized by Ghirardi et al. \cite{Ghi02}, though this is no product state, it does not mean entanglement, with the property of nonlocality. The single particles  all have definite properties of their own. True,  in function (3.11) neither particle 1 nor particle 2 have definite spin values; nevertheless there is one particle in $R$, \emph{whichever of the two it is}, with definite spin up  and another particle    in $L$ with definite spin down. This does not suffice to violate the Bell inequality (Appendix C, Eq. (C12) and following).

An entangled wavepacket is
\begin{displaymath}(3.12)\hsp
\Psi=\varphi_1(R\uparrow)\,\varphi_2(L\downarrow) \pm
\varphi_1(L\uparrow)\,\varphi_2(R\downarrow) \hspace{10pt}.
\end{displaymath}
This function indeed means that the particle in region $R$, \emph{whichever of the two it is}, has neither the property $\uparrow$ nor the property $\downarrow$, and the same holds for the particle in region $L$. Eq. (3.12) is the type of function which usually is the base  of the discussions of the experiments designed to verify a violation of the Bell inequality (e.g. \cite{Kwi93}, \cite{Bow} , \cite{Shi}). In fact, Eq. (3.12) represents 2 of the 4 ``Bell states'', which can lead to maximal violation (cf. Sec.~4.3). The 2 others are
 \begin{displaymath}(3.13)\hsp
\Psi=\varphi_1(R\uparrow)\,\varphi_2(L\uparrow) \pm
\varphi_1(R\downarrow)\,\varphi_2(L\downarrow) \hspace{10pt}.
\end{displaymath}

\noindent
Thus, (anti)symmetrizing a product of two similar one-quantum wavepackets by itself does not guarantee entanglement with its nonlocality. This is an exception of the statements made in Sec.~3.2.
\vspace{20pt}

\medskip

\noindent
Another interesting special case are the condensed wavepackets, which are formed when the similar wavepackets, such as $\varphi_1$ and $\varphi_2$ in Eqs. (3.12) or (3.13), are equal in all respects. Any configuration space function, we recall,  can be  expanded  in
terms of a complete set of one-particle functions in ordinary space $\varphi _{r_{i}}({\bf x},t)$ \cite[Sec.~VII.6]{Mes}.
\begin{displaymath}(3.14)\hspace{40pt}
\Psi _{\rm{SA}}({\bf x}_{1},\ldots 
,{\bf x}_{N},t) = \sum^{}_{r_{1},\ldots 
,r_{N}}c(r_{1},\ldots 
,r_{N},t)\; \varphi _{r_{1}}({\bf x}_{1},t)\cdots \varphi _{r_{N}}({\bf x}_{N},t) .
 \end{displaymath}
\noindent The ${\bf x}_{i}$ may include the spin components. The functions $\varphi _{r_{i}}$ may be taken  as
functions  of   time  too,  as  in  the  Dirac  or  interaction  picture.
$c(r_{1},\ldots 
,r_{N},t)$ is then the transformed wave  function.  In  the  case  of  similar
wavepackets  it is completely \mbox{determined} if we specify
the number of times each of the arguments $r_{1}, r_{2}, r_{3},\ldots 
$ occurs in it.  These
 are the occupation numbers $n_{r_{i}}$,  and  the  set $| n_{1},n_{2},n_{3},\ldots 
,t\rangle $  is
\mbox{another}  representation  of  the  wave  function $\Psi _{\rm{SA}}({\bf x}_{1},\ldots 
,{\bf x}_{N},t)$.  The  set \linebreak[4]
$| n_{1},n_{2},n_{3},\ldots 
,t\rangle $ is the state (vector,  wave  function)  in  the  occupation-number, $N$, or Fock representation \cite{Foc32b} - \cite{Schi68}. The change in occupation  numbers
in the course of time, due to interactions, is described by  the
creation and annihilation operators {\emph a}$^{\dag}$  and {\emph a}. This works for bosons $(n_{r_{i}} =
0,\ldots 
,\infty )$ and, with minor additions, for fermions $(n_{r_{i}}= 0, 1)$.

The occupation-number representation is well suited for our interpretation
because numbering of  quanta  within one wavepacket $\varphi_{r_i}(\bf x_i,t)$ is  not  even  mentioned  in  it.  The  term ``occupation number'' is, however, likely to mislead one to  think  that  the quanta and the wavepackets filled with them  are two different things. In  our
interpretation there are only wavepackets and therefore  we  choose  a  different
formulation: the one-particle basis functions are wavepackets, and occupation
numbers of 2 or more mean that two or more wavepackets have \emph {condensed} to form
one single wavepacket, even though this is still normalized to 1 in the current formalism. Thus boson wavepackets can condense, but fermi packets
are excluded from doing so. This is our formulation of  the  Pauli  exclusion
principle.

The inverse process we call  \emph{decondensation}.  The  change  in
occupation numbers will then be described  as  condensation  and
decondensation of wavepackets. Condensation  and  decondensation  occur  only
between Bose but not between Fermi packets. In Chap.~5 we shall show how these processes lead to a new derivation
of the quantum-statistical Bose and Fermi distributions and to the corresponding fluctuations in the wavepacket picture.

Another manifestation of the multi-quantum condensed wavepackets is the `photon bunching' in thermal radiation; that is, the observation that the photon coincidence rate in small  temporal coincidence windows is higher than can be explained by random coincidences \cite{Han}, \cite{Pur}, \cite{Man95}. By now, multi-quantum wavepackets have also been isolated experimentally: in the Bose-Einstein condensates of atoms \cite{Dav}, \cite{And} and of photons \cite{Kla}.  

In Sec.~5.2 we shall also see that there are many other objects, conceived long ago in radiation theory, that are very similar to the condensed packets and may be taken to be  just other aspects of them.
\newpage

\noindent {\bf  4~~NONLOCALITY AND SUPERLUMINAL SIGNALING}
\bigskip

\noindent
In this chapter we return to the nonlocality described in Secs.~ 3.1 and 3.2 and  proof that the observed spacelike correlations cannot be ascribed to common causes in the past, and that they do not allow superluminal signaling.
\medskip

\noindent {\bf 4.1~~The EPR Problem and Nonlocality}
\medskip

\noindent
As an introduction we consider the problem of ``simultaneous elements of reality'' and of action at a distance formulated by
Einstein, Podolsky and Rosen (EPR) in 1935 \cite{Ein35}. This paper in its time had raised new interest in the nonlocality problem.
The title of the EPR  paper  is 

 ``Can  Quantum-Mechanical  Description  of
Physical Reality Be Considered 

 \hspace{5pt} Complete?''

\noindent
The authors wanted  to  demonstrate
that the answer is ``no''. For a physical theory to be complete it  is  necessary
that 

``every element of the physical reality must have a  counterpart  in  the
physical 

\hspace{5pt}theory'', 

\noindent
and reality is characterized by  the  following  sufficient
criterion: 

``If, without in any way disturbing a system, we can  predict  with
certainty (i.e., 

 \hspace{5pt}with probability equal to unity) the  value  of  a  physical
quantity, then there 

\hspace{5pt}exists an element of physical reality  corresponding  to
this physical quantity.''
\smallskip

\noindent
EPR consider two systems, 1 and 2 (imagine two  protons) that  have  interacted
from time {$t=0$}     to   {$t=T$}, after which time they are well separated, so, EPR assume, there is  no  longer  any  interaction
between them. Let the (exactly calculable)  wave  function  of  the  combined
system 1+2 after $T$ be $\Psi(1,2)$. The number 1 stands for all variables  used  to
describe the first system and 2 for those of the second  system.  In  general
the function $\Psi (1,2)$ cannot be  written  as  a  product  of one function $\varphi _{1}(1)$ of the variables 1 and one function $\varphi _{2}(2)$  of
the variables 2, and hence we cannot describe the state in which  either  one
of the two systems is left after the interaction. This  state,  according  to
the Copenhagen interpretation, can only be known by a subsequent measurement:
Let $m_{1}, m_{2}, m_{3},\ldots 
$ be the eigenvalues of some physical quantity
$M$ pertaining to system  1  and $u_{1}(1), u_{2}(1), u_{3}(1),\ldots 
$  the  corresponding
orthonormal eigenfunctions; then $\Psi (1,2)$ can be  expanded  into  a  series  of
these eigenfunctions with coefficients that are functions of the variables 2
\begin{displaymath}(4.1)\hsp
\Psi (1,2) = \sum^{\infty }_{r=1}\zeta _{r}(2) \, u_{r}(1) .
 \end{displaymath}
\noindent Although not necessary for the argument,  we
assume for  simplicity  of  presentation,  that the  eigenvalues  are  discrete.  The  functions $\zeta _{r}(2)$  are  not
normalized and in general are not orthogonal to each other, but this  is  not
relevant here. Suppose that the quantity $M$ is measured on system 1  and  that
the value $m_{7}$ is found. According to reduction,  the first system after  the
measurement  is left in  the  state $u_{7}(1)$ [i.e.  the  first
wavepacket assumes the form $u_{7}(1)$]. Hence the sum (4.1)  is  reduced  to  the
single term $\zeta _{7}(2)u_{7}(1)$, and due to the  product form of this  term  the
second system is left in the state $\zeta _{7}(2)$, apart from normalization.

The set of functions $u_{n}(1)$ is determined by the  choice  of  the  physical
quantity {\emph M}. If, instead of $M$, we had chosen a different quantity $N$,  with  eigenvalues $n_{1}, n_{2}, n_{3},\ldots 
$ and orthonormal eigenfunctions $\upsilon _{1}(1), \upsilon _{2}(1), \upsilon _{3}(1),\ldots 
$
we would have obtained a different expansion
\begin{displaymath}(4.2)\hsp
\Psi (1,2) = \sum^{\infty }_{s=1}\eta _{s}(2) \, \upsilon _{s}(1),
 \end{displaymath}

\noindent where the $\eta _{s}(2)$ are the new coefficient functions. If  the quantity $N$  is now
measured and the value $n_{5}$ is found, then  system 1 is  left  in  the  function
$\upsilon _{5}(1)$ and system 2 in the function $\eta _{5}(2)$.

Therefore, as a consequence of two different measurements performed on the
first system, the second system may be left in states  with  two  essentially
different wave functions. On the other hand, at the time of  measurement  the
two systems, according to EPR, no longer interact, that is, no  real  change  can
take place in the second system as a result of anything that may be  done  to
the first system. Thus it is possible to assign two different types  of  wave
functions, $\zeta$ and  $\eta$, to the same physical reality, namely to system 2  after  the
interaction.

\vspace{5pt}
 \begin{figure}[h]
\begin{center}
\includegraphics[width=0.58\textwidth]{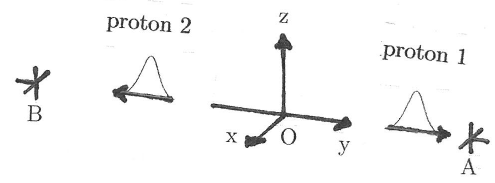}
\caption{Arrangement for a proton spin correlation experiment.}
\end{center}
\end{figure}

It is even possible to choose noncommuting operators $M$ and $N$, operating in
system 1, in such a way that the two sets of wave functions $\zeta _{r}, \eta _{s}$ of  system
2 are discrete eigenfunctions of two noncommuting operators, for  example  of
the two operators $s_{x}$ and $s_{z}$ of the spin component of a proton in $x$  direction
and in $z$ direction  respectively.  Such  a  case  was  first  considered  by
Bohm \cite[p.~614]{Bohm51}. Let the two protons interact at O (Fig.~3) and let the  scattering proceed through an intermediate state of zero total  spin  (singlet  spin
state). The general expression (4.1) in this particular case becomes \cite[p.~562]{Mes}
\begin{displaymath}(4.3)\hsp
\Psi (1,2) =  \frac{1}{\sqrt2}\left( |  + \rangle ^{(1)} |  - \rangle^{(2)}
\; -\;  |  - \rangle^{(1)} |  + \rangle^{(2)} \right)  A(1,2),
\end{displaymath}

\noindent where the spin projections up $|  +\rangle$ and down $|  - \rangle$ refer to an arbitrary axis. {\emph A}(1,2)
is the spatial part and the bracket is the  spin  part  of  the  wave
function. Notice  that  the  spin  part  follows  solely  from  spin  algebra
(Clebsch-Gordon coefficients) and happens to be antisymmetric in the particle
labels, regardless of whether the particles are similar. In the  case
of similar particles the spatial part may take care of the correct symmetry.

After the interaction the protons propagate  with  opposite  momentum $|  {\bf p|  }$
towards the observers A and B respectively. Each observer is equipped with a
Stern-Gerlach-type apparatus (e.g.  a  scattering  device  with  counters;  a
Stern-Gerlach magnet with registering screen would not do in this  case,  see
Sec.~2.2). The (spin-reference) axes of the apparatuses can be oriented in
any direction. Observer A may thus put the axis of his apparatus either in $x$
direction or in $z$ direction, thereby obtaining either the $x$  or  the $z$  spin
component of the first proton. He is then  in  a  position  to  predict  with
certainty, and without in any way disturbing the \emph{second} proton,  either  the
value of the $x$ or the value of the $z$ component of  the  spin  of  the  second
proton. According to the EPR  criterion  of  reality  both  components  must be elements of physical reality. Therefore, the values of both must  enter  into
the complete description of reality. On the other hand, in the formalism of quantum  mechanics
no wave function can contain both an eigenvalue of some  operator $M$  and  an
eigenvalue of an operator $N$ that does not  commute  with  {\emph M}.  Therefore  EPR
conclude that the quantum-mechanical  description  of  reality  by  the  wave
function is not complete.

Einstein seems to consider the elementary particles like dice, where all of their  numbers are permanent properties even if we can see only one number at a time. Bohr, on the other hand, considers the particular properties as coming into existence in interaction with the environment. Our conception, specified in Sec.~2.1, in this point is in line with Bohr's view (cf. the analogy with dough in Sec.~1.4).

In the disputes following the publication of the EPR paper  the  adherents
of orthodox quantum mechanics pointed out that the conclusions  of  EPR  were
only valid  provided  the  two  systems  after  the  interaction  are  truly
independent of each other  in  every  respect.  Present  quantum  mechanics,
however,  conceives the  two  particles  to  be entangled, that is, 
inseparably incorporated into the single wave  function  (4.1),  so  that  we
cannot operate on the one particle ``without in any way disturbing'' the other, and
 only a reduction of the sum to one of its terms by means  of  a reduction can achieve a separation of the two particles.

This is independent of whether $\zeta , u, \eta $ and $\upsilon $ in  expressions (4.1)  and
(4.2) re\-pre\-sent Schr\"odin\-ger  scalar  wave  functions  or  relativistic  Dirac
spinors or other tensors. We may consider the functions $\zeta , u$ in (4.1) or $\eta , \upsilon $
in (4.2) as functions not only of  the  space  coordinates  but  of  time  as
well, with the same time variable $t$ in all functions (cf. Sec.~3.2). Thus,
the moment of the observation at the  one  place  achieves  the  simultaneous
reduction at the other place. Since the time of this  subsequent  observation
is at the observer's disposal, he or she may perform it  with an  arbitrarily  long
delay after the interaction, so that the wave functions of the two systems can
be taken to be separated from one another by an arbitrarily large distance.

On the other hand, things may be arranged so that observer A  operates  on
system 1 such a short time before B  operates  on  system  2  that  no  light
signals could connect these two events. What happens at B  then  depends  on
what A has done in a region that is separated from B by a  spacelike  distance
in spacetime. 

Thus  in  regarding  the  two
systems as independent, EPR are not in  accordance  with  quantum  mechanics;
hence they cannot maintain that it provides only an incomplete description of
physical reality. But then  the  original  question  ``Can  quantum-mechanical
description of physical reality  be  considered  complete?''  is  replaced  by
another question: ``Are these spacelike correlations a feature of physical  reality?''.  Einstein
agreed that the EPR conclusion rests on the assumption of  complete  independence of the two systems after the interaction, but the  assumption  of  non-independence   in  the  form  suggested  by  orthodox  quantum
mechanics appeared to him an unacceptable ``spooky action at a distance''  \cite{Ein27}, 
\cite{Ein48}, \cite{Bor71}, \cite[p.~84, 682, 683]{Schi}.

The question is so important for our  conception  of  nature  that,  in
spite of the fact that the spacelike correlations mentioned are predicted by quantum mechanics and that
quantum mechanics has been confirmed in  innumerable  situations,  one  would
wish  this particular prediction to be tested  in  specific  experiments.  We
shall come to these experiments in Sec.~4.4.

\vspace{20pt}
\noindent {\bf 4.2~~Superluminal Signaling}
\medskip

\noindent
Here we want to consider  the  question  whether the spacelike correlations permit the transmission of  signals or messages with superluminal speed from  one
person to another. Such a transmission would  mean  a  drastic  violation  of
relativistic causality because we may consider A's sending a  message ``the
cause'' and B's receiving it ``the effect'',  and  with  superluminal  transmission
these cause and effect could appear in reverse order  of  time  in  a different
Lorentz system. 

Let us try to construct an early-warning  system.  Consider
Fig.~3 of Sec.~4.1. Imagine B to be the Earth and O  and  A  two  space
stations. Invaders (the Borg) from a distant star are expected  to  approach  the  Earth
from the direction of A. The task of A is to inform the Earth as soon as the invaders have been seen (emergency case).  For  this  purpose
the auxiliary space station O continually emits pairs of  scattered  protons,
say at a rate of 1 pair per second, and  the  protons  are  to  pass  through
Stern-Gerlach-type apparatuses on station A and on the  Earth  respectively.
The distance between O and the Earth is made only a little  larger  than  the
distance between O and A, so that A receives its proton  such  a  short  time
(which still may amount to some hours) before the Earth receives  its  proton
that no light signal could have informed the Earth of A's  operation.  In routine cases, the apparatuses at both A and B  have  their  axes  in $+z$
direction, and in the emergency case A turns the axis of its  apparatus  into
the $+x$ direction. One might think that this changes the probability of an  up
or down result in apparatus B on Earth, so from the changes in the ratio  of
up and down results the physicists on Earth  would  soon  learn  (before  any
light signal could be sent from A to B) that the invaders had been seen.
Now, the joined probability that proton 1 in  the Stern-Gerlach-type apparatus A 
 becomes an $r_{\rm A}$-proton (=  up  proton  if $r_{\rm A}=+1,$  down
proton if $r_{\rm A}=- 1)$ and that proton 2  in  apparatus B  becomes  an $r_{\rm B}$-proton,
according to the approved formulas of quantum mechanics, is (Appendix C)
\begin{displaymath}(4.4)\hsp
P(r_{\rm A},r_{\rm B}|  {\bf a},{\bf b}) = \mbox{$\frac{1}{4}$} (1 -  r_{\rm A}r_{\rm B}\cos \vartheta ) ,
 \end{displaymath}
\noindent where the unit vector {\bf a}  specifies  the  axis  of  apparatus A,  $ {\bf b}$  that  of
apparatus B, and $\vartheta\;  (0~\le \vartheta \le ~\pi )$  is  the  angle  between  {\bf a}  and  {\bf b}.  Hence  the
probability that B observes the result $r_{\rm B}$, whatever the result $r_{\rm A}$, is just
\begin{displaymath}\hspp
P(r_{\rm B}|  {\bf a},{\bf b}) = \sum^{}_{r_{\rm A}}P(r_{\rm A},r_{\rm B}|  {\bf a},{\bf b}) = \mbox{$\frac{1}{2}$} ,
 \end{displaymath}
\noindent regardless of the axis {\bf a} (as well as of ${\bf b})$,  and  in  this  way  the  early-warning system will not work. In fact, the above arrangement cannot  transmit
any message, superluminal or subluminal; the superluminal case  is  only  the  most
interesting aspect of this general incapability.

We may try to exploit the fact that the change of A's axis {\bf a}, if  it  does
not change B's probabilities, will at least change the  correlations  between
A's and  B's  results $r_{\rm A}$  and $r_{\rm B}$  respectively. 
However, the physicists on Earth do not know this.  They  only  know
the results of their own apparatus, that is, the lower line of Fig.~4, but
not the upper line. Thus, they do not know the correlation of  their  results
with those of A, still less can they realize that there had been any  change  in  those  correlations. Either of the two lines of Fig.~4  is  just  a  random  series;  the
probability of an up result is equal to that of a  down  result,  before  and
after the emergency case. So, the early-warning system does not work this way
either.

\vspace{5pt}
 \begin{figure}[h]
\begin{center}
\includegraphics[width=0.83\textwidth]{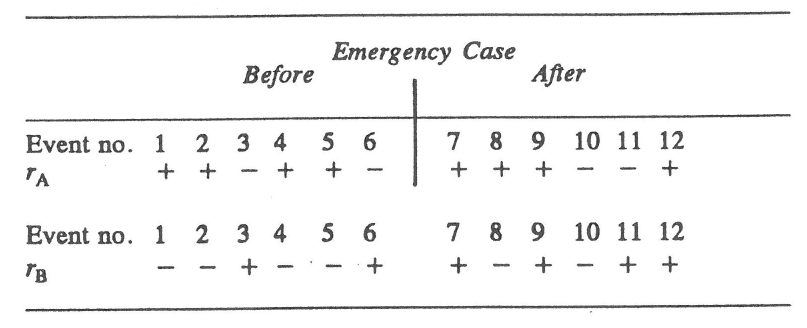}
\caption{Records of results of observers A and B before and after the emergency case.}
\end{center}
\end{figure}
One may try more general apparatuses than  just  Stern-Gerlach-type  ones.
These also will not work. It can be shown quite  generally  that  no  faster-than-light warning system can be built with devices obeying the  formulas  of
quantum mechanics. The proof is shown in Appendix D.

On the other hand we note that the formulas of quantum mechanics only give
\emph{probabilities} for the various possible  results.  If  the physicist at A  could  arrange  with
certainty that his proton always goes into the up state (say) with respect
to his axis, that is, if there were macro-determinism (predictability) superluminal messages would be possible. For then, with A's  and
B's axes parallel, the physicist at B would register only down protons. In the emergency  case
let A turn his apparatus upside down. From then on B  would  obtain  only  up
protons, and the first of these would tell B that A has  seen  the  invaders.
For the construction of an  early-warning  system  one  might,  therefore,  try
situations in the grey zone between quantum and classical physics, hoping that
here the probability features of quantum mechanics have already  sufficiently
approached classical deterministic  behaviour  while  the  superluminal  features
persist. An attempt has been made in that direction by Herbert \cite{Herb}, who  used
the amplification of a weak beam of light. But it was  soon  shown  that  the
proposal would not work because the amplification of arbitrary states by  one
and the same apparatus is impossible because it is at variance with the linearity of the  quantum
mechanical operators \cite{Mil} - \cite{Ghi}. Thus,  the  very  theory,  quantum  mechanics,
that predicts superluminal features also predicts that these features cannot be used
for transmitting superluminal messages from one person to another.

What, then, is the remarkable feature of the quantum mechanical formula  (4.4)  expressing  superluminal features, i.e. spacelike correlations? To see this we have to consider Bell's inequality.

\vspace{20pt}
\noindent {\bf 4.3~~Bell's Inequality}
\medskip

\noindent
A priori, spacelike correlations, i.e. correlations between spacelike separated events,  may be thought of as caused by arrangements in the past, for example by the searchlight spots on the Moon mentioned at the end of Section 3.2, or the letters running over the lights of a billboard. Such effects are brought about by common causes in the past, i.e. by events in the overlap of the past light-cones of the correlated events. The remarkable feature of formula (4.4) is that the correlations described by it cannot be accomplished with the above mentioned arrangements. This is shown by means of the Bell inequality. Spacelike correlations with no common causes in the past exhibit what we have called multiparticle nonlocality (Section~3.2) or just \emph{nonlocality}.

The Bell test of whether spacelike correlations exhibit nonlocality is that  the joint-prob\-ability formula leads to expectation values of the product $r_{\rm A}r_{\rm B}$  of
the dichotomic variables $r_{\rm A}$ and $r_{\rm B}$
\begin{displaymath}\hspace{40pt}
E(a,b) := P(+,+|  a,b) + P(-,- |  a,b) - P(+,- |  a,b) - P(-,+ |  a,b)
\end{displaymath}
\noindent that violate Bell's inequality
\begin{displaymath}(4.5)\hsp
K := |  E(a,b) + E(a,b^\prime ) + E(a^\prime ,b) -  E(a^\prime ,b^\prime )|   \le  2
 \end{displaymath}
\noindent for some choice of the parameters $a, b, a^\prime , b^\prime .$ (For convenience  we  write  {\emph a}
and {\emph b} instead of {\bf a} and ${\bf b})$. Actually, there are many versions of the Bell inequality; the particular version (4.5) was first written down in \cite{Cla69}. It is easy to see that formula (4.4) leads to $E(a,b) = - \cos \vartheta $, and this  may  violate  (4.5).
For example, choose vectors {\bf a} and ${\bf b}$ that lie in planes normal to  the  direction of propagation of the protons, and let {\bf a} form the angle $0^{\circ}, {\bf b}\quad 45^{\circ}, {\bf a}^\prime \quad 
90^{\circ}$, and ${\bf b}^\prime \;  - 45^{\circ}$ relative to some standard direction. This choice results  in
$K = 2\sqrt 2 = 2.83 > 2.$  

Thus  the quantum  mechanical formula (4.4) for spacelike correlations implies  nonlocality. In order to see why Bell's inequality accomplishes this we must look at the assumptions that are made in  its  derivation.  In
Appendix B it is shown that Bell's  inequality  is  satisfied  if  the  joint
probability can be written in the form
\begin{displaymath}(4.6)\hsp
P(r_{\rm A},r_{\rm B}|  a,b) = \int  P_{1}(r_{\rm A}|  a,\lambda ) \, P_{2}(r_{\rm B}|  b,\lambda ) \, f(\lambda )\, d\lambda  .
\end{displaymath}
\noindent This form means that we first go to a more detailed level of  description  by
introducing the additional variable $\lambda $  into  the  joint  probability  in  the
integrand, and that on this level we  write  the  joint  probability  in  the
special form \cite{Cla74}
\begin{displaymath}(4.7)\hsp
P(r_{\rm A},r_{\rm B}|  a,b,\lambda ) = P_{1}(r_{\rm A}|  a,\lambda ) \, P_{2}(r_{\rm B}|  b,\lambda ) .
 \end{displaymath}
\noindent The variable $\lambda $ fluctuates with the probability density
\begin{displaymath}(4.8)\hsp
f(\lambda ) \ge  0,\qquad \int f(\lambda )\, d\lambda  = 1 ,
 \end{displaymath}
\noindent and $f(\lambda )$ and the range of $\lambda $ do not depend on $a, b, r_{\rm A}$, and $r_{\rm B}$.  Actually, $\lambda $
stands for any set of variables that might  be  relevant in determining the probabilities. 
The  product  form
(4.7) is more than  just  the  separability $P(r_{\rm A},r_{\rm B}|  a,b,\lambda ) = P_{1}(r_{\rm A}|  a,b,\lambda )\times 
P_{2}(r_{\rm B}|  a,b,\lambda )$ since in (4.7) the first factor does not depend on $b$  nor  the
second on $a$. Thus, not only are the events $r_{\rm A}$ and $r_{\rm B}$  statistically  independent for given $\lambda $, but  the probability that A obtains the result $r_{\rm A}\,  [ \mbox{i.e.,  }
 \sum^{}_{r_{\rm B}}P(r_{\rm A},r_{\rm B}|  a,b,\lambda ) = P_{1}(r_{\rm A}|  a,\lambda ) ]$  is   also independent of B's  parameter $b$;  and
similarly $P_{2}$ is independent of $a$. There is however still a link between A and B, namely the common variable $\lambda $ and after the integration the  probability
(4.6)  need  no  longer  have  the  product  form $P(r_{\rm A},r_{\rm B}|  a,b) = P_{1}(r_{\rm A}|  a)
P_{2}(r_{\rm B}|  b)$,~and correlations between $r_{\rm A}$ and $r_{\rm B}$ may arise that depend on $a$  and
 $b$. 
 
Nevertheless, in order to calculate the probabilities $P_{1}(r_{\rm A}|  a,\lambda )$  at A it is sufficient to take into account simply the common parameter $\lambda$ and the local parameter $a$, but not the remote parameter $b$; and analogously for B. This is why the correlations based on the probability formula (4.6), satisfying the Bell inequality, are called locally explicable \cite{Bel90}.

\vspace{5pt}
Compare formula (4.4) with the formula
\[(4.9)\hspace{50pt}
P(r_{\rm A},r_{\rm B}|  {\bf a},{\bf b}) = \mbox{$\frac{1}{4}$} (1 - \mbox{$\frac{1}{3}$} r_{\rm A}r_{\rm B}\cos \vartheta ).
\]
All considerations in Sec.~4.2 would remain unaltered if instead of formula (4.4) we  used formula (4.9). Formula (4.9) results if we assume, in a ``semiclassical'' model,  that  the  two  protons in Fig.~4 were
completely independent after their interaction at O (cf. \cite{Fur}, \cite{Schr36}, \cite{BohmAha}), one having the spin $\vec{\sigma }$ and the other $-\vec{\sigma }$ (total spin zero) \emph{already at} O, and
that the probability
formulas for  the  single  protons  were  still  those  of  standard  quantum
mechanics ((C.9) of Appendix C)
\begin{displaymath}  \hsp
P(r_{\rm A}| {\bf a},\vec{\sigma }) = \mbox{${1\over 2}$} (1 +r_{\rm A}\vec{\sigma }{\bf a}) = \mbox{${1\over 2}$} (1 +r_{\rm A}\cos \alpha ) . 
\end{displaymath}
Then
 proton 1 would arrive at A with $\vec{\sigma }$, proton 2 at B  with
$- \vec{\sigma }$, and the conditional joint probability would be
\begin{displaymath}
\hsp
P_{\rm SC}(r_{\rm A},r_{\rm B}| {\bf a},{\bf b},\vec{\sigma }) = \mbox{$\frac{1}{2}$} (1 +r_{\rm A}\vec{\sigma }{\bf a})\, \mbox{$\frac{1}{2}$} (1 - r_{\rm B}\vec{\sigma }{\bf b}). 
\end{displaymath}
\noindent The integration over an isotropic distribution of $\vec{\sigma }$ would then result in
\begin{displaymath}   (4.10)  \hspace{10pt}
P_{\rm SC}(r_{\rm A},r_{\rm B}| {\bf a},{\bf b}) = \frac{1}{4\pi} \int  d\varphi  \int  \sin \alpha  d\alpha  \mbox{$\frac{1}{2}$}(1 +r_{\rm A}\vec{\sigma }{\bf a}) \mbox{$\frac{1}{2}$}(1 - r_{\rm B}\vec{\sigma }{\bf b})
\end{displaymath}
\begin{displaymath}\hspace{40pt}
= \frac{1}{4\pi} \int d \varphi  \int \sin\alpha d \alpha  \mbox{$\frac{1}{4}$} (1+r_{\rm A} \cos \alpha) \; (1-r_{\rm B}[\sin\vartheta \cos\varphi\sin\alpha + \cos \vartheta \cos \alpha ])
\end{displaymath}
\begin{displaymath}  \hspace{40pt}
= \mbox{${\frac{1}{4}}$} (1-  \mbox{${\frac{1}{3}}$} r_{\rm A}  r_{\rm B} \cos \vartheta ) .  
\end{displaymath}
\noindent Expression (4.10) differs from the quantum-mechanical formula (4.4) only by the  factor
1/3 inside the last bracket. This has however the consequence that the expectation
is now
\begin{displaymath}\hsppp
E(a,b) =- \mbox{${\frac{1}{3}}$} \cos \vartheta 
\end{displaymath}
\noindent and the Bell inequality (4.5)
\begin{displaymath}\hsppp
| E(a,b) +E(a,b^\prime ) +E(a^\prime ,b) - E(a^\prime ,b^\prime )|  \le  4/3 < 2
\end{displaymath}
\noindent is always satisfied. The variable $\vec{\sigma }$ in the semiclassical model corresponds to the $\lambda $ introduced in formula (4.6) because it is independent  of  {\bf a}  and ${\bf b}$. In  the  quantum-mechanical case it is not so, because there the value of $\vec{\sigma }$ for proton  2  before  it
enters B's apparatus is influenced by A's variable ${\bf a},\; \vec{\sigma }$ for  proton  2  being
either +{\bf a} or $-${\bf a}.

Thus correlations described by formulas such as (4.4), which do not satisfy the Bell inequality, mean local inexplicability, i.e. nonlocality \cite{Bel90}.

It is always assumed that the choice of the parameters $a$ and $b$ can be made at any time at will by the experimenters. If we accept a strictly deterministic view, where free  will  is  an illusion, it is possible to assume that \emph{all} correlations arise from common causes in the past. Nevertheless, it is also possible to assume that there are particular correlations which do not arise in this way. Some support of this view is seen in the superluminal contraction of the one-particle wavepackets discussed in Sec.~3.1  and in the arguments given in \cite{Jab12} (cf. also \cite{Simul}).

Finally, we note that in the above considerations about the Bell inequality the question of determinism is not involved because the parameters $a, b$ and $\lambda$ only determine the probability of an outcome, not the outcome itself. Whether this probability is reducible to some underlying constellations of additional variables is left open. -- Neither is the question of realism involved because it is left open  whether the outcomes come into existence by our observation or arise independently of the observer. The Bell inequality is just about nonlocality.

\vspace{20pt}
\noindent {\bf 4.4~~Experiments}  
\medskip

\noindent
Are the formulas of quantum mechanics  that  
lead to a violation of the Bell inequality
 confirmed in specific experiments?  Many  experiments  have  been
performed  \cite{Rar}, \cite{Ou}, \cite{Shi}, \cite{Dun},  \cite{Giustina13} and the result  is  that  they  generally  confirm
quantum mechanics. Most experiments were concerned with the  Bell  inequality
in its different but essentially  equivalent  forms.  There  are  also  other
experiments confirming the nonlocal features \cite{Pau80}, \cite{Spa},  but  the  experiments
related to the Bell inequality seem to be the most stringent  ones  and  have
been subjected to the closest scrutiny. In all of them, except in  two  early
cases which are now considered unreliable, a violation of the respective variant of the Bell inequality has  been
found. Moreover, the particular type of violation was exactly that predicted by the formulas  of
quantum mechanics. The  probability
$P(r_{\rm A},r_{\rm B}|  a,b)$ in formula (4.6) or the average $E(a,b)$ are 
measured by means of normalized coincidence rates and appropriate average values. Not all experiments
were absolutely conclusive  because  simplifying  though  very  plausible
assumptions had to be made. These  assumptions  were  necessary  because  the
filters and detectors employed were not ideal, because the two photons in the
atomic-cascade experiments are not  strictly  antiparallel,  and  because  of
other  reasons.  Clauser  and  Horne \cite{Cla74},  for   example,   introduced   the
``no-enhancement assumption'', which means that the photon detection  probabilities [$P_{1}$ and $P_{2}$ in (4.7)], for every value of the variable $\lambda$, can at most be
reduced but not enhanced by a polarization filter  placed  in  front  of  the
detector (``detection loophole'').   In subsequent experiments most of the simplifying assumptions have been gradually  eliminated
or reduced in their influence. 

For example, in
all pre-1982 experiments the spin-(polarization)-reference axes were fixed
before the individual measurements were done, hence  the  measurements  were  not
separated by spacelike intervals (``locality loophole''). Therefore the result of A could,  in  principle, have been transmitted to B with light (or even  under-light)  velocity
before the measurement by B had taken place and so could have influenced  B's
result. Of course, in  the  actual  experiments  any  mechanism  that  might,
according to current knowledge, have permitted this was excluded;  still  the
possibility  was  only  excluded  technically,  not  in  principle. In  1982
Aspect et al. \cite{Asp82a} performed an experiment in  which  this  was  excluded  in
principle. They used variable polarizers that jumped  between  two  orientations in a time that was short compared with the  photon  transit  time.  In
this experiment, too, a violation of Bell's inequality and a confirmation  of
the quantum mechanical formulas was found (see also \cite{Asp82b}, \cite{Wei}, \cite{Hensen},  \cite{Shalm}, \cite{Giustina}). Thus, the experiments provide overwhelming evidence that
nonlocality is indeed a feature of physical reality.

Finally, let us have a look at the spatial separations of the  wavepackets
between which spacelike correlations have been observed in the experiments.

(1) In  the  proton-proton  scattering  experiment  of  Lamehi-\-Rachti  and
Mit\-tig \cite{Lam} the distance $\overline{\rm OA}$ in Fig.~4 was about 5 cm.  After scattering at O the  protons  had  a
kinetic energy  of 6 MeV, and the length of  the  proton
packets was calculated from the lifetime of the intermediate  singlet $s$-wave
state to be $4\times 10^{-15}$ m. A proton packet of the above energy, for which $4\times 10^{-15}$ m
is the minimum width, spreads out to an extension of 2.3 cm while  its  centre
traverses the distance of 5 cm [Appendix  A,  Eq.~(A23)].  Thus,  the
separation between the two proton packets is about 4  times  their  width  as
measured by the standard deviation.

(2) Some experiments for  testing  Bell's  inequality  employ photons  from  a
cascade decay of excited atoms are employed. The length of the photon packets
is estimated from the mean lives of the decaying levels, which gives values of
the order of 1.5 --  3 m. This  is  comparable  to  or  even  larger  than  the
dimensions ($\overline{\rm OA}$) of the apparatuses used up to 1980. In the subsequent  experiments  \cite{Asp81}, \cite{Asp82b} the apparatuses A and B are separated by  about  13
m. This is 8 times the estimated length of the photon packets.

(3)   In   correlation   experiments   with   photon   pairs   from ${\rm e}^{+}{\rm e}^{-}$ 
an\-ni\-hi\-la\-tion \cite{Far} - \cite{Bru} the individual photon packets are  usually
assumed to have lengths of the order of $7 - 15$ cm while the distance  between
O and A was  up to 2.5 m. This is 16 to 35 times the packet length.

Admittedly, the lengths ascribed to the individual wavepackets may  be  larger
than assumed, in particular they may  be  larger  than  the  usually  adopted
standard deviation $\Delta ${\emph y}. The value of $\Delta y$ is often calculated from $\Delta p_{y}$ by  means
of the Heisenberg relation with the equality sign $\Delta y\Delta p_{y} = \hbar /2.$  However,  the
equality sign can only hold for a Gaussian form of the wavepacket,  and  even
for a Gaussian form it holds only at one instant of time; at other times  the
length may have spread out to values considerably  larger  than  the  minimum
value. Also, the length $\Delta y$ of a wavepacket is often taken to be the coherence
length of the  beam  in  which  it  takes  part \cite{Lam}, \cite{Far}, \cite{Wil}.  Actually,  the
coherence length of the beam is of the order of a lower bound for the  length
of the constituent wavepackets (Appendix A).

Brendel et al. \cite{Bre} used
pairs of parametrically down-converted photons and measured the  correlations
in coincidence counts over distances of 55 cm. At the same time they measured
the length of the individual photon wavepackets and obtained values of less  than
10 cm. There is thus very little overlap between the photon wavepackets.  However,
at the same time they also obtained the high value of $87\%$ for   visibility
in the coincidence rate as a function of wavepacket separation.  This  cannot
be explained as interference of wavepackets in ordinary  space.  And
there are by now many other photon-correlation experiments that point in the
same direction \cite{Kwi92}, \cite{Lar}, \cite{Zou}.  In fact, spacelike correlations between photons (length of the order of $30\,\mu$m) entangled over more than 10 km \cite{Tit}, \cite{Zbi}, 16 km \cite{Yin}, and 143 km \cite{Her14} have been observed.

\newpage
\noindent {\bf 5~~QUANTUM STATISTICS WITH WAVEPACKETS}
\bigskip

\noindent {\bf 5.1~~Field Quantization}
\medskip

\noindent
The Schr\"odinger equation in  ordinary  (3+1)-dimensional  space  is  a
classical field equation, and  the  discrete  eigenfunctions  following  from
imposing the usual normalizability and uniqueness conditions are no more than
the standing waves of classical physics. This is just  a  consequence  of  de
Broglie's idea of matter waves. We would thus not call that appearance of  discrete  eigenvalues  and  eigenfunctions real quantum effects but would reserve this denomination to effects
that cannot be explained in the mentioned way. At first, Schr\"odinger seems to
have believed that there are no such other effects, but  he  was  opposed  by
Heisenberg \cite{Hei26a}, \cite{Hei26b} who pointed, among other  things,  to  the  photoelectric effect and to the Planck radiation law.
 
Schr\"odinger's equation in its general form is not an equation in  ordinary
space but in $(3N+1)$-dimensional configuration space, and this goes beyond  de
Broglie's conception. The  Schr\"odinger  or  de  Broglie  function $\psi ({\bf x},t)$  in
ordinary space in itself does not tell us whether it refers to one or to more
particles or to particles at all. It is just a field and expresses  only  the
wave aspect.  With  the  introduction  of  the  configuration-space  function
$\Psi ({\bf x}_{1},{\bf x}_{2},\ldots 
,{\bf x}_{N},t)$, however, the number $N$ of particles or quanta is explicitly
introduced. 

In 1927 in his fundamental paper on the quantum theory of the emission and
absorption of radiation Dirac \cite{Dir27} derived, among other things, the  Einstein
{\emph A} and $B$ coefficients and hence Planck's law by  means  of  a  new  procedure,
which had first been introduced by Born and Jordan \cite{Bor25}  and  in  the  famous
three-man work on matrix mechanics by Born, Heisenberg and  Jordan \cite{Bor26}. The
procedure consisted in turning some canonically conjugate  variables  of  the
Hamilton  formalism  into  operators  satisfying  the  canonical  commutation
relation, such as between  position  and  momentum.  Here  the  now  familiar
creation and annihilation operators $a^{\mbox{\dag }}$ and $a$ (in present notation) showed  up
for the first time. In the same paper Dirac, and later Tomonaga \cite{Tom}, showed that  this  procedure  was
equivalent with Schr\"odinger's configuration space treatment with  symmetrical
wave functions. The equivalence was subsequently elaborated and  extended  by
Jordan, Klein and Wigner \cite{Jor27a} - \cite{Jor28} to include antisymmetric  wave  functions, i.e. fermions, in which  case  anticommutators  were  to  replace  the
commutators. In 1932 Fock \cite{Foc32b} gave a lucid summary  of  these  developments,
and he showed that the restriction to a configuration space of  fixed  dimension can easily be overcome. This  he  made  particularly  clear  by  casting
Schr\"odinger's configuration-space  formalism  into  the  form  of  the  Fock,
occupation-number,  or $N$  representation \cite{Foc32b}, \cite{Schw66}. In this representation the total number  of  particles no longer appears explicitly, and this makes it possible  to  apply  the
formalism to systems in which the total number of particles is not conserved,
as for example to the photons in a cavity.

The method of commutation relations was  then  further  developed  into  a
comprehensive scheme by Heisenberg and Pauli \cite{Hei29}, \cite{Hei30}. They no longer derived
the  commutation  relations  from  configuration  space  but  set  out   from
ordinary-space fields $\psi ({\bf x},t)$ and introduced the commutation relations by  way
of postulate. In this way the quanta of the fields (i.e. the particles) arise
from interpreting the operator $a^{\mbox{\dag }}a$,  which  has  only  non-negative  integer
eigenvalues, as a particle-number operator.  Moreover,  and  most  important,
they extended the formalism to include  Lorentz  invariant  interactions  and
hence retardation between the similar particles. Retardation  effects  cannot
be taken into account  in  Schr\"odinger's  configuration  space,  so  the  two
schemes are no longer equivalent. Heisenberg's  and  Pauli's  scheme is 
canonical quantization. In it the well known difficulties with the  diverging
integrals, irreparable by simple normal ordering, began. Thus, in my  opinion, 
this is  where something went wrong with the  relativistic  formulation of quantum theory, and I suspect that this is related to  the  general
negative attitude  towards  nonlocality  at  that  time,  as  reflected,  for
example, by Pauli's classification \cite{Pauli33a}  of  Landau's  and  Peierls'  nonlocal
density as ``unnatural''. In fact, some features, which had been thought to be explicable only in the formalism of canonical quantization, could by now be shown to be explicable within the formalism of quantum mechanics \cite{Jab14}, \cite{Jab08}.

Thus  here  we  stop,  and  we  conclude  our  treatise  with   the re-interpretation of the  nonrelativistic 
field-quantization formalism as far as
it is equivalent with Schr\"odinger's configuration-space formalism,  including
symmetrization and particle non-conservation. Our
treatment still includes quantization of the electromagnetic radiation field, though not in the way of canonical quantization, but  in the spirit of Einstein. There are no retardation  effects  between  the  photons, because  
there are no direct interactions between them.

In the following sections we shall  derive  the  Bose,  Planck  and  Fermi
distributions as well as  the  corresponding  fluctuations,  by  speaking  of
alteration, condensation and decondensation, introduced in Sec.~3.3, of  the realist quantum wavepackets, rather than of distributing particles over phase-space cells.

\vspace{20pt}
\noindent {\bf 5.2~~The Many Aspects of the Condensed Wavepackets}
\medskip

\noindent
We consider similar particles of mass $m$  in  a  cavity  of  volume $V$  at
temperature $T$, and we write  the  general  quantum  statistical  distribution
function in the well-known form
\begin{displaymath}(5.1)\hsp
N(p,T)dp = {4\pi Vp^{2}dp\over h^{3}} \times \left\lbrace \exp\left[ \left( \sqrt{p^2c^2 +m^2c^4} - \mu
\right) {1\over kT}\right] \pm 1 \right\rbrace^{-1} 
\end{displaymath}
\noindent where $N(p,T)dp$ means, in the usual interpretation, the time  averaged  number
of particles in $V$ whose absolute value of momentum lies in  the  interval $dp$
about $p$. The plus sign refers to fermions and the minus sign to bosons. $\mu $  is
the chemical potential [fugacity $z=\exp (\mu /kT)$]. In the special case of photons
we have $m=0,\; p=h\nu /c,\; \mu =0$, and formula (5.1) with the minus  sign  reduces  to
the Planck distribution for polarised radiation.

We have written the distribution (5.1) as the product of two factors.  The
first factor is
\begin{displaymath}(5.2)\hsp
g_{p} = {4\pi V\over h^{3}}p^{2}dp ={4\pi V\epsilon \sqrt{\epsilon^2 - (mc^2)^2} \over h^3c^3} d\epsilon
\end{displaymath}
\noindent
where $\epsilon  = \sqrt{p^2c^2 + (mc^2)^2}$    is the total energy of a particle. In this section  we
will consider only this factor; the second factor will be considered  in  the
next section. In the special case of photons (5.2) becomes
\begin{displaymath}(5.3)\hsp
g_{\nu } = 4\pi V\nu ^{2}d\nu /c^{3} ,
 \end{displaymath}
\noindent and in this case it has a long history:

In 1899 it was calculated by Planck \cite{Pla99}  as  the  proportionality  factor
between the mean energy of electromagnetic radiation in $V$ and $d\nu $ and the mean
energy of a charged oscillator with  radiation  damping.  In  1900  and  1905
Rayleigh \cite{Ray} and Jeans \cite{Jea} considered the factor as the number  of  degrees
of freedom of the ether inside the cavity, this number in turn being  considered equal to the easily  calculable  number  of  eigenvibrations  (modes  of
vibration) within $d\nu $ of the ether.  In  1914  von  Laue \cite{Lau14} decomposed  the
cavity radiation into mutually independent radiation bundles, each converging
to its focal region and  then  diverging.  To  these  bundles  he  attributed
degrees of freedom and obtained (5.3) as the sum of the degrees of all  these
bundles (see below). Bose, in his famous paper of  1924 \cite{Bos}, considered  the
factor (5.3) as the number of phase-space cells of size $h^{3}$.  Such  cells  had
already been considered by  Planck  in 1906 \cite{Pla06}  in  the  special  case  of
harmonic oscillators. In 1925 Land\'e \cite{Lan} called those of von Laue's radiation
bundles that had just one degree of freedom elementary light-quantum  bundles
or just quantum bundles, and  he  proposed  to  identify  these  with  Bose's
quantum phase-space cells.

In present-day  quantum  mechanics  (5.2)  is  the
number of eigenvalues of the Hamilton operator for a free particle in $V$ that
fall into the energy interval $d\epsilon $ which corresponds to $dp$. Each  eigenvalue  is multiply counted according to its order of degeneracy. In other words,  (5.2)
is the number of eigenstates in $V$ and $dp$. For photons we are thus effectively
back at Rayleigh's and  Jeans'  determination.  In  quantum  mechanics  (5.2)
holds, however, for any kind  of  particle,  not  just  photons,  because  de
Broglie waves  are  associated  with  each  kind  of  particle.  Finally,  in
canonically quantized radiation theory (5.3) is the  number  of  oscillators.
But in contrast to Planck's oscillators, which  represent  atoms  (`resonators'') interacting
with the radiation field,  these  oscillators  are  to  represent  the  field
itself, a point of view that had  already  been  indicated  by  Ehrenfest  in
 1906 \cite{Ehr06}.

Now we want to show that (5.2) or (5.3) can also be taken as the number of
(condensed) wavepackets in the cavity covering the momentum interval $\Delta p$.  For
this purpose we employ the fact that (5.3) is the total number of degrees  of
freedom of von Laue's radiation bundles and that a bundle  of $F$  degrees  of
freedom may be taken to consist of $F$ wavepackets. Von Laue defines the number
$F$ of degrees of freedom of a bundle of length $l$ (from wall  to  wall  of  the
cavity), spectral range $d\nu$  (equal to the  spectral  range  of  the  radiation
considered), convergence half angle $\alpha $ and focal cross section {\emph A} with the help
of the theory of optical resolving power and the counting of Fourier  coefficients. He arrives at the expression

\begin{displaymath}(5.4)\hsp
F = \frac{\emph Ald\nu}{a c}
 \end{displaymath}
\noindent where $a$ is the minimum focal area that is possible for a  bundle  of  convergence angle $\alpha$  (cf. Appendix  A formula (A16), with $\alpha  = \upsilon _{{\rm s}x\infty } /c$).  Now
imagine that the bundle of $F$ degrees of  freedom  consists  of  a  stream  of
wavepackets, all moving parallel to the axis of the bundle and going side  by
side through its focal area. The convergence of the bundle to the focal plane
and its subsequent divergence comes about by the contraction  and  subsequent
spreading in the transverse direction of each one of these wavepackets,  assuming
that all packets have their minimum transverse extension in the focal  plane.
In front of and behind the focal plane the wavepackets  may  overlap  in  the
lateral direction. We then write von Laue's  degrees  of  freedom $F$  as  the
product of three factors: $F = N_{1}N_{2}N_{3}$ where $N_{1} = d\nu /\Delta \nu , N_{2} = A/a$,  and $N_{3} =
l/(\Delta y)$. Each factor is the ratio of some quantity  relating  to  the  bundle
divided by the corresponding quantity relating to  the  packets. $\Delta \nu $  is  the
frequency range of a wavepacket of total length $\Delta y$. $\Delta \nu $ is related to $\Delta y$  by
$\Delta \nu  = rc/(4\pi \Delta y),\; (r\ge 1)$,  which  follows  from   the   Fourier   reciprocity
(Heisenberg) relation $\Delta y\Delta p_{y} = r\hbar /2$ with $\Delta p_{y} = (h/c)\Delta \nu $. With this relation  we
obtain $N_{1}N_{2}N_{3} \le  (4\pi /r)${\emph F}. The number $r$ may be set  equal  to $4\pi $  because  the
packets in the cavity do not all have  their  minimum  phase-space  extension
$(r=1)$. Moreover, there is always some degree of arbitrariness  in  the  exact
definition of the widths $\Delta y, \Delta p_{y}$ etc., resulting in some arbitrariness in  {\emph r}.
Thus we take $r=4\pi $ to mean the average extension of  the  wavepackets  in  the
cavity with an appropriate definition of the widths, and with this we obtain

\begin{displaymath}(5.5)\hsp
N_{1}N_{2}N_{3} = F .
 \end{displaymath}
\noindent Now, by the definitions given above $N_{1}$ is the number  of  spectral  types  of
wavepackets in the bundle, as defined by their individual frequency ranges $\Delta \nu$ 
(colors),  $N_{2}$ is the number of wavepackets of a particular spectral type  that
go side by side and $N_{3}$ those that go one after the other  through  the  focal
area of the bundle. It follows that the product $N_{1}N_{2}N_{3}$ is equal to the  total
number of wavepackets that make up the bundle, and the relation  (5.5)  means
that this number is equal to the number of degrees of freedom of the  bundle.
Thus the total number of wavepackets can be identified with the total  number
(5.3) of degrees of freedom, or of von  Laue's  bundles,  if  we  imagine  the
radiation to consist only of elementary bundles.  This  not  only  holds  for
photon packets but for packets of any kind. The considerations  by  von  Laue
can easily be extended to matter waves. One just has to replace $\nu/c$ by $1/\lambda$ in (5.4) and the ensuing text, and then to replace $1/\lambda$ by $p/h$, and with $r=4\pi$ the same result obtains.

Now, the discrete energy values $nh\nu $ may be attributed to each of  Planck's
oscillators or Jeans' degrees of  freedom \cite{Deb},  and $n$  quanta  may
occupy each of Bose's cells. These are then our \emph{condensed} wavepackets  representing $n$ quanta. Empty wavepackets, without a quantum, are also included  in
formula (5.2). This is a convenient means of indicating that there  is  space
left for more wavepackets  to  show  up  in $V$  and $dp$ (cf.  Bose \cite{Bos} and
Schr\"odinger \cite{Schr24}) Equivalently,  one  may  say  that  (5.2)  is  the  maximal
possible number of existing (non-empty) wavepackets.

The condensed wavepackets also resemble the  degenerate  light  pulses  of
Man\-del \cite{Man}.  Man\-del  introduced  the  degeneracy  parameter $\delta $,  meaning  the
average number of photons in a light beam that are to be found in  the  same
cell of phase space. He expressed the phase space volume with the help  of  a
certain coherence volume defined in  the  theory  of  optical  coherence.  We
obtain the same formulas when we  take  the  phase-space  volume  to  be  the
product of the ranges of the wavepacket at the moment of its  minimum  extension (Appendix A).

Finally  we  want  to  point  out  to  the  ``light  molecules'',   ``quantum
multiples'', ``$n$-quantum rays'' and ``radiation bundles'' first  mentioned  parenthetically by Joff\'e (1911) \cite{Jof} and  considered  more  closely  by  Ishiwara
(1912) \cite{Ish} , Wolfke  (1921) \cite{Wol}, de  Broglie  (1922) \cite{Bro22},     Bothe (1923,\allowbreak 
1924) \cite{Bot24}  and   especially   by     Schr\"odinger  (1924) \cite{Schr24} and  Bothe  (1927) \cite{Bot23}.
These authors noticed that the Planck distribution can be written in the form
of a sum and that the $n$-th term can be interpreted  as  a  contribution  from
objects that are composed of $n$ light quanta (see formula  (5.16)  below).  It
seems that these ideas retreated under the blow of  Dirac's  quantization  of
radiation in 1927 \cite{Dir27}, but it is seen that they also strongly  resemble  the
condensed wavepackets.

The wavepackets, not the single quanta (if we were to take these  for a moment 
as entities of their own) are  the  statistically  independent  objects,  and
condensation of wavepackets is our means of expressing the ``mutual  influence
of the molecules [i.e. quanta] which  for  the  time  being  is  of  a  quite
mysterious nature'' mentioned by Einstein in 1925 \cite{Ein25}.

\vspace{20pt}
\noindent {\bf 5.3~~The Balance Relation}
\medskip

\noindent
Now we turn to the second factor in (5.1)
\begin{displaymath}\hspp
\left\lbrace \exp\left[ \left(\sqrt{p^2c^2+m^2c^4} - \mu \right) {\frac{1}{kT}}\right] \pm 1\right\rbrace^{-1}.
  \end{displaymath}
\noindent We want to derive this factor  by  means  of  Einstein's  method  of  balance
relations between transition rates \cite{Ein17},  formulated in terms of wavepackets. 
Einstein's original treatment of 1917 did  not  explicitly  take  into  account  that
photons are bosons and not fermions; it  would  give  the  same  distribution
function in both cases. Of course, the Fermi distribution was published  only
 in 1926 \cite{Fer}. The fact that photons are bosons can only be taken  into
account when phase-space regions or  energy  intervals  are  subdivided  into
those fundamental units that are counted by formula (5.2). It is  not  enough
to consider the  number  of  photons  in  a  given  energy  interval,  as  in
Einstein's procedure of 1917, but one must further specify  how  the  photons
are distributed over the various fundamental units within this interval. This was done by Einstein later in 1924  in Bose's  quite different phase-space cell approach.
 The fact that  Einstein
did obtain the Planck distribution by his 1917 method, in  spite  of  not  accounting  for  the
subdivision into fundamental units, is due to the special way  he  formulated
the balancing equations. We shall return to this point below.

With the subdivision into fundamental units we shall  be  able  to  derive
both the Bose and the Fermi distribution on an equal footing by means of  the
method of balancing  equations between wavepackets.  The  Fermi  distribution  has  already  been
obtained in the framework of this method by  several  authors,  though  in  a
different way \cite{Blo}, \cite{Jor27} - \cite{Cra}. Our  procedure  is inspired  by  the
comprehensive  treatment  by  Oster \cite{Ost} and   the  remarkable  paper  by
Bothe \cite{Bot23}. The general mechanism in all these works is  exchange  of  quanta
between fundamental units. These units have sometimes been taken  to  be  the
discrete energy states of atoms or oscillators which emit and absorb photons.
We emphasize, however, that the  fundamental  units  are  not  restricted  to
discrete energy states. That they may well be small but finite energy  intervals, centred about any energy values,  had  already  been  pointed  out  by
Pauli \cite{Pauli23} and  by Einstein  and  Ehrenfest \cite{Ein23}  when  considering  photons
scattered by free electrons in the Compton effect.

In our interpretation the fundamental units are the wavepackets and we are
going to consider processes taking place between these. As we have  discussed
in Sec.~3.3 this is  our  interpretation  of  the  change  in  ``occupation
numbers'', which in the occupation-number representation is described by means
of the creation and annihilation operators. The one-particle basis  functions
in the expansion (3.14) now are energy eigenfunctions or narrow superpositions
of these. Specifically, we consider two types of elementary processes. Either
of them is decomposed in alteration, condensation and decondensation. Alteration changes the energy of a wavepacket, whereas condensation and decondensation change the
number of quanta it represents. Thus the first type is:

(1) An $s$-quantum wavepacket of kind 1 which represents $s$  quanta  in  the
energy interval $\epsilon ^{\rm i}_{1} \ldots 
 \epsilon ^{\rm i}_{1}+d\epsilon ^{\rm i}_{1}$ (an $s$-packet in $d\epsilon ^{\rm i}_{1}$,  for  short)  decondenses
into an $(s-n)$-packet and an $n$-packet in $d\epsilon ^{\rm i}_{1}$. The $n$-packet  exchanges  energy
and momentum with an $n^\prime $-packet of kind 2 in $d\epsilon ^{\rm i}_{2}$ whereby  it  is  altered  and
goes into the energy interval $d\epsilon ^{\rm f}_{1}$ and then condenses with an $r$-packet in $d\epsilon ^{\rm f}_{1}$
to form an $(r+n)$-packet in $d\epsilon ^{\rm f}_{1}$. Simultaneously an $s^\prime $-packet of kind 2 in $d\epsilon ^{\rm i}_{2}$
decondenses into an $(s^\prime - n^\prime )$-packet plus an $n^\prime $-packet in $d\epsilon ^{\rm i}_{2}$. The $n^\prime $-packet is
altered in an interaction with the $n$-packet in $d\epsilon ^{\rm i}_{1}$ whereby it goes  into  the
interval $d\epsilon ^{\rm f}_{2}$ and  then  condenses  with  an $r^\prime $-packet  in $d\epsilon ^{\rm f}_{2}$  to  form  an
$(r^\prime +n^\prime )$-packet in $d\epsilon ^{\rm f}_{2}$. A  graphical scheme is presented in Fig.~5. Conservation of energy requires

 \vspace{5pt}
 \begin{figure}[h]
\begin{center}
\includegraphics[width=0.72\textwidth]{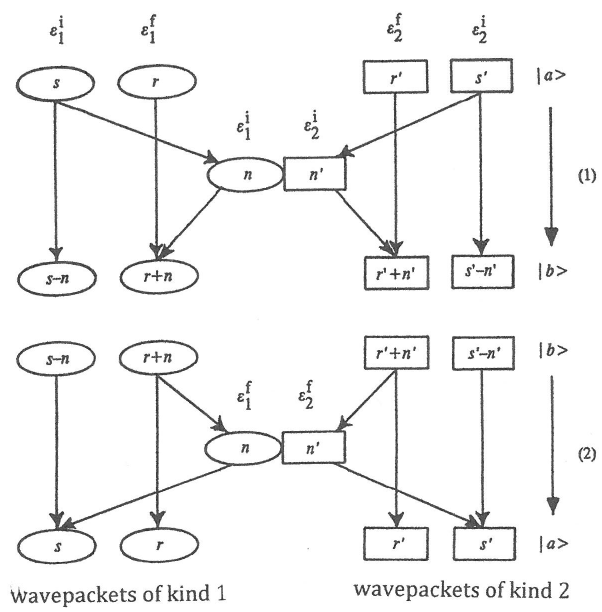}
\caption{Scheme of the considered processes (1) and (2) between the wavepackets of kind 1 and kind 2.}
\end{center}
\end{figure}

\begin{displaymath}(5.6)\hsp
n~(\epsilon ^{\rm i}_{1}- \epsilon ^{\rm f}_{1}) = n^\prime~(\epsilon ^{\rm f}_{2}- \epsilon ^{\rm i}_{2}) .
 \end{displaymath}
\noindent The energy intervals $d\epsilon $ are chosen so that they correspond to the  respective
intervals $dp$ in absolute value of momentum used in formula (5.1), i.e. $d\epsilon  =
(d\epsilon /dp)dp$. Effectively, if we may use here the picture of quanta as  standard
portions of water, an $s$-packet in $d\epsilon ^{\rm i}_{1}$ gives $n$ quanta to an $r$-packet  in $d\epsilon ^{\rm f}_{1}$,
and an $s^\prime $-packet in $d\epsilon ^{\rm i}_{2}$ gives $n^\prime $ quanta to an $r^\prime $-packet in $d\epsilon ^{\rm f}_{2}$. Thus, process
(1) leads from the state $|  a\rangle$ characterized by the 4 packets  which  represent
$s, r, s^\prime $ and $r^\prime $ quanta respectively, to some state $|  b\rangle$ characterized by the 4
packets that represent $s-n, r+n, s^\prime -n^\prime $ and $r^\prime +n^\prime $  quanta  respectively.  The
probability of such a transition is denoted by $W_{1}$. The process comprises most
particular physical situations that lead to Fermi or  Bose  distributions  as
special cases.

There is no interaction between photons, so here we need the  second  kind
of wavepacket (atoms, electrons etc.). Electrons interact  with  each  other,
and kind 1 and kind 2 may be the same. In the  Boltzmann  case  there  is  no
change in the number of quanta but only an alteration in the energies of  the
wavepackets. This can be described by putting $r=r'=0, n=s, n'=s'$  in  in  the
scheme of Fig.~5 and dropping $p(0,\epsilon )$ and $q(0,\epsilon )$ from  the Eqs.~(5.7)
and (5.8) below, which then  lead  to  the  Boltzmann  distribution  for  the
wavepackets (of energy $\epsilon s)$.

(2) The second type of processes is this: an $(r+n)$-packet of kind 1 in $d\epsilon ^{\rm f}_{1}$
(which need not be the same packet as that at the end of process (1))  decondenses into an $r$-packet and an $n$-packet. The $n$-packet  exchanges  energy  and
momentum with an $n^\prime $-packet of kind 2 whereby it goes into the energy interval
$d\epsilon ^{\rm i}_{1}$ and then condenses with an $(s- n)$-packet of that interval.  Simultaneously
an $(r^\prime +n^\prime )$-packet of kind 2 in $d\epsilon ^{\rm f}_{2}$  decondenses  into  an $r^\prime $-packet  and  an
$n^\prime $-packet. The $n^\prime $-packet is altered in an interaction with  the $n$-packet  of
kind 1 whereby it goes into $d\epsilon ^{\rm i}_{2}$ and then condenses  with  an $(s^\prime - n^\prime )$-packet.
Conservation of energy is again guaranteed by Eq.~(5.6).

The initial (final) $s, r, s^\prime $ and $r^\prime $-packets of process (2) have  the  same
momenta etc. as the final (initial) $s, r, s^\prime $ and $r^\prime $-packets  of  process  (1)
and differ from those only by spatial translations. Thus,  process  (2)  goes
back from state $|  b\rangle$ to state $|  a\rangle$.  The  probability  of  this  transition  is
denoted by $W_{2}$.

Process (2) is not the time reversed (``converse'') process to process  (1),
but may be called the ``reverse'' process, after Dirac \cite{Dir24}. Only  the  reverse
process can lead to  statistical  equilibrium \cite{Ein23}, but  only  the  converse
process is suggested to exist and to occur at the same rate as  the  original
process on account of the general principle of  time-reversal  invariance  of
basic processes. Now, in an isotropic  medium  the  reverse  process  can  be
obtained from the converse one by successive reflections  in  three  mutually
perpendicular mirrors at rest relative to the system as a  whole,  and  their
frequencies of occurrence must be equal \cite{Dir24}. Thus,  in  statistical  equilibrium the two processes (1) and (2) also occur at equal rates.  We  now  consider these rates. Rate 1 is the mean number of processes (1) that occur  per
second in the volume $V$. According to the above-given description it should be
equal to
\begin{displaymath}(5.7)\hsp
p(s,\epsilon ^{\rm i}_{1})d\epsilon ^{\rm i}_{1}\; p(r,\epsilon ^{\rm f}_{1})d\epsilon ^{\rm f}_{1}\; q(s^\prime ,\epsilon ^{\rm i}_{2})d\epsilon ^{\rm i}_{2}\; q(r^\prime ,\epsilon ^{\rm f}_{2})d\epsilon ^{\rm f}_{2}\; W_{1}
 \end{displaymath}
\noindent where $p(s,\epsilon ^{\rm i}_{1})d\epsilon ^{\rm i}_{1}$ is the mean (time averaged) number of $s$-packets of kind 1 in
$V$ that represent quanta in $d\epsilon ^{\rm i}_{1}, q(s^\prime ,\epsilon ^{\rm i}_{2})d\epsilon ^{\rm i}_{2}$ is the mean number of $s^\prime $-packets
of kind 2 in $V$ and $d\epsilon ^{\rm i}_{2}$, and so on. Analogously, for process (2) the rate is
\begin{displaymath}(5.8)\hsp
p(s- n,\epsilon ^{\rm i}_{1})d\epsilon ^{\rm i}_{1}\; p(r+n,\epsilon ^{\rm f}_{1})d\epsilon ^{\rm f}_{1}\; q(s^\prime - n^\prime ,\epsilon ^{\rm i}_{2})d\epsilon ^{\rm i}_{2}\; q(r^\prime +n^\prime ,\epsilon ^{\rm f}_{2})d\epsilon ^{\rm f}_{2}\; W_{2} ,
 \end{displaymath}
\noindent and the two rates (5.7) and (5.8) have to be equated.

Now, according to the preceding section  our  wavepackets  and  hence  the
states $|  a\rangle$ and $|  b\rangle$ mean pure states of quantum mechanics, and since the  probability of a transition in quantum mechanics is the same for  a  process that
goes from $|  a\rangle$ to $|  b\rangle$ as for a process that goes from $|  b\rangle$  to $|  a\rangle$ (Hermitean
operators), the probabilities $W_{1}$ and $W_{2}$ are  equal  and  disappear  from  the
balancing equation. We shall thus obtain the statistical  distribution  functions \emph{ without using any special property of the transition probabilities}. The
differentials $d\epsilon $ also cancel, and our balance relation  acquires  the  simple
and symmetric form
\begin{displaymath}(5.9)\hsp
 p(s,\epsilon ^{\rm i}_{1})~ p(r,\epsilon^{\rm f}_1)~q(s^{\prime},\epsilon^{\rm i}_2)~q(r^{\prime},\epsilon^{\rm f}_2)
\end{displaymath}
\begin{displaymath}\hspp
= p(s-n,\epsilon^{\rm i}_1)~p(r+n,\epsilon^{\rm f}_1)~q(s^{\prime}-n^{\prime},\epsilon^{\rm i}_2)~q(r^{\prime}+n^{\prime},\epsilon^{\rm f}_2).
\end{displaymath} 
\noindent The relation is reminiscent of the relation for chemical equilibrium  between
several kinds of molecules. It still comprises the Bose and Fermi cases. There
is no spontaneous emission term, i.e. one that would be  independent  of  the
number of wavepackets. A general solution is
\begin{displaymath}(5.10)\hsp
p(s,\epsilon )=a(\epsilon ) \exp [-(b\epsilon -c)s] 
\end{displaymath}
\begin{displaymath}\hspp~
q(s,\epsilon )=a^{\prime}(\epsilon)\exp[-(b\epsilon-c^{\prime})s].
\end{displaymath} 
\noindent [Insert and use (5.6)]. Notice that only the parameter $b$ (which shortly  will
be identified with $1/kT)$ is the same for the two kinds  of  packets.  In  any
other respect the distribution function for kind-1 packets is independent  of
the distribution function for kind-2 packets.

The parameters $c$ and $b$  are  obtained  via  the  thermodynamic  relations
$\partial S\allowbreak /\partial E_{\mbox{tot}}\allowbreak = \allowbreak 1/T$  and $\partial S/\partial N_{\mbox{tot}}= - \mu /T$  where
\begin{displaymath}(5.11)\hsp
S = k \ln \prod_{\{d\epsilon_i\}} \frac{g_{p}!}{ [p(0,\epsilon _{i})d\epsilon _{i}]!~ [p(1,\epsilon _{i})d\epsilon _{i}]! \cdots }
 \end{displaymath}
\noindent is Bose's or Natanson's \cite{Nat} formula (in our notation) for the entropy of the
total system, with $N_{\mbox{tot}}=\sum _{\{d\epsilon _{i}\}} Ndp$  and $E_{\mbox{tot}}=\sum _{\{d\epsilon _{i}\}}\epsilon _{i} Ndp$ . The  total  energy
here is thought to be subdivided into a set of intervals $\{d\epsilon _{i}\}$.  The  entropy
(5.11) does not depend explicitly on  the  numbers  {\emph s}.  Formula  (5.11)  also
implies that the wavepackets (as units representing $s$  quanta)  are  thermodynamically independent \cite{Wol}, \cite{Ein05}. The number of all wavepackets in $V$ and $d\epsilon $,
including the empty ones, is given by (5.2), so we have
\begin{displaymath}(5.12)\hsp
\sum _{\{s\}} p(s,\epsilon ) d\epsilon  = g_{p} .
 \end{displaymath}
\noindent The total number of quanta in $V$ and $d\epsilon $ is {\emph Ndp}, so
\begin{displaymath}(5.13)\hsp
\sum _{\{s\}} s~ p(s,\epsilon ) d\epsilon  = Ndp.
 \end{displaymath}
\noindent With (5.10), (5.13) and  Stirling's  approximation $\ln p!\approx p\ln p$  the  entropy  in
thermal equilibrium may  be  written  as $S = k \left( bE_{\mbox{tot}}- cN_{\mbox{tot}} - \sum _{\{d\epsilon _{i}\}} g_{p}
\ln \left [ a(\epsilon _{i})d\epsilon _{i}/g_{p}\right ]\right)$  hence the above-mentioned thermodynamic relations lead to
\begin{displaymath}\hspp
b=1/kT,\qquad  c=\mu /kT .
\end{displaymath}
So far all mathematical operations  could  be  carried  out  even  if  the
numbers $s, s', r, r'$ were not integers.

\vspace{20pt}
\noindent {\bf 5.4~~The Bose and Fermi Distributions}
\medskip

\noindent
Now we go to special cases. First we take the kind-1 packets  to  be  \emph{boson}
packets. In this case the numbers $s, r$ and $n$ are  non-negative  integers  and
(5.12) becomes
\begin{displaymath}(5.14)\hsp
\sum^{\infty }_{s=0}p(s,\epsilon )\, d\epsilon  = g_{p} = {\mbox{{\emph ad}}\epsilon \over 1- \exp [-(\epsilon -\mu )/kT]} .
 \end{displaymath}
\noindent From this we obtain  {\emph ad}$\epsilon  = g_{p}(1-\exp [-(\epsilon-\mu )/kT])$ and
\begin{displaymath}(5.15)\hsp
p_{\rm B}(s,\epsilon )d\epsilon  = g_{p}\; (1-\exp [-(\epsilon -\mu )/kT])\; \exp [-s(\epsilon -\mu )/kT] .
 \end{displaymath}
\noindent This formula coincides with Bose's expression for the number  of  phase space
cells occupied with $s$ quanta. The total number of quanta (5.13) becomes
\begin{displaymath}(5.16)\hsp
Ndp = \sum^{\infty }_{s=0}s~ p(s,\epsilon ) \, d\epsilon  = {g_{p}\over \exp [(\epsilon -\mu )/kT] - 1}
 \end{displaymath}
\noindent and we identify these quanta, not the wavepackets, with the particles in  the
usual interpretation of formula (5.1). Thus we have arrived  at  the  desired
Bose-Einstein distribution function, Eq.~(5.1) with the minus sign.

Let us further consider photons, as a special kind of bosons, and  let  us
consider the processes where photons are absorbed and emitted  by  atoms.  In
this case we take the packets of kind 1 to be  the  photons and  the
packets of kind 2 to be the atoms. In one respect the situation  goes  beyond
the scheme of Fig.~5, in that the number of photons is no longer conserved.
In its interaction with the atom the photon is absorbed and exists no longer.
Thus, in Fig.~5 the arrow that points from the $n$-packet  in $d\epsilon ^{\rm i}_{1}$ (second
line) to the $(r+n)$-packet in $d\epsilon ^{\rm f}_{1}$ (third  line)  no  longer  exists,  and  the
$(r+n)$-packet remains an $r$-packet. Equivalently, one may say that the $n$-packet
turns into an empty packet $(n=0)$.

 Thus, in the balance  relation  (5.9)  the
function $p(r+n,\epsilon ^{\rm f}_{1})$ on the right-hand  side  becomes  equal  to  the  function
$p(r,\epsilon ^{\rm f}_{1})$ on the left-hand side  and  disappears  from  the  equation.  In  the
energy-conservation relation (5.6) we have to put  $\epsilon ^{\rm f}_{1}$=0. What the atom
does beyond satisfying the energy conservation in absorbing  and  emitting  a
photon is irrelevant. Likewise, in the reverse process the arrow that points
from the $(r+n)$-packet in $d\epsilon ^{\rm f}_{1}$ (fourth line) to  the $n$-packet  in $d\epsilon ^{\rm f}_{1}$ (fifth
line) no longer exists, or equivalently, refers to an empty packet. This does
not, however, affect the balance  equation,  which  thus  is
\begin{displaymath}(5.17)\hsp
p(s,\epsilon ^{\rm i}_{1})\; q(s^\prime ,\epsilon ^{\rm i}_{2})\; q(r^\prime ,\epsilon ^{\rm f}_{2}) = p(s-1,\epsilon ^{\rm i}_{1})\; q(s^\prime -n^\prime ,\epsilon ^{\rm i}_{2}) \; q(r^\prime +n^\prime ,\epsilon ^{\rm f}_{2}) .
 \end{displaymath}
\noindent The solution is again given by (5.10), but only if $c=0$ in $p(s,\epsilon )$ there, so we  have
obtained Planck's law.

There is no spontaneous emission in our treatment.  Let  us  compare  this
with Einstein's treatment. Einstein's balancing  equation  (Sec.~3  in  his  1917
paper \cite{Ein17}) is
\begin{displaymath}(5.18)\hsp
\exp (-\epsilon _{n}/kT)\, \rho  = \exp (-\epsilon _{m}/kT)\, (\rho +A^{n}_{m}/B^{n}_{m})
 \end{displaymath}
\noindent where $\rho  = h\nu ${\emph Ndp}$/(Vd\nu )$ is $h\nu $ times the (time averaged) number of  photons  per
unit volume and per unit frequency interval in  the  cavity.  This  equation,
unlike our Eq.~(5.9) or (5.17), is concerned with the number of quanta,
not with the number of wavepackets. The second term in the bracket, $A^{n}_{m}/B^{n}_{m} =
4\pi h\nu ^{3}/c^{3}$ (polarised radiation), is independent of $\rho $ and is  the  spontaneous 
emission term. One may obtain Einstein's Eq.~(5.18) from our Eq.~(5.17) if one  puts
back the transition probabilities $W_{1}$ and $W_{2}$ into this  equation  and  uses  a
special property of them. The left-hand side of (5.17)  effectively  means  a
process where an atom absorbs a photon  from  an $s$-photon  packet,  and  the
right-hand side a process where an atom emits a photon into an $(s-1)$-packet.
In order to compare with Einstein's  1917  treatment,  which  disregards  the
wavepacket structure of the radiation, one has to sum Eq. (5.17),  with
$W_{1}$ and $W_{2}$ restituted, over all photon packets, i.e. over all values of $s$ (with $n$=1):
\begin{displaymath}\hsppp
q(s^\prime ,\epsilon ^{\rm i}_{2})\; q(r^\prime ,\epsilon ^{\rm f}_{2}) \sum^{\infty }_{s=0}p(s,\epsilon ^{\rm i}_{1})\; W_{1}(s,\alpha ) =
\end{displaymath}
\begin{displaymath}(5.19)\hsp
= q(s^\prime - n^\prime ,\epsilon ^{\rm i}_{2})\; q(r^\prime +n^\prime ,\epsilon ^{\rm f}_{2}) \sum^{\infty }_{s=0}p(s-1,\epsilon ^{\rm i}_{1})\; W_{2}(s-1,\beta ) . 
 \end{displaymath}
\noindent $\alpha $ and $\beta $ are the other arguments in $W_{1}$ and $W_{2}$, which do not depend on  {\emph s}.  Now
we use the special property
\begin{displaymath}(5.20)\hsp
W_{1}(s,\alpha ) = W_{2}(s- 1,\beta ) = f\cdot s
 \end{displaymath}
\noindent where $f$ may depend on anything but $s$. This fits with Dirac's  statement  that
the probability of a transition in which a boson is  absorbed  from  (emitted
into) state $x$ is proportional to the number of bosons originally in  state $x
$  (in state $x$, plus one) \cite{Dir27}.

With Eq.~(5.20) one may write (5.19) in the form
\begin{displaymath}\hsp
\underbrace{ 
{q(s^\prime ,\epsilon ^{\rm i}_{2})\over q(s^\prime -n^\prime ,\epsilon ^{\rm i}_{2})}}_{C_1}d\epsilon ^{\rm i}_{1} \sum^{\infty }_{s=1}s~p(s,\epsilon ^{\rm i}_{1})
=\underbrace{
{q(r^\prime +n^\prime ,\epsilon ^{\rm f}_{2})\over q(r^\prime ,\epsilon ^{\rm f}_{2})}}_{C_2}d\epsilon ^{\rm i}_{1} \sum^{\infty }_{s=1}s~p(s-1,\epsilon ^{\rm i}_{1})
\end{displaymath}
\begin{displaymath}\hsp
= C_2 \left[d\epsilon ^{\rm i}_{1} \sum^{\infty }_{s=1} (s-1)~p(s-1,\epsilon ^{\rm i}_{1}) + d\epsilon ^{\rm i}_{1} \sum^{\infty }_{s=1}p(s-1,\epsilon ^{\rm i}_1)\right] ,
\end{displaymath}
\noindent and with (5.14) and (5.16) one obtains $C_{1}${\emph Ndp} $= C_{2}(${\emph Ndp}$+g_{p})$. If one  multiplies
this by $h\nu /(Vd\nu )$ and observes that $C_{1}/C_{2} = \exp [n'(\epsilon ^{\rm f}_{2}- \epsilon ^{\rm i}_{2})/kT]$  and $n'=1$  one
obtains Einstein's balancing equation (5.18). We note that  it  is  only  the
special form (5.18) of the  balancing  equation  that  requires  the  special
property (5.20) of the transition probabilities in order  to  arrive  at  the
Planck distribution.

Second, we take the kind-1 packets to be \emph{fermion} packets. In  this  case  we
have only zero- and one-quantum packets, and the sums in  (5.14)  and  (5.16)
range only from 0 to 1. Thus
\begin{displaymath}\hsppp
\sum^{1}_{s=0}p(s,\epsilon )d\epsilon  =\mbox{ {\emph ad}}\epsilon (1 + \exp [- (\epsilon -\mu )/kT]) = g_{p}.
\end{displaymath}
\noindent Hence
\begin{displaymath}(5.21)\hsp
p_{\rm F}(s,\epsilon )d\epsilon  = g_{p}\; (1 + \exp [-(\epsilon -\mu )/kT])^{-1}~\exp [-s(\epsilon -\mu )/kT]
 \end{displaymath}
\noindent and the total number of quanta is
\begin{displaymath}\hsppp
Ndp = \sum^{1}_{s=0}s\,p(s,\epsilon )d\epsilon  = {g_{p}\over \exp [(\epsilon -\mu )/kT]+1},
\end{displaymath}

\noindent which is the desired Fermi-Dirac distribution, i.e. formula  (5.1)  with  the
plus sign.

The balance relation (5.9) in the Fermi case may  be  specified  to  read
$(s=n=1, r=0)$
\begin{displaymath}\hspace{20pt}
p(1,\epsilon ^{\rm i}_{1})~p(0,\epsilon^{\rm f}_1)~q(s^{\prime},\epsilon^{\rm i}_2)~q(r^{\prime},\epsilon^{\rm f}_2) = p(0,\epsilon^{\rm i}_1)~p(1,\epsilon^{\rm f}_1)~q(s^{\prime}-n^{\prime},\epsilon^{\rm i}_2)~q(r^{\prime}+n^{\prime},\epsilon^{\rm f}_2)
 \end{displaymath}
\noindent and may also be given a specific interpretation: a fermion 1-packet in $d\epsilon ^{\rm i}_{1}$  is
altered in an interaction with a wavepacket of kind 2 whereby  it  goes  into
$d\epsilon ^{\rm f}_{1}$, conservation of energy requiring $\epsilon ^{\rm i}_{1}- \epsilon ^{\rm f}_{1} = n^\prime (\epsilon ^{\rm f}_{2}-\epsilon ^{\rm i}_{2})$. Then it goes into  a
region of phase space within $d\epsilon ^{\rm f}_{1}$ that is not yet occupied by  a  (non-empty)
wavepacket. The transition rate is proportional to the size of  this  region,
expressed in fundamental units, that is, to the number of  empty  packets  in
$d\epsilon ^{\rm f}_{1},\; p(0,\epsilon ^{\rm f}_{1})d\epsilon ^{\rm f}_{1}$.

\vspace{20pt}
\noindent {\bf 5.5~~Quantum Count Fluctuations}
\medskip

\noindent
Finally we extend our considerations on wavepackets in a cavity to include
fluctuations. To be definite, we consider a small subvolume $\upsilon $  of  the  total
cavity volume {\emph V}. We imagine that the subvolume is homogeneously  filled  with
detectors (groups of sensitive  atoms).  The  detectors  are  sensitive  only
within the interval $p\ldots 
p+dp$ of the absolute value of momentum or the corresponding energy interval $\epsilon \ldots \epsilon +  
d\epsilon $, and we assume that within this  interval  the
sensitivity is constant. These detectors are switched on during the  interval
$\Delta t$, and the number of counts is registered.  This  procedure  is  repeated  a
great many times, where the time intervals between the repetitions are  large
compared with $\Delta ${\emph t}. We then ask for the mean  square  deviation,  or  variance,
$(\Delta m)^{2}$ of the number of counts. Again we shall treat both  boson  and  fermion
wavepackets on an equal footing.

The subvolume $\upsilon $ together with the interval $dp$ define a certain  volume  of
phase space and with this a certain  number $g_{\upsilon }$  of  (empty  plus  non-empty)
wavepackets, given by formula (5.2) with $V$  replaced  by $\upsilon $.  The  number  of
wavepackets with which the counter can interact during $\Delta t$ is larger  than $g_{\upsilon }$
because (1) the switch-on time $\Delta t$ may be so long  that  many  sets  of  wavepackets, each set filling the counter volume at one time,  may  pass  through
the counter during $\Delta t$, and (2) the counter can also interact with wavepackets
that only partially extend into it. The (integer) number of wavepackets that
partially and/or totally cover the phase-space region of the  counter  during
$\Delta t$ is denoted by $g$, where $g \ge  g_{\upsilon }$ and $g \ge  1$. A count is always an  interaction
of the counter with a wavepacket, not with a quantum. The  number  of  quanta
represented by the $g$ packets fluctuates because between two measurements some
few-quantum packets may have replaced  some  many-quantum  packets  and  vice
versa. This is the only source of fluctuations. Fluctuations that arise  from
a non-empty wavepacket leaving the region without  another  non-empty  packet
entering it are already accounted  for  because  our  number  of  wavepackets
includes empty packets, so that a non-empty  packet  leaving  the  region  is
equivalent with an empty packet entering it.

First we want to consider a special property of a  condensed  wavepacket
which we shall use below.  As  the
condensed packet arises from the  process  where  some  of  the  one-particle
functions in the expansion of $\Psi _{\rm S}$ (3.14) become equal we  may  describe  it  by
means of the product
\begin{displaymath}(5.22)\hsp
\Psi _{\rm S} = \varphi ({\bf x}_{1},s_{1},t)\,\varphi ({\bf x}_{2},s_{2},t)\cdots \varphi ({\bf x}_{N},s_{N},t)
 \end{displaymath}
\noindent where the $s_{i}$ signify the spin variables, and  the  same $\varphi $  is  used  in  all
factors. $\varphi ({\bf x},s,t)$ here does  not  necessarily  mean  the  lowest-energy  one-particle state, as it does in Bose-Einstein condensation proper \cite{Bla}. We  may
mention that in the treatment of laser coherence \cite{Ern69}, \cite{Ern68} the  wave  function
of a stationary $N$-photon state can also be written as a product of  the  type
(5.22). The condensed packet is thus effectively described by the one function
$\varphi ({\bf x},s,t)$ in ordinary space. Consider the expression
\begin{displaymath}(5.23)\hspace{30pt}
P_{1} =\int_{ {\bf x}_1 \in D^3}  \int_{{\bf x}_i \in R^3\; (i\ge2)} \cdots \int
 | \Psi _{\rm S}({\bf x}_{1},{\bf x}_{2},\ldots 
,{\bf x}_{N},t)| ^{2}d^{3}x_{1}d^{3}x_{2}\ldots 
d^{3}x_{N}
\end{displaymath}
\begin{displaymath}\hsppp
=\int_{{\bf x} \in D^3} | \varphi ({\bf x},t)| ^{2}d^{3}x .
 \end{displaymath}
\noindent ${\bf x} \in D^3$  means $x_{a}\le x\le x_{b},\, y_{a}\le y\le y_{b},\, z_{a}\le z\le z_{b}.\; R^3$  means   total   space.   If
necessary  ${\bf x}, D^{3}$ and $R^{3}$ are to include the spin variables. For $D^{3} = d^{3}x$   the
expression reduces to $P_{1} = | \varphi ({\bf x},t)| ^{2}d^{3}${\emph x}.

In the Copenhagen interpretation expression (5.23)  means  the  probability
that particle 1 of a system of $N$ similar particles is found in the spatial
region $D^{3}$, irrespective of where the other $N-1$ particles are  found.  In  our
interpretation it is the probability that wavepacket 1 acts in $D^{3}$,  irrespective of where the other $N-1$ wavepackets act. Actually, the wavepackets are
equal, so what we really want is an expression for the probability  that  any
one packet acts in $D^{3}$, irrespective of where the other $N-1$ packets act, which
is $N\cdot P_{1}$. In the same way the standard formalism gives the probability $P_{2}$ that
any $m$ wavepackets of the $N$ act in $D^{3}$ while the other $N-m$ do not act in $D^{3}$
\begin{displaymath}(5.24)\hspace{30pt}
P_{2} = \mbox{$ N \choose m$ } \int_{{\bf x}_1 \in D^3} \cdots \int_{{\bf x}_m \in D^3 }\int_{{\bf x}_{m+1} \in R^3}^{\prime} \cdots \int_{{\bf x}_N \in R^3}^{\prime} |\Psi_{\rm S}({\bf x}_{1},{\bf x}_{2},\ldots 
,{\bf x}_{N},t)| ^{2}
\end{displaymath}
\begin{displaymath}\hsp
\hspace{50pt}\times  d^{3}x_{1}d^{3}x_{2}\cdots 
d^{3}x_{N}
 \end{displaymath}
\noindent where a prime at the integral sign means that  in  the  integration  over  the
variables ${\bf x}_{i}$ the region $D^{3}$ has to be excluded. The combinatorial  factor $ N \choose m $ is just the number of ways $m$ billiard  balls  (or  particle  labels)  can  be
chosen from {\emph N}. In the case of the condensed packet of the product form  (5.22)
$P_{2}$ reduces to
\begin{displaymath}(5.25)\hsp
P_{2} =\mbox{$ N \choose m$ } \eta ^{m}(1- \eta )^{N-m}
 \end{displaymath}
\noindent with
\begin{displaymath}(5.26)\hsp
\eta  = \int_{{\bf x}\in D^3}| \varphi ({\bf x},t)| ^{2}d^{3}x .
 \end{displaymath}
\noindent This is the binomial distribution. It is just the probability that $N$ independent trials with probabilities $\eta $ for success and $1-\eta$ for failure result in $m$ successes and $N-m$ failures. We may thus say that $P_2$ is the probability that the condensed packet acts with $m$ of its $N$ quanta in $D^3$.

Now, of all the quanta of a wavepacket  only  some  fraction  will  be  counted. The  probability  of $m$
counts from an $n$-quantum wavepacket is formula (5.25) with $n$ instead of $N$:\begin{displaymath}(5.27)\hsp
b(m;n,\eta ) =\mbox{$n \choose m$}  \eta ^{m}(1 -  \eta )^{n-m}
 \end{displaymath}
\noindent where $\eta  = \bar m /n \;  (0 \le  \eta  \le  1, \;(\Delta m)^2=\bar m(1-\eta))$ now is the average fraction of  quanta  of  the  packet
that are counted during the interval $\Delta ${\emph t}. In (5.26) we would have  $\eta =1$ if $D^{3}=R^{3}$. Now,  the  detector  volume $\upsilon $
may happen to be much larger than a wavepacket, and in this case $\upsilon $ is  effectively equivalent to $R^{3}$, leading to $\eta =1$ in Eq.~(5.26). Eq.~(5.27)  would  then  always
give zero for $m\neq n$ and be useless. It is, nevertheless, possible  to  maintain
formula (5.27) even in this situation if we take into account that the efficiency $\eta$ is not only limited by the finite counter volume, as accounted for in Eq.~(5.26), but also by the intrinsic counter efficiency and short interval $\Delta t$, so that even when $D^{3}\rightarrow R^{3}$  the efficiency $\eta$ may be less than unity, and we can maintain formulas (5.25)  and  (5.27),  independently  of  whether  the
counter volume covers the wavepacket totally or partially.  The
difference between these two cases is absorbed in the numerical value
of $\eta$, that value suffering an additional decrease when we go from the
case of total to that of partial spatial covering.

What, then, is the
probability $W(m;g)$ of counting $m$ quanta from $g$ wavepackets? To
answer this question we first evaluate the probability $w(n;g)$ that
the $g$ packets represent $n$ quanta and then the probability $B(m;n)$
that of these $n$ quanta $m$ are counted, and we write $W(m;g) =
\sum^{\infty }_{n=m} w(n;g)~ B(m;n)$.

We first consider \emph{boson} packets. The probability that a randomly
chosen boson packet is an $s$-quantum packet is given by (5.15). From this we obtain 
the average number $\bar s$
of quanta represented by one packet averaged over all packets in $V$
and $dp$ (or the corresponding $d\epsilon$)
\begin{displaymath}(5.28)\hsp
\bar s = \sum^{\infty
}_{s=0} s~ p(s)d\epsilon /g_{p}=\exp[-(\epsilon-\mu)/kT](1-\exp[-(\epsilon-\mu)/kT])^{-1}
\end{displaymath}
so that $\exp[-(\epsilon-\mu)/kT]=\bar s/(1+\bar s)$ and
 \begin{displaymath}(5.29)\hsp
 p(s)d\epsilon /g_{p} = {1\over (1+\bar s)}
{1\over (1+1/\bar s)^s}.
\end{displaymath}

 \noindent
The probability that of the $g$ packets in $\upsilon $ the first one
represents $s_{1}$, the second $s_{2},\ldots $ and the $g$-th $s_{g}$
quanta, with $\sum ^{g}_{i=1} s_{i} = n$, is the product
 \begin{displaymath}(5.30)\hsp
\prod_{i=1}^g p(s_{i})d\epsilon /g_{p} = {1\over (1+\bar s)^{g}} {1\over (1+1/\bar s )^{n}} .
 \end{displaymath}
\noindent We are, however, not interested in the particular distribution (which  quanta
are represented by which packets), so we have to form a sum of
expressions (6.30), one for each distribution. Since (6.30) is the same
for any distribution we need only to multiply (6.30) by the number of
possible distributions, given by the well-known combinatorial
expression
 \begin{displaymath}(5.31)\hsp
 {g+n-1 \choose n} \equiv {(g+n-1)! \over (g-1)!\; n!}. 
 \end{displaymath}
\noindent Thus the probability that the $g$ packets represent $n$ quanta is
\begin{displaymath}(5.32)\hsp
 w(n;g) 
= \mbox{$g+n-1 \choose n$}  {1\over (1+\bar s)^{g}} {1\over (1+1/\bar s)^{n}} .
 \end{displaymath}
\noindent This formula was already obtained by Mandel \cite{Man59} in a related context.

Next we evaluate the probability $B(m;n)$ that if $n$ quanta are present $m$ are
counted. Here we take advantage of the fact that this probability is independent of how these quanta are represented by the different wavepackets. This 
is seen in the following way: assume that all the $n$ quanta are from  one  and
the same wavepacket. Then the probability of $m$ counts is  given  by  the  binominal distribution (5.27). Next assume that the quanta are from  two  wavepackets, one with $n_{1}$ and one with $n_{2}=n- n_{1}$ quanta. The probability of $m$ counts
would then be
\begin{displaymath}\hsppp
\underbrace
{\sum_{m_1}\sum_{m_2}}_{m_1+m_2=m}  \mbox{$ n_1 \choose m_1$}  \eta ^{m_{1}} (1-\eta )^{n_{1}-m_{1}} \;  \mbox{$n_2 \choose m_2$}  \eta ^{m_{2}} (1-\eta )^{n_{2}-m_{2}} .
\end{displaymath}
\noindent But due to a special folding property of the binominal  distribution \cite[p.~173, 268]{Fel}, this is equal to $b(m;n,\eta )$ of (5.27). And this would remain so if  the  quanta
were from any number of packets. We may thus assume that  all $n$  quanta  are
from one and the same wavepacket whence $B(m;n) = b(m;n,\eta )$  of  (5.27).  In
deriving this result we have used the same value of $\eta $  for  all  wavepackets.
This requires a justification  because  when  the  phase-space  volume  of  a
wavepacket $\Delta ^{3}x\Delta ^{3}p$ and that of the counter $4\pi \upsilon p^{2}dp$ with which it interacts are
comparable, their overlap and with this the value of $\eta $ may  vary  appreciably
from one wavepacket to the next, even if, as we assume,  all  wavepackets  in
the cavity have very nearly the same size in phase space. In  this  case  the
above folding theorem in fact no longer  holds  generally,  although  still  in
special cases $(\eta \ll 1, \eta \approx 1,$ Poisson approximation,  normal  approximation \cite[Chap.~6]{Fel}).
When, however, the phase-space volume  of  the  counter  is  large  (in  each
direction) compared with that of a wavepacket the counter covers almost  each
wavepacket completely, and since we consider a counter with constant sensitivity over its whole phase-space volume, $\eta $ is still the same for all  packets.
When, in the opposite limit, the counter is small  compared  with  the  wavepacket, it is true that $\eta $ may vary considerably because the counter may cover
regions of the wavepacket with varying $|  \psi |  ^{2}$ (in $x$ space or in $p$ space) since  we
have not assumed that $|  \psi |  ^{2}$ is constant over the wavepacket. But  now,  whichever region of a wavepacket is covered by  the  counter, $\eta $  will  always  be
small, and under this condition the folding theorem still holds in the form
\begin{displaymath}\hspace{70pt}
\underbrace
{\sum_{m_1}\sum_{m_2}}_{m_1+m_2=m}  b(m_{1};n_{1},\eta _{1})~ b(m_{2};n_{2},\eta _{2}) -  b\!\!\left( m;n_{1}+n_{2},{n_{1}\eta _{1}+n_{2}\eta _{2}\over n_{1}+n_{2}}\right) 
\end{displaymath}
\centerline{
$\propto  (\eta _{1}-\eta _{2})^{2}$  +  terms of higher order in $\eta _{1}$ and $\eta _{2}$ ,}
\bigskip

\noindent so that $B(m;n) = b(m;n,\eta )$ still holds with $\eta $ representing some  average  over
different regions of the wavepacket. Therefore we consider $B(m;n) =
b(m;n,\eta )$ as an acceptable approximation.

With this the probability $W(m;g)$ of counting $m$ quanta when $g$
packets are in $4\pi \upsilon p^{2}dp$ is
\begin{displaymath}(5.33)\hspace{50pt}
W(m;g) =\sum^{\infty }_{n=m}w(n;g)~b(n;m,\eta )
 =\sum^{\infty }_{n=m}w(n;g)\; \mbox{$n\choose m$} \eta ^{m}(1- \eta )^{n-m} 
\end{displaymath}
\begin{displaymath}\hsppp=\sum^{\infty }_{n=m}
\mbox{$g+n-1\choose n$}  {1\over (1+\bar s)^{g}} {1\over (1+1/\bar s)^{n}} \mbox{$n\choose m$} \eta^{m}(1-\eta )^{n-m} .
\end{displaymath}
\noindent Formula (5.33) is the same as that obtained  via  the  standard  quantization
for\-mal\-ism \cite{Mol}, as it should be. Substituting $l = n- m$ and using the  binomial
identity  ${{l+g-1+m} \choose l}\allowbreak  = \allowbreak (- 1)^{l} {{-g-m} \choose l}$ and Newton's binomial formula  $\sum^{\infty }_{l=0}
 \allowbreak {{-g-m}\choose l} \allowbreak \left (- {1-\eta \over 1+1/s} \right )^l \allowbreak  = \allowbreak \left(1-{1-\eta \over 1+1/s}\right)^{-g-m}$  we obtain
\begin{displaymath}(5.34)\hsp
W(m;g) = \mbox{$g+m-1 \choose m$}  {1\over (1+\eta \bar s)^{g}} {1\over (1+1/(\eta \bar s))^m} .
 \end{displaymath}
\noindent From this we can  calculate  the  variance $(\Delta m)^{2} := \sum^{\infty }_{m=0}(m- \bar m)^{2}\; W(m;g) =
\eta \bar s g (1+\eta \bar s)$ [use the generating function \cite[p. 266]{Fel} of $W(m;g)$]. Observing that $\eta \bar s g =
\bar m$ is the average number of counted quanta from $g$ packets we may finally write
the variance of the number $m$ of counts as
\begin{displaymath}(5.35)\hsp
(\Delta m)^{2} =\bar m \left( 1+ {\bar m \over g} \right) .
 \end{displaymath}
\noindent Comparing (5.34)  with  (5.32)  we  see  that  the  distribution,  and  hence
variance, of the counted quanta has the same form as  that  of  the  existing
quanta.

Our procedure is easily carried over to \emph{fermion} packets. We  only  have  to
observe that there are only zero- and one-quantum packets. Of  course  (5.22)
no longer holds, but (5.27) still does since for $m\le n\le 1$ it  reduces  to  three
trivial expressions. Further, instead of (5.29) we have to take (5.21)  which
we write in the form (cf. (5.28))
\begin{displaymath}(5.36)\hsp
p(s)d\epsilon /g_{p} = (1-\bar s) {1\over (1/\bar s - 1)^{s}} ,
 \end{displaymath}
\noindent and instead of (5.31) we have to take
\begin{displaymath}\hsppp
{g \choose n} = {g!\over (g-n)!\,n!} .
 \end{displaymath}
\noindent (5.34) is then replaced by
\begin{displaymath}(5.37)\hsp
W(m;g) =\mbox{$ g \choose m$} (1-\eta \bar s)^{g} {1\over (1/(\eta \bar s)-1)^m }
\end{displaymath}
\noindent and (5.35) by
\begin{displaymath}\hsppp
(\Delta m)^{2} =\bar m \left( 1-{\bar m \over g} \right) .
 \end{displaymath}
For $g=1$ Formula (5.34) becomes $(\bar m = \eta \bar s g)$
\begin{displaymath}(5.38)\hsp
W_{\rm B}(m;1) = {\bar m^m \over (1+\bar m)^{m+1}}\qquad \mbox{(Bose)}
 \end{displaymath}
\noindent and (5.37)
\begin{displaymath}\hsppp
W_{\rm F}(m;1) = {\bar m^m \over (1- \bar m)^{m-1}}\qquad \mbox{(Fermi)} .
 \end{displaymath}
For $g\rightarrow \infty $ both (5.34) and (5.37) approach the Poisson distribution
\begin{displaymath}(5.39)\hsp
W_{\rm BM}(m;\infty ) = {\bar m^m \over m!} e^{-\bar m}\qquad \mbox{(Boltzmann)} .
 \end{displaymath}

It seems that our results not only hold for cavities but also  for  beams,
at least in some situations. This we infer  from  the  observation  that  the
distributions (5.38) and (5.39) are also obtained  for  stationary  beams  of
chaotic light in standard quantized radiation theory \cite{Lou} and  are  confirmed
by experiment \cite{Are}. Formula (5.38) obtains  when  the  counting
time $\Delta t$ is short compared with the  coherence  time $\tau_{c},\; \Delta t\ll \tau_{c}$,  and  (5.39)
obtains in the opposite limit, $\Delta t\gg \tau_{c}$. These limits can be compared  with  the
limits $g=1$ and $g\gg 1$ of our treatment because the wavepacket structure  of  the
radiation field reflects its coherence properties, in that the  spatial  size
of a photon wavepacket is a measure of the size of the coherence region  with
length $L_{c}=c\tau_{c}$ (Appendix A). The limit $\Delta t\gg \tau_{c}$ means that many  wavepackets  can
interact with the counter during $\Delta t$, either consecutively, if the packets are
large, or simultaneously, if they are small compared with the counter  volume
$\upsilon $. Thus $g\gg 1$ in this limit. In the opposite limit, $\Delta t\ll \tau_{c}$, the wavepackets  can
interact with the counter only during such a short time interval  that  their
movement is negligible. Whether we have $g\gg 1$ or $g=1$  now  depends  on  further
specifications. When the packets are large and  pass  over  the  counter  one
after the other we have $g=1.$ This was  in  fact  implicitly  assumed  in  the
(plane-wave) calculations referred  to  by  Loudon \cite{Lou}  and  was  explicitly
stated in the experimental verification \cite{Are}  of  Formula  (5.38).  So,  here
$\Delta t\gg \tau_{c}$ means $g\gg 1,$  and $\Delta t\ll \tau_{c}$  means $g=1,$  and  the  situations  are  those
covered by the above formulas.

\newpage
\noindent
{\bf  AKNOWLEDGEMENT}
\medskip

\noindent
I thank my wife Marianne for her understanding and support over all these years.
\vspace{ 40pt}

\noindent
{\bf APPENDIX~~A:     Wavepacket Spreading Formulas}
\medskip

\noindent
The wavepackets considered here are the usual free packets of  Schr\"odinger
or de Broglie waves. Since the formulas are rather spread out in the  literature
and are often restricted to special cases we have collected  here  some  more
general results for easy reference.

The general wavepacket is written in the form
\begin{displaymath}     (\mbox{A}1) \hsp
\psi ({\bf x},t) = (2\pi )^{-3/2} \int^{+\infty }_{-\infty }\tilde{\psi }({\bf k}) \exp [i({\bf kx} - \omega ({\bf k})t)] d^{3}k  
\end{displaymath}
\noindent where $\tilde{\psi }({\bf k})$, and hence $\psi ({\bf x},t)$, is normalized
\begin{displaymath}\hspp
\int^{+\infty }_{-\infty }| \tilde{\psi }({\bf k})| ^{2}d^{3}k = 1 . 
\end{displaymath}
\noindent The (three-dimensional) Fourier transform of $\psi ({\bf x},t)$ is
\begin{displaymath}\hspp
\tilde{\psi }({\bf k},t) = \tilde{\psi }({\bf k}) \exp [-i\omega ({\bf k})t] .
\end{displaymath}
\noindent This is the wavepacket in $k$ (momentum) space. Mathematically  the  wavepacket
need not have a sharp boundary in $x$ space (or in $k$ space), but for  practical
purposes it may be considered to have a finite extension given, for  example,
by the standard deviation
\begin{displaymath}           (\mbox{A}2)       \hsp
\Delta x := \langle (x - \langle x\rangle )^{2}\rangle ^{1/2}  
\end{displaymath}
\noindent with
\begin{displaymath}\hspp
\langle x\rangle (t) = \int^{+\infty }_{-\infty }\psi^*({\bf x},t)\; x\; \psi ({\bf x},t)\; d^{3}x ,  
\end{displaymath}
\noindent and with analogous expressions for the widths $\Delta y$ and $\Delta ${\emph z}. In the special  case
of the Gaussian form (and $t=0)$
\begin{displaymath}\hspp
| \psi (x)| ^{2} = (2\pi \sigma ^{2})^{-1/2} \exp [-(x -\langle x\rangle )^{2}/(2\sigma ^{2})]  
\end{displaymath}
\noindent $\Delta x$ of (A2) is equal to the parameter $\sigma $  and  is  the  distance  between  the position of the
maximum and the point where the distribution has fallen off  to $\exp (- 1/2) =
0.61$ of the maximum value. Sometimes the form of $\psi ({\bf x})$ or of $\tilde{\psi }({\bf k})$ is such that
the integrals in (A2) etc. diverge, for example for $\tilde{\psi }({\bf k}) \propto  \sin (ak)/k$ and for
$\tilde{\psi }({\bf k}) \propto  (1 +a^{2}k^{2})^{-1/2}$. In such cases one uses other definitions of the  width,
for instance the distance between the maximum and the first  zero,  the  half
width, the equivalent width \cite{Bau} or the overall width \cite{Hil}.

The dispersion law $\omega ({\bf k})$ is determined by the relation between the energy $E$
and the momentum ${\bf p}$ of the object the wavepacket is to represent  provided  we
make use of the Einstein-Planck relation
\begin{displaymath}\hspp
\nu  = E/h\mbox{      or     }\omega  = E/\hbar   
\end{displaymath}
\noindent and the de Broglie relation
\begin{displaymath}\hspp
{\bf p} = \hbar {\bf k} 
\end{displaymath}
\noindent where
\begin{displaymath}\hspp
k \equiv  | {\bf k| } = 2\pi /\lambda  .  
\end{displaymath}
\noindent Thus, the relativistic relation for a free particle
\begin{displaymath}\hspp
E = E_{\mbox{tot}} = \pm (p^{2}c^{2} +m^{2}c^{4})^{1/2} 
\end{displaymath}
\noindent leads to
\begin{displaymath}            (\mbox{A}3)             \hsp
\omega ({\bf k}) = \pm c(k^{2} +\kappa ^{2})^{1/2} 
\end{displaymath}
\noindent where
\begin{displaymath}\hspp
\phantom{n}^{-} \llap{$\lambda$}_{\rm C} = 1/\kappa  = \hbar /(mc) 
\end{displaymath}
\noindent is the Compton length belonging to the rest-mass parameter {\emph m}.

The most general solution of a  relativistic  wave  equation  would  be  a
function of the type (A1) in which $\omega ({\bf k})$ from (A3) has  the  positive  sign
plus a function (A1) in which $\omega ({\bf k})$ has the negative sign. That is, the  most
general form would be a superposition of waves with positive as well as those
with negative frequencies viz. energies. Although we shall consider the general
relativistic formula (A3) we will  restrict  ourselves  to  positive-energy
wave functions in order to have a  simple  connection  with  the  results  of
nonrelativistic quantum mechanics.

Let us consider the time dependence of the wavepacket's ``centre'' $\langle {\bf x\rangle }$  and
``width''
\begin{displaymath}\hspp
\sigma (t) = ( [\Delta x(t)]^{2} + [\Delta y(t)]^{2} + [\Delta z(t)]^{2} )^{1/2}  
\end{displaymath}
\noindent where
\begin{displaymath}\hspp
[\Delta x(t)]^{2} := \int (x- \langle x\rangle )^{2}| \psi ({\bf x},t)| ^{2}d^{3}x 
\end{displaymath}
\noindent and analogously for the $y$ and $z$ components. A  very  general  calculation  of
these quantities has been given by  Bradford \cite{Bra}.  His  treatment  works  in
three dimensions and does not require a special form either for $\tilde{\psi }({\bf k})$  or  for
$\omega ({\bf k})$, except for the usual convergence requirements  of  the  integrals  that
appear in the averaging procedures. In particular, the treatment is valid  in
the nonrelativistic as well as in the relativistic domain. The result for the
centre is
\begin{displaymath}               (\mbox{A}4)\hsp
\langle {\bf x\rangle }(t) = \langle {\bf x\rangle }(0) + \langle \vec{\upsilon }_{g}\rangle t  
\end{displaymath}
\noindent with
\begin{displaymath}\hspp
\langle {\bf x\rangle }(0) = - \int \mbox{Im}\left\{ \tilde{\psi}^*\nabla _{k}\tilde{\psi }\right\}d^{3}k  
\end{displaymath}
\begin{displaymath}  (\mbox{A}5)\hsp
\langle \vec{\upsilon }_{\rm g}\rangle  = \int | \tilde{\psi }| ^{2}\nabla _{k}\omega \; d^{3}k   
\end{displaymath}
\begin{displaymath}\hspp
\tilde{\psi } = \tilde{\psi }({\bf k}) .
\end{displaymath}
\noindent According to (A4) and (A5) the centre moves  at  constant  velocity $\langle \vec{\upsilon }_{\rm g}\rangle $
which is the mean group velocity.

The result for the width is
\begin{displaymath}(\mbox{A}6)\hsp
\sigma ^{2}(t) = \sigma ^{2}(t_{0}) + [\Delta \upsilon _{\rm g}]^{2}(t- t_{0})^{2}  
\end{displaymath}
\noindent where
\begin{displaymath}\hspp
\sigma ^{2}(t_{0}) = \sigma ^{2}(0) - [\Delta \upsilon _{\rm g}]^{2}t^{2}_{0}  
\end{displaymath}
\begin{displaymath}  (\rm A.7)\hsp
\sigma ^{2}(0) = - \int \mbox{Re}\left\{ \tilde{\psi}^*\nabla ^{2}_{k}\tilde{\psi} \right\} d^3k - [ \langle {\bf x\rangle }(0) ]^2 
\end{displaymath}
\begin{displaymath}\hspp
t_{0} = \left( \int \mbox{Im} \left\{ \tilde  {\psi }^* \nabla _{k}\tilde{\psi} \right\} \nabla_k \omega \; d^{3}k + \langle {\bf x} \rangle (0)\langle \vec{\upsilon }_{\rm g}\rangle \right) [\Delta \upsilon _{\rm g}]^{-2}  
\end{displaymath}
\noindent and
\begin{displaymath}\hspp
[\Delta \upsilon _{\rm g}]^{2} := \int | \tilde{\psi }| ^{2}(\nabla _{k}\omega - \langle \vec{\upsilon }_{\rm g}\rangle )^{2} \; d^{3}k .  
\end{displaymath}
\noindent Formula (A6) shows that the width varies  hyperbolically  with  time.  This
type of variation is even independent of the form of the dispersion law $\omega ({\bf k})$,
except when $\Delta \upsilon _{\rm g}$ is zero: there is either a hyperbolic dependence or none. The
minimum extension of the wavepacket occurs at $t=t_{0}$; before $t_{0}$ the  wavepacket
shrinks, after $t_{0}$ it spreads out.

When we write $\tilde{\psi }({\bf k})$ in the polar form
\begin{displaymath}\hspp
\tilde{\psi }({\bf k}) = \rho ({\bf k}) \exp [i\alpha ({\bf k})] 
\end{displaymath}
\noindent we are led to
\begin{displaymath}    (\mbox{A}8)\hsp
\langle {\bf x\rangle }(0) = - \langle \nabla _{k}\alpha \rangle  
\end{displaymath}
\begin{displaymath}  (\mbox{A}9)\hsp
t_0 = \left( \langle \nabla _{k}\alpha~  \nabla _{k}\omega \rangle - \langle \nabla _{k}\alpha \rangle \langle \nabla _{k}\omega \rangle \right)\;  [\Delta \upsilon _{\rm g}]^{-2} .  
\end{displaymath}
\noindent We thus can fix the time and  the  place  of  the  minimum  extension  by  an
appropriate choice of the phase $\alpha ({\bf k})$,  that  is,  by $\langle \nabla _{k}\alpha \rangle $  and $\langle \nabla _{k}\alpha ~\nabla _{k}\omega \rangle $.
Resolving the two Eqs. (A8) and (A9) for  these  two  quantities  we
obtain
\begin{displaymath}(\mbox{A}10)\hsp
\langle \nabla _{k}\alpha \rangle  = \langle \nabla _{k}\omega \rangle \, t_{0} - \langle {\bf x\rangle }(t_{0}) 
\end{displaymath}
\begin{displaymath}\hsppp
\langle \nabla _{k}\alpha  \nabla _{k}\omega \rangle  = \langle [\nabla _{k}\omega ]^{2}\rangle \, t_{0} - \langle \nabla _{k}\omega \rangle \, \langle {\bf x\rangle}\! (t_{0}) .  
\end{displaymath}
\noindent For example, the especially simple form with three parameters $\vec{\xi }_{0}, \tau_{0}$ and $\eta _{0}$
\begin{displaymath}\hsppp
\alpha ({\bf k}) = - {\bf k} \vec{\xi }_{0} + \omega ({\bf k}) \tau_{0} + \eta _{0}  
\end{displaymath}
\noindent leads to
\begin{displaymath}\hsppp
\langle \nabla _{k}\alpha \rangle  = - \vec {\xi _{0}} + \tau_{0}\langle \nabla _{k}\omega \rangle   
\end{displaymath}
\noindent which, by comparison with  (A10),  shows  that  the  parameters $\vec{\xi }_{0}$  and $\tau_{0}$
coincide with the initial values
\begin{displaymath}\hsppp
\vec{\xi }_{0} = \langle {\bf x\rangle }(t_{0}), \qquad \tau_{0} = t_{0} , 
\end{displaymath}
\noindent and $\eta _{0}$ is an arbitrary constant.

In what follows we will always assume 
$\langle \nabla _{k}\alpha \rangle  =\langle \nabla _{k}\alpha  \nabla _{k}\omega \rangle=0$,  so  that
the minimum extension occurs at $t=t_{0}=0$, and the centre of the packet at  that
time is $\langle {\bf x\rangle }(0)=0$. Formulas (A6) and (A7) then simplify to
\begin{displaymath}\hsppp
\sigma ^{2}(t) = \sigma ^{2}(0) + [\Delta \upsilon _{\rm g}]^{2}t^{2}  
\end{displaymath}
\noindent and
\begin{displaymath}\hsppp
\sigma ^{2}(0) = - \int \mbox{Re}\left\{ \tilde{\psi}^* \nabla ^{2}_{k}\tilde{\psi } \right\} d^{3}k .
\end{displaymath}

Let us now consider the spreading velocity
\begin{displaymath}  (\mbox{A}11)  \hsp
\upsilon _{\rm s} := \partial \sigma (t)/\partial t = [\Delta \upsilon _{\rm g}]^{2}\; t/\sigma (t)  
\end{displaymath}
\noindent which for large $t$ tends to the asymptotic spreading velocity
\begin{displaymath}\hsppp
\upsilon _{{\rm s}\infty } := \lim_{t\rightarrow \infty } \upsilon _{\rm s} = \Delta \upsilon _{\rm g} .  
\end{displaymath}
\noindent The packet spreads out to double its initial $(t=0)$ extension in a time $ \tau_{2}  =
\sqrt 3 \sigma (0)/\Delta \upsilon _{\rm g}$. After this time the spreading velocity is $\upsilon _{\rm s}
 = (\sqrt 3 /2) \Delta \upsilon _{\rm g} = 0.87
\upsilon _{{\rm s}\infty }$. That is, $87\%$ of the asymptotic spreading  velocity  is  already  reached
when the packet has doubled its initial extension. The asymptotic  value  may
thus be used in all practical estimates. At the time $t_{1}$ when  the  asymptotic
spreading  velocity  has  (practically)  been  reached,  the  time  for   the
wavepacket to further double its extension is $\sigma (t_{1})/\Delta \upsilon _{\rm g}$  which  is  somewhat
smaller than $\tau_{2}$ provided we identify $\sigma (t_{1})$ with $\sigma (0)$ in this comparison.
\smallskip

To proceed further we must make specific assumptions about $\tilde{\psi }({\bf k})$. We  shall
assume a nearly unidirectional and quasimonochromatic packet, that is, a  {\emph{narrow}}
packet in $k$ space; that is, $\tilde{\psi }({\bf k})$ is assumed to be appreciably different  from
zero only in a narrow region concentrated about the point ${\bf k}_{0} = (0,k_{0},0)$  so
that
\begin{displaymath}  (\mbox{A}12) \hsp
\Delta k_{x}, \Delta k_{y}, \Delta k_{z} \ll  | {\bf k}_{0}|  \equiv  k_{0}=2\pi/\lambda_0 .  
\end{displaymath}

\noindent With the help of the Fourier reciprocity (Heisenberg) relations for the considered wavepackets
\begin{displaymath}   (\mbox{A}13) \hsp
\Delta x(0) \Delta k_{x} = \frac{1}{2}\,r, \mbox{  etc. ~~~~~with $r \ge 1$}.  
\end{displaymath}

\noindent where $r$ depends on the form of the wavepacket, and $r$=$1$ can only occur for a  Gaussian  form  for $\tilde{\psi }({\bf k})$,  the
condition (A12) can be written as a condition in ordinary space
\begin{displaymath}       (\mbox{A}14)  \hsp
\Delta x(0), \Delta y(0), \Delta z(0) \gg  r/(2k_{0}) .  
\end{displaymath}
\noindent It is then possible to express $\Delta \upsilon _{\rm g}$ as a function of $\Delta k_{x}$ and $\Delta k_{y}$. In the  case
where $k_{0}\neq 0$  it  is  reasonable  to  consider  separately  the   longitudinal
spreading, along the direction $y$ of the centre, and the transverse spreading,
normal to that direction, say in the $x$ direction. We then expand $\nabla _{k}\omega  = c^{2}{\bf k}/\omega ({\bf k})$
in a three-dimensional Taylor series about ${\bf k}_{0}$ and break the series off  after
the quadratic terms. After a  straightforward  but  tedious  calculation  one
arrives at
\begin{displaymath}\hsppp
\upsilon _{{\rm s}x\infty } := \lim_{t\rightarrow \infty } \partial \Delta x(t)/\partial t
\end{displaymath}
\begin{displaymath}\hsppp
= \Delta \upsilon _{{\rm g}x} :=  \left( \int | \tilde{\psi }| ^{2}(\partial \omega /\partial k_{x} -\langle \upsilon _{{\rm g}x}
\rangle )^{2}d^{3}k \right)^{1/2}
\end{displaymath}
\begin{displaymath}   (\mbox{A}15)  \hsp
= {c^{2}\over \omega _{0}} \Delta k_{x} = {c\over (k^{2}_{0}+\kappa ^{2})^{1/2}} \Delta k_{x} 
\end{displaymath}
\begin{displaymath}  (\mbox{A}16) \hsp
 ={r\,c\over 2(k^{2}_{0}+\kappa ^{2})^{1/2}\Delta x(0)},  
\end{displaymath}
\noindent where (A13) has been used for obtaining (A16). Likewise we obtain 
\begin{displaymath}\hsppp
\vec{\upsilon }_{0} \equiv  \langle \vec{\upsilon }_{\rm g}\rangle  = (0,k_{0}c^{2}/\omega _{0},0)  
\end{displaymath}
\begin{displaymath}\hsppp
\upsilon _{0} = k_{0}c^{2}/\omega _{0} ,\quad \omega _{0} = c(k^{2}_{0}+\kappa ^{2})^{1/2} , 
\end{displaymath}
\noindent and with this we may write (A15) as
\begin{displaymath}  (\mbox{A}17)\hsp
\upsilon _{{\rm s}x\infty } = {c\over \kappa }\Delta k_{x} \left( 1 - (\upsilon _{0}/c)^{2}\right)^{1/2}  
\end{displaymath}
\begin{displaymath}  (\mbox{A}18)  \hsp
 \ge  {c\left( 1 - (\upsilon _{0}/c)^{2} \right)^{1/2} \over 2\kappa \Delta x(0)} .  
\end{displaymath}
\noindent In the same way we arrive at the longitudinal asymptotic spreading velocity
\begin{displaymath} (\mbox{A}19)  \hsp
\upsilon _{{\rm s}y\infty } = \left( {c\kappa \over \omega _{0}}\right)^{2} {c^{2}\over \omega _{0}} \Delta k_{y} = {c\over \kappa } \Delta k_{y}\left( 1- (\upsilon _{0}/c)^{2} \right)^{3/2}   
\end{displaymath}
\begin{displaymath}  (\mbox{A}20)  \hsp
\ge  {c\left( 1- (\upsilon _{0}/c)^2 \right)^{3/2} \over 2\kappa \Delta y(0)} 
\end{displaymath}
\noindent and
\begin{displaymath}      (\mbox{A}21) \hsp
\upsilon _{{\rm s}y\infty } /\upsilon _{0} = \Delta k_{y}/k_{0}\times \Big(1 +(k_{0}/\kappa )^{2}\Big)^{-1}= \Delta k_{y}/k_{0}\times \Big(1 - (\upsilon _{0}/c)^{2}\Big) .
\end{displaymath}
We thus have the interesting result that the  spreading  velocities  of  a
narrow packet in $k$ space do not depend on the detailed form of the wavepacket
over and above its second  central  moments $\Delta k_{x}, \Delta k_{y}$.  In  particular, 
formulas (A15) to (A21) hold in the  nonrelativistic  as  well  as  in  the
relativistic domain.

 \vspace{5pt}
 \begin{figure}[h]
\begin{center}
\includegraphics[width=0.7\textwidth]{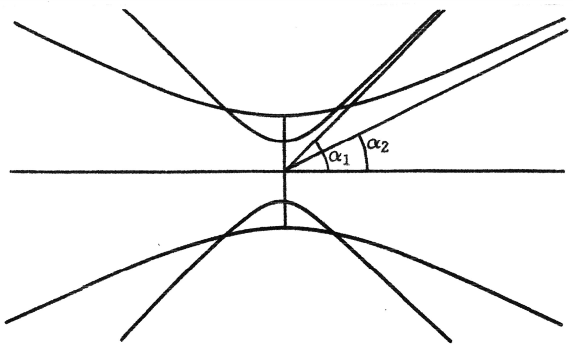}
\caption{Relation between minimum extension and spreading angle.}
\end{center}
\end{figure}

Fig.~6 pictures $\sigma(t)$ according to (A6), and  according to (A16) shows that the smaller the minimum extension $\Delta x(0)$, at the waist of the packet, the larger is the spreading angle $\alpha$:
\begin{displaymath}    (\mbox{A}22)\hsp
\tan \alpha: = \upsilon _{{\rm s}x\infty } /\upsilon _{0} =r{c\over 2(k^{2}_{0}+\kappa ^{2})^{1/2}\Delta x(0)\upsilon_0}\,,\hspace{20pt}
\end{displaymath}
where $r\ge1$ (Heisenberg relation) depending on the form of the wavepacket ($r=1$ for Gaussian form).

In the nonrelativistic domain it is $k_0^2\ll \kappa^2$ and (A22) becomes
\begin{displaymath}      (\mbox{A}23)  \hsp
\tan\alpha=r\frac{\lambda_{\rm deBr}}{4\pi\Delta x(0)},\hspace{30pt}\lambda_{\rm deBr}=\frac{h}{mv_0}. \hspace{20pt}
\end{displaymath}

For photons it is $\kappa=m=0, v_0=c,$ and (A22) becomes
\begin{displaymath}   (\mbox{A}24) \hsp
\tan\alpha=r\frac{\lambda_0}{4\pi\Delta x(0)},
\end{displaymath}
where $\lambda_0$ is the centre wavelength of the photon packet. Relation (A24) coincides, except for some numerical factors $\approx1$, with Verdet's condition for the cone of coherence \cite{Ber56}, \cite{BoWo}, and also with the angular distance between the maximum and the first zero in diffraction at a slit of width $2\Delta x(0)$, though the above  formulas were obtained without any use of holes, slits, or microscopes.

From (A20) it might appear that $\upsilon _{{\rm s}x\infty }\rightarrow \infty $  if $\Delta x(0)\rightarrow 0$. This  is  not  true
because by (A13) $\Delta x(0)\rightarrow 0$ would  imply $\Delta k_{x}\rightarrow \infty $,  and  our  assumption  of  a
narrow wavepacket would no longer hold. In fact, closer inspection shows that
the spreading velocity (A11) is always limited when $\Delta x(0)\rightarrow 0$ \cite{Alm}. For  equal
widths $\Delta k_{x} = \Delta k_{y}$ the transverse spreading is always larger than the  longitudinal one since by (A17) and (A19) it is then
\begin{displaymath}\hsppp
{\upsilon _{{\rm s}x\infty }\over \upsilon _{{\rm s}y\infty }} = {1\over 1 - (\upsilon _{0}/c)^{2}} \ge  1 .  
\end{displaymath}

In the {\emph nonrelativistic domain} we have
\begin{displaymath}\hsppp
k^{2}_{0} \ll  \kappa ^{2} ,
\end{displaymath}
\noindent and condition (A12) now takes the form
\begin{displaymath}\hsppp
\Delta k_{x}, \Delta k_{y}, \Delta k_{z} \ll  \kappa  . 
\end{displaymath}
\noindent The integrals involved in the averaging procedures may then be restricted  to
regions where the dispersion law (A3) can be approximated by $\omega ({\bf k}) = mc^{2}/\hbar  +
\hbar k^{2}/(2m)$ from which one obtains the nonrelativistic relation $\Delta \vec{\upsilon }_{\rm g} = \hbar \Delta {\bf k}/m =
\Delta {\bf p}$/{\emph m}. With (A13) the last condition in ordinary space reads
\begin{displaymath}\hsppp
\Delta x(0), \Delta y(0), \Delta z(0) \gg  1/(2\kappa ) = \mbox{$\frac{1}{2}$} \phantom{n}^{-} \llap{$\lambda$}_{\rm C} . 
\end{displaymath}

In the {\emph relativistic domain} the spreading becomes slower as the velocity $\upsilon _{0}$
of the packet (centre) approaches {\emph c}. In the  zero-mass  limit, $\kappa \rightarrow 0,\, \omega _{0}\rightarrow ck_{0}$,
formulas (A15) and (A21) lead to
\begin{displaymath}\hsppp
\upsilon _{{\rm s}x\infty } \rightarrow  {c\over k_{0}}\Delta k_{x}  
\end{displaymath}
\begin{displaymath}\hsppp
\upsilon _{{\rm s}y\infty } \rightarrow  0 . 
\end{displaymath}
\noindent Thus for photons there is a finite transverse spreading but  no  longitudinal
spreading, in accordance with the well known absence  of  spreading  of  one-dimensional electromagnetic pulses composed of unidirectional waves.
\smallskip

Let us finally consider another length, the  {\emph coherence  length} $\Delta_{\rm c}y$  {\emph of  a
wavepacket} (in the $y$ direction).  We want to show that this length is equal to the
coherence length of the beam in which the wavepackets take part and that $\Delta_{\rm c}y$
does not spread out in time.

The  coherence  length  of  a  {\emph wavepacket}  is  defined  by  means  of  the
autocorrelation function
\begin{displaymath}   (\mbox{A}25) \hsp
\gamma (b) = \int  \psi^* (x,y,z,t)\, \psi (x,y+b,z,t)\, d^{3}x .  
\end{displaymath}
\noindent The function $| \gamma (b)| $ is maximal (=1) at $b=0,$ and $\Delta_{\rm c}y$ is defined as that  value
of $b$ where $| \gamma (b)| $ has decayed for the first time to $\exp (-1/2).\;   \Delta _{c}y$ is closely
related to the ``mean peak width'' of Hilgevoord and Uffink  \cite{Hil}, \cite{Uff84}, \cite{Uff85}.

The coherence length $L_{\rm c}$ of a {\emph beam} is defined by means of the ``contrast'' or
``visibility''
\begin{displaymath}\hsppp
V(b) = (I_{\rm max } - I_{\rm min })/(I_{\rm max }+ I_{\rm min }) 
\end{displaymath}
\noindent where $I_{\rm max }\; (I_{\rm min })$ is the maximal  (minimal)  intensity  in  the  interference
pattern obtained by dividing the beam into two sub-beams,  delaying  the  one
sub-beam by the distance $b$, and then reuniting the two. The length $L_{\rm c}$ is defined as that
value of $b$ where $V(b)$ has decayed to $\exp(-1/2)$. It is  a  quantity  that  is
easy to measure.

In order to see that $\Delta_{\rm c}y$ equals $L_{\rm c}$ suppose that the beam is  a  stream  of
equal wavepackets and that the subdivision of the beam means the  subdivision
of each single wavepacket. The reunited beam then means reunited  wavepackets
and for each wavepacket we have $\psi _{\rm f} = \psi (x,y+b,z,t) + \psi (x,y,z,t)$.  The  probability that it will cause a count in the final counter, at any moment of time
and at any place within the large counter (assuming $100\%$ detection  efficiency) is proportional to
\begin{displaymath}      (\mbox{A}26) \hsp
W= \int^{+\infty}_{-\infty} | \psi_{\rm f} | ^2\,d^3x
\end{displaymath}
\begin{displaymath}\hsppp
=\int | \psi (x,y,z,t)| ^{2}d^{3}x +\int | \psi (x,y+b,z,t)| ^{2}d^{3}x
\end{displaymath}
\begin{displaymath}  \hsppp
+\; 2 {\rm Re}\int \psi^*(x,y,z,t)\, \psi(x,y+b,z,t)\, d^{3}x . 
\end{displaymath}
\noindent The function $\psi _{\rm f}$ need not be normalized  because  the  original  beam  may  be
divided into more than the two sub-beams considered.  The  integrals  in  the
second line of (A26) are equal and we denote them by {\emph A}. The third  line  may
then be written as 2{\emph A}Re$\gamma (b)$ when we use (A25) and observe that $\psi $  in  (A25)
is normalized but in (A26) perhaps not. Writing Re$\gamma  = | \gamma | \cos \alpha $ we have
\begin{displaymath}\hsppp
W(b) = 2A + 2A| \gamma (b)| \cos (\alpha (b)) .  
\end{displaymath}
\noindent Now, the detection probability $W(b)$ of the wavepacket is proportional to  the
registered time averaged intensity $I(b)$ of  the  final  beam,  and  when  one
assumes that $\cos (\alpha (b))$ in typical cases varies much more  rapidly  than  does
$| \gamma (b)| $ one obtains $V(b) = | \gamma (b)| $ and with this $\Delta _{\rm c}y = L_{\rm c}$.

$\gamma (b)$ and with it $\Delta_{\rm c}y$ is independent of time because $\gamma (b)$ is just the  mean
value of the $y$-translation  operator  in  the  state $\psi (x,y,z,t)$,  and  this
operator commutes with the Hamilton operator  for  free  packets \cite{Kle}.  Thus,
whereas the length $\Delta y(t)$  of  the  packet  may  spread  out  with  time,  the
coherence length $\Delta_{\rm c}y$ does not and is  usually  proportional  to  the  minimum
length $\Delta y_{\rm min } = \Delta y(0)$ (assumed to occur at $t=0)$. For a  Gaussian  packet,  for
example, one obtains $\Delta _{\rm c}y = 2\Delta y_{\rm min }$ \cite{Jab87}.

There is no strict relation between  the  coherence  length $\Delta_{\rm c}y$  and  the
length $\Delta y(t)$ of the wavepacket. Any beam, with long or with  short  coherence
length, may be considered to be one very long wavepacket, simply by superposing  the  shorter  wavepackets that  originally  were  conceived   as   its
constituents. All that can be said is that the coherence  length  is  of  the
order of a lower bound for the length of  the  constituent  wavepackets.  The
difference between $\Delta y(t)$ and $\Delta_{\rm c}y$ may always be small in the  case  of  photon
wavepackets which propagate in $y$  direction,  since  photon  packets  do  not
spread out in longitudinal but only in transverse directions. In fact, it  is
often found that for photon packets the coherence length as measured by means
of the interference pattern extends over  almost  the  whole  length  of  the
packet as measured by means of the distance the light travels during the mean
life of the decaying state. In these cases the packet length is always  close
to its lower bound. In the case of wavepackets describing massive  particles,
however, there is considerable spreading even in the  direction  of  propagation. In the neutron-interference  experiment  of  Kaiser  et  al. \cite{Kai},  for
example, the length $\Delta y(t)$ of the neutron packet at the time of its  registration may be larger than the measured coherence length $\Delta _{\rm c}y = 20$ {\AA}       
by more than
a factor of $10^{5}$.

\vspace{20pt}
\noindent {\bf APPENDIX~~B: Derivation of the Bell Inequality}
\medskip

\noindent
In this appendix we derive the  inequality  (4.5)
\begin{displaymath}  (\mbox{B}1) \hsp
K := | E(a,b) +E(a,b^\prime ) +E(a^\prime ,b) - E(a^\prime ,b^\prime )|  \le  2  
\end{displaymath}
\noindent of Sec.~4.3 following Bell \cite{Bel81}.  $E(a,b)$ is the expectation of the  product
$r_{\rm A}r_{\rm B}$ of two dichotomic variables, $r_{\rm A}$ and $r_{\rm B}$, either of which can take on  the
values +1 and $-1$ only:
\begin{displaymath}   (\mbox{B}2)\hsp
E(a,b) := P(+,+| a,b) +P(-,-| a,b) - P(+,- | a,b) - P(-,+| a,b)  
\end{displaymath}
\noindent and
\begin{displaymath}\hspp
P(r_{\rm A},r_{\rm B}| a,b) = \int P(r_{\rm A},r_{\rm B}| a,b,\lambda )\, f(\lambda )\,d\lambda 
\end{displaymath}
\begin{displaymath}\hspp
r_{\rm A}, r_{\rm B} \in  \{-1,+1\} 
\end{displaymath}
\begin{displaymath}   (\mbox{B3), (58})\hspace{35pt}
P(r_{\rm A},r_{\rm B}| a,b,\lambda ) = P_{1}(r_{\rm A}| a,\lambda ) \;  P_{2}(r_{\rm B}| b,\lambda ) 
\end{displaymath}
\begin{displaymath} (\mbox{B4), (59})  \hspace{35pt}
 f(\lambda ) \ge  0,\qquad  \int f(\lambda )\,d\lambda  = 1 . 
\end{displaymath}
\noindent Since $f(\lambda )$ does not depend on $r_{\rm A}$ and $r_{\rm B}$ it is possible to write (B2) as
\begin{displaymath}\hspp
E(a,b) = \int d\lambda \, f(\lambda ) 
\end{displaymath}
\begin{displaymath}\hspp
\times \{ P_1(+| a,\lambda) \; P_2(+| b,\lambda)+ P_1(-| a,\lambda)\;P_2(-| b,\lambda)
\end{displaymath}
\begin{displaymath}\hspp
- \; P_1(+ | a, \lambda) \; P_2(-| b,\lambda) - P_1(-| a,\lambda) \; P_2(+| b,\lambda) \} ,
\end{displaymath}
\noindent and since the first factor on the right-hand side of condition (B3) does not
depend on $r_{\rm B}$, nor the second on $r_{\rm A}$ (outcome independence), we may factorize the integrand
\begin{displaymath}    (\mbox{B}5) \hspace{35pt}
E(a,b) = \int d\lambda \, f(\lambda)\, \{ P_1(+| a,\lambda) - P_1(-| a,\lambda)\} \; \{ P_2(+| b,\lambda) - P_2(-| b,\lambda) \}  
\end{displaymath}
\begin{displaymath}  \hspace{35pt}
\hspace{60pt}= \int d\lambda \, f(\lambda ) \,\bar A(a,\lambda) \,\bar B (b,\lambda )  
\end{displaymath}
\noindent where
\begin{displaymath}\hspp
\bar A (a,\lambda ) = P_{1}(+| a,\lambda ) - P_{1}(-| a,\lambda )
\end{displaymath}
\begin{displaymath}\hspp
\bar B(b,\lambda) = P_2(+| b,\lambda) - P_2(-| b,\lambda).  
\end{displaymath}
\noindent The probability nature of $P_{1}$ and $P_{2}$ means that
\begin{displaymath}\hsppp
0 \le  P_{1} \le  1 , \qquad  0 \le  P_{2} \le  1 , 
\end{displaymath}
\vspace{-5pt}
\noindent hence
\begin{displaymath}   (\mbox{B}6) \hsp
| \bar A(a,\lambda )|  \le  1 ,\quad  | \bar B(b,\lambda )|  \le  1 . 
\end{displaymath}
\noindent Using (B5) and the fact that $\bar A(a,\lambda )$ does not depend on $b$ nor $\bar B(b,\lambda )$  on  $a$ (para\-meter independence)
we have
\begin{displaymath}\hspp
E(a,b) \pm  E(a,b^\prime ) = \int d\lambda \, f(\lambda )\,\bar A (a,\lambda ) \,  [\bar B (b,\lambda ) \pm \bar B (b^\prime ,\lambda )] , 
\end{displaymath}
\noindent and using (B6) for $\bar A (a,\lambda )$, (B4) for $f(\lambda )$  and the fact that $f(\lambda )$  does  not
depend on $a$ and $b$ we obtain
\begin{displaymath}    (\mbox{B}7) \hsp
| E(a,b) \pm  E(a,b^\prime )| \; \le \; \int d\lambda \,  f(\lambda ) \,| \bar B (b,\lambda ) \pm \bar B (b^\prime ,\lambda )|  . 
\end{displaymath}
\noindent Likewise
\begin{displaymath}  (\mbox{B}8) \hsp
| E(a^\prime ,b) \mp  E(a^\prime ,b^\prime )| \; \le \; \int d\lambda \,  f(\lambda )\, | \bar B (b,\lambda ) \mp \bar B (b^\prime ,\lambda )|.  
\end{displaymath}
\noindent Now, we have
\begin{displaymath}   (\mbox{B}9) \hspace{40pt}
| \bar B (b,\lambda ) \pm  \bar B (b^\prime ,\lambda )|  + | \bar B (b,\lambda ) \mp  \bar B (b^\prime ,\lambda )|  
= 2 \max ( | \bar B (b,\lambda )| , | \bar B (b^\prime ,\lambda )|  ) . 
\end{displaymath}
\noindent This can be seen by observing that
\begin{displaymath}\hspp
| x\pm y|  + | x\mp y|  = 2 \max (| x| ,| y| ) , 
\end{displaymath}
\noindent which in turn may be obtained by considering separately the various  possible
cases of positive and negative $x$ and $y$. For example, for $x$  positive  and $y$
negative it is $x=| x| , y=- | y| $ and $| x+y| =|\, | x| - | y| \, |  = \max (| x| - | y| , | y| - | x| )$,
and $\quad | x - y|  = | \, | x| +| y| \, |  = | x| +| y| $,   so   that $| x+y|  + | x - y|  =
\max (| x| - | y| +| x| +| y| , | y| - | x| +| x| +| y| ) = \max ( 2| x| , 2| y|  )$.  Then  (B6)
means    $| x| \le 1, | y| \le 1$ so that (B9) with (B6) can be written as
\begin{displaymath}\hspp
| \bar B (b,\lambda ) \pm \bar B (b^\prime ,\lambda )|  + | \bar B (b,\lambda ) \mp \bar B (b^\prime ,\lambda )|  \le  2 .
\end{displaymath}
\noindent With the normalization (B4) in (B7) and (B8) we arrive at
\begin{displaymath}\hspp
| E(a,b) \pm  E(a,b^\prime )|  + | E(a^\prime ,b) \mp  E(a^\prime ,b^\prime )|  \le  2  
\end{displaymath}
\noindent and this includes the desired Bell inequality (B1), (4.5).

\vspace{20pt}
\noindent {\bf APPENDIX~~C: EPR Joint Probability Formulas}
\medskip

\noindent
Here we derive formula (4.4) of Sec.~4.2
\begin{displaymath}(4.4)\hsp
P(r_{\rm A},r_{\rm B}| a,b) = \mbox{${\frac{1}{4}}$}(1 - r_{\rm A}r_{\rm B}\cos \vartheta )
 \end{displaymath}
\noindent from the rules of quantum mechanics. We have two spin-${1\over 2}$ particles (similar or
not) in a state of zero total spin (spin singlet state). The  particles  move
in opposite directions and each enters a Stern-Gerlach-type  apparatus  where
it is deflected upwards or downwards with respect to the axis of its  respective apparatus (cf. Fig.~4 in Sec.~4.1). Particle 1 enters apparatus  A,
which has its axis in the direction of the unit  vector  {\bf a},  and  particle  2
enters apparatus B with axis {\bf b}. We  first  calculate  the  joint  probability
$P(+,-| a,b)$ that A obtains an up deflection and B a down deflection.

The wave function of the two-particle system in the spin singlet state  is
(4.5)
\begin{displaymath}   (\mbox{C}1)\hsp
\Psi (1,2) = \frac{1}{\sqrt2}\Big[ | +\rangle^{(1)}{| - \rangle^{(2)}} - \,  | - \rangle^{(1)}{| +\rangle^{(2)}} \Big] A(1,2),
\end{displaymath}
\noindent where the up $| +\rangle $ and down $| -\rangle $ eigenfunctions of the spin component refer to a
fixed but arbitrary axis. This is an entangled wave function. Only the spin part needs to be considered here.  It
happens to be antisymmetric, independent of whether it refers to  similar  or
dissimilar particles. 

Let the two eigenfunctions of the one-particle spin component operator  of
apparatus A be $| a+\rangle $ and $| a -\rangle $, and those of apparatus B    $| b+\rangle $  and $| b -\rangle $.  The
operation of a Stern-Gerlach-type apparatus on either of  the  two  particles
leads to either   $| a+\rangle^{(1)}\!| b - \rangle^{(2)}$   or   $| b - \rangle ^{(1)}\!| a+\rangle^{(2)} $ with  equal  probability.
Hence
\begin{displaymath}  (\mbox{C}2)\hsp
P(+,- | a,b) = \mbox{$\frac{1}{2}$} \Big |  \Big(  | a+\rangle^{(1)}\!| b - \rangle^{(2)}     , \Psi (1,2) \Big) \Big | ^{2}
\end{displaymath}
\begin{displaymath}
\hspace{142pt}+ \; \mbox{$\frac{1}{2}$} \Big | \Big(   | b - \rangle ^{(1)}\!| a+\rangle^{(2)}  , \Psi (1,2) \Big)  \Big | ^{2} ,  
\end{displaymath}
\noindent whether the particles are similar or not. In the case of dissimilar particles
the factors 1/2 reflect the fact that we are not interested in distinguishing
the particles but only in the average result. In the case of similar  particles  the probability expressions should  be  invariant  under  particle-label permutation (Sec.~3.3), and this is now the reason for  the  factors
1/2.

To be able to evaluate expression (C2) we  need  the  eigenfunctions $| a+\rangle , | b -\rangle $ etc. in terms of the eigenfunctions $| +\rangle , | - \rangle $, which refer to
the fixed but arbitrary axis employed in (C1). We  let  this  axis  coincide
with A's axis {\bf a}, so that
\begin{displaymath}  \ (\mbox{C}3)\hsp
| a+\rangle  = | +\rangle , \quad | a -\rangle  = | - \rangle  . 
\end{displaymath}
\noindent Let the axis of B form an angle $\vartheta $ with the axis of A We then have to express
the eigenfunctions $| b+\rangle , | b -\rangle $ in  the  rotated  system B  in  terms  of  the
eigenfunctions of system A \cite{Fey65}, \cite[p.~1073]{Mes}
\begin{displaymath}\hspp
| b+\rangle  = \cos {\vartheta \over 2} \exp [i(\beta +\gamma )/2]\, | +\rangle\;  +\; i \sin {\vartheta \over 2} \exp [-i(\beta - \gamma )/2]\, |-\rangle \mbox{~~~~~~~}
\end{displaymath}
\begin{displaymath}  (\mbox{C}4) \hsp
| b -\rangle  = i \sin {\vartheta \over 2} \exp [i(\beta - \gamma )/2]\, | +\rangle \; + \; \cos {\vartheta \over 2} \exp [-i(\beta +\gamma )/2]\, |-\rangle . 
\end{displaymath} 
The angles $\beta $ and $\gamma $ define possible rotations of the other axes but  will  not
appear in the final formulas. Inserting (C1), (C3) and (C4) into (C2) and
observing the orthonormality of the functions $| +\rangle , | - \rangle $  for  the  respective
particles we obtain
\begin{displaymath}    (\mbox{C}5)\hsp
P(+,- | a,b) = {1\over 2} \left( \cos {\vartheta \over 2}\right)^{2} = \mbox{${\frac{1}{4}}$}(1 +\cos \vartheta ) . 
\end{displaymath}
\noindent Proceeding in the same way in the other cases (A up, B up; A down, B  up;  A
down, B down) we obtain
\begin{displaymath}  (\mbox{C}6)  \hsp
P(-,+| a,b) = P(+,-| a,b) , 
\end{displaymath}
\begin{displaymath}   (\mbox{C}7)\hsp
P(+,+| a,b) = P(-,- | a,b) = {1\over 2} \left( \sin {\vartheta \over 2} \right)^{2} = \mbox{${\frac{1}{4}}$} (1 - \cos \vartheta ) . 
\end{displaymath}
\noindent (C5), (C6) and (C7) may be summarized in the form
\begin{displaymath}   (\mbox{C}8) \hsp
P(r_{\rm A},r_{\rm B}| a,b) = \mbox{${\frac{1}{4}}$} (1 - r_{\rm A}r_{\rm B}\cos \vartheta ) 
\end{displaymath}
\noindent where $r_{\rm A}, r_{\rm B} \in  \{-1,+1\}$. This is formula (4.4) of Sec.~4.2.

\vspace{10pt}
It is amusing  to  notice  that  we  can  also  obtain  formula  (4.4)  by
proceeding \emph{as if} the following situation  were  to  hold:  after  any  single
interaction either of the two protons of Fig.~4 in Sec.~4.1 has a definite direction of
spin (i.e. is a spin-up eigenfunction of some $s_{z^\prime }$, cf. Sec.~2.2),  say $\vec{\sigma }$
and $- \vec{\sigma }$, respectively, where $\vec{\sigma }$ is a unit vector, and the total spin  is  zero.
The direction of $\vec{\sigma }$ varies from one interaction to the other  in  such  a  way
that there is spherical  symmetry  on  the  average.  The  Stern-Gerlach-type
apparatus which obtains its proton first, say A, turns the spin of its proton
into either up or down direction with respect to its axis, say into direction
+{\bf a}, and at the same time turns the spin of the other proton into the opposite
direction $-{\bf a}$. Thus here angular momentum is conserved within  the  system  of
the two protons, and the apparatuses are  not  involved  in  angular-momentum
conservation. Then B's apparatus turns the spin of its proton from  direction
$-{\bf a}$ into either up or down direction with respect  to  B's  axis ${\bf b}$,  without,
however, influencing the spin of proton 1 any  more.  Here,  angular-momentum
conservation involves proton 2 and apparatus B, as mentioned in  Sec.~2.2
on the Stern-Gerlach experiment.

 The spin direction $\vec{\sigma }$ plays the role  of  the
parameter $\lambda $ in the general consideration of Sec.~4.3. From  the  point  of
view of quantum mechanics the parameter $\vec{\sigma }$, interpreted in the above  way,  is
hidden. Of course, in the one-particle states $| +\rangle $ etc. $\vec{\sigma }$ is  not  hidden  but
explicitly specifies the spin direction, but the two-particle state $\Psi (1,2)$ of
(C1), which is the only state existing  after  the  interaction  at O,  has
spherical symmetry, and there can be no parameter specifying any direction in
such a state in quantum mechanics. This is why we used the proviso \emph{as if}.

In order to prove our above assertion regarding the {\emph{as-if}} derivation of (4.4) we
observe that the conditional probability of obtaining  the  result $r_{\rm A}$ (i.e.
either + or $-$) with respect to the axis {\bf a}, given that the spin of the  proton
before it enters the apparatus points in the direction ${\bf b}$, is,  by  (C3)  and
(C4),
\begin{displaymath}\hspp
P(r_{\rm A}| {\bf a},{\bf b}) = | \langle ar_{\rm A}| b+\rangle | ^{2} = | \langle r_{\rm A}| b+\rangle | ^{2} = \mbox{${1\over 2}$} (1 +r_{\rm A}{\bf ba}) ,
\end{displaymath}
\noindent and if we replace the axis ${\bf b}$ by the axis $\vec{\sigma }$ and the angle $\vartheta $ between {\bf a} and ${\bf b}$ by
the angle $\alpha $ between {\bf a} and $\vec{\sigma }$ we get the probability that A obtains the  result
$r_{\rm A}$ given that the spin of the proton before it entered the apparatus  pointed
in the direction $\vec{\sigma }$
\begin{displaymath} (\mbox{C}9) \hsp
P(r_{\rm A}| {\bf a},\vec{\sigma }) = \mbox{${1\over 2}$} (1 +r_{\rm A}\vec{\sigma }{\bf a}) = \mbox{${1\over 2}$} (1 +r_{\rm A}\cos \alpha ) . 
\end{displaymath}
\noindent However, the probability of B obtaining the result $r_{\rm B}$ is  not  the  analogous
formula $(1 +r_{\rm B}(- \vec{\sigma }){\bf b})/2,$ because A has turned not only the spin  of  proton  1
into the direction $r_{\rm A}{\bf a}$ but also the spin of proton 2 into direction $- r_{\rm A}{\bf a}$  and
then separated the two protons. Thus here we have to replace $- \vec{\sigma }$ by $- r_{\rm A}{\bf a}$,  and
B's probability is
\begin{displaymath}    (\mbox{C}10)\hsp
P(r_{\rm B}| {\bf b},- r_{\rm A}{\bf a}) = \mbox{${1\over 2}$} (1 +r_{\rm B}(- r_{\rm A}{\bf a}){\bf b}) . 
\end{displaymath}
\noindent The conditional joint probability of A obtaining $r_{\rm A}$ and B  obtaining $r_{\rm B}$  is
given by the product of (C9) with (C10)
\begin{displaymath}\hsppp
P(r_{\rm A},r_{\rm B}| {\bf a},{\bf b},\vec{\sigma }) = \mbox{${1\over 2}$} (1 +r_{\rm A}\vec{\sigma }{\bf a})\, \mbox{${1\over 2}$} (1- r_{\rm A}r_{\rm B}{\bf ab}) .  
\end{displaymath}
\noindent If we integrate over all directions of $\vec{\sigma }$, assuming an isotropic distribution,
we obtain
\begin{displaymath}\hspace{20pt}
P(r_{\rm A},r_{\rm B}| {\bf a},{\bf b}) = {\frac{1}{4}\pi } \int^{+\pi }_{- \pi }d\varphi  \int^{\pi }_{0} \sin \alpha\,  d\alpha \; {\frac{1}{4}}(1 +r_{\rm A}\cos \alpha )\, (1 - r_{\rm A}r_{\rm B}\cos \vartheta )
\end{displaymath}
\begin{displaymath}    (\mbox{C}11)\hsp
= {1\over 8}(1 - r_{\rm A}r_{\rm B}\cos \vartheta ) \int^{\pi }_{0}\sin \alpha\;  (1 +r_{\rm A}\cos \alpha )\, d\alpha 
\end{displaymath}
\noindent where the system of coordinates $(x,y,z)$ for the integration  is  chosen  such
that the $z$ axis is in the direction {\bf a}, and ${\bf b}$ lies in the $x$-$z$ plane:
\begin{displaymath}\hsppp
{\bf a} = (0,0,1),\qquad {\bf b} = (\sin \vartheta ,0,\cos \vartheta ) ,
\end{displaymath}
\noindent so that
\begin{displaymath}\hsppp
\vec{\sigma } = (\sin \alpha  \cos \varphi , \sin \alpha  \sin \varphi , \cos \alpha ) ,
\end{displaymath}
\begin{displaymath}\hsppp
{\bf ab} = \cos \vartheta ,\quad \vec{\sigma }{\bf a} = \cos \alpha ,\quad  \vec{\sigma }{\bf b} = \sin \vartheta  \cos \varphi  \sin \alpha  + \cos \vartheta  \cos \alpha  .
\end{displaymath}
\noindent It is not difficult to verify that (C11) leads to formula  (C8)  or  (4.4).
Such an \emph{as-if} procedure is in fact possible in any  EPR
situation, not only in that of two spin-${1\over 2}$ particles in the singlet state.

\vspace{20pt}
\noindent {\bf APPENDIX~~D: EPR Probabilities in Different Systems of Eigen\-functions}
\vspace{1pt}

\noindent
We shall show here that not only Stern-Gerlach-type devices (as in Sec.~4.2), but any devices that obey the formulas of quantum mechanics  are incapable of building   faster-than-light warning systems
\cite{Ebe} - \cite{Jor83}. The  proof
uses the fact that in quantum mechanics the apparatuses  are  represented  by
operators. Different apparatuses  mean  different  operators,  and  different
operators in general mean different systems of eigenfunctions, and these  can
be transformed into  one  another.  The  normalized  quantum-mechanical  wave
function for a system of two similar particles can be written in the form

\begin{displaymath}  (\mbox{D}1) \hsp
\Psi _{\rm SA}(1,2) = C \sum^{\infty }_{k=1} \left[ \,\zeta _{k}(1)\;u_{k}(2)\;\pm \zeta _{k}(2)\; u_{k}(1)\, \right]  ,  
\end{displaymath}
\noindent which is formula (4.1) from Sec.~4.1 with a properly symmetrized function.
The $u_{k}(x)$ form a complete set of orthonormal eigenfunctions of some  operator
representing the apparatus of experimenter B.  $C$ is a  real  overall  normalization constant, which need not be equal to $1/\sqrt 2$  because the functions $\zeta _{k}(x)$,
which describe the particle at A, are not presupposed to be orthonormal. With
the expansion
\begin{displaymath}    (\mbox{D}2)\hsp
\zeta _{k}(x) = \sum_{l} a_{lk}\, w_{l}(x) , 
\end{displaymath}
\noindent where the $w_{l}(x)$ form a complete set of  orthonormal  eigenfunctions  of  some
operator representing the apparatus of experimenter A, (D1) can  be  written
as
\begin{displaymath}    (\mbox{D}3)\hsp
\Psi _{\rm SA}(1,2) = C \sum_{lk}a_{lk}\left[ w_{l}(1)u_{k}(2) \pm w_{l}(2)u_{k}(1) \right] . 
\end{displaymath}
We first calculate the probability $P_{1}(u_{n},w_{m})$ of  a  transition  where  the
state  (D3)  changes  into  either  the  state $w_{m}(1)u_{n}(2)$   or  the  state
$w_{m}(2)u_{n}(1)$  with equal probability. This  is  the  probability  that  in  B's
apparatus there will be a particle (``whichever of the two it is'') with  state
$u_{n}$ and in A's apparatus a particle with state $w_{m}$. It is [see  the  remark  on
(C2) in Appendix C]
\begin{displaymath}    \hsp
P_{1}(u_{n},w_{m}) =
 {1\over 2}\bigg | \Big(  C \sum_{kl}a_{lk}\left( w_{l}(1)u_{k}(2) \pm w_{l}(2)u_{k}(1) \right), w_{m}(1)u_{n}(2) \Big) \bigg |^2 
\end{displaymath}
\begin{displaymath}\hsp
\mbox{~~~~~~~~~~~~~~~}+ {1\over 2} \bigg | \Big(  C \sum_{kl}a_{lk} \left( w_{l}(1)u_{k}(2)\pm w_{l}(2)u_{k}(1) \right), w_{m}(2)u_{n}(1) \Big) \bigg |^2
\end{displaymath}

\begin{displaymath}\hspace{55pt}
= {C^{2}\over 2}\bigg | \sum_{kl} a^*_{lk} \Big[ \Big( w_{l}(1)u_{k}(2),w_{m}(1)u_{n}(2)\Big) 
\pm  \Big( w_{l}(2)u_{k}(1),w_{m}(1)u_{n}(2)\Big) \Big] \bigg |^2
\end{displaymath}

\begin{displaymath}\hspace{60pt}
+ {C^{2}\over 2}\bigg |  \sum_{kl}a^*_{lk}\Big[ \Big( w_{l}(2)u_{k}(1),w_{m}(2)u_{n}(1)\Big) 
\pm  \Big( w_{l}(1)u_{k}(2),w_{m}(2)u_{n}(1)\Big) \Big] \bigg | ^{2} .
\end{displaymath}

\noindent With $\Big( w_{l}(1)u_{k}(2),w_{m}(1)u_{n}(2)\Big)  = \Big( w_{l}(2)u_{k}(1),w_{m}(2)u_{n}(1)\Big)  = \delta _{lm}\delta _{kn}$    this
becomes
\begin{displaymath}    (\mbox{D}4)      \hsp
P_1(u_n,w_m) = 
{C^{2}\over 2} \bigg | a^*_{mn} \pm  \sum_{kl}a^*_{lk} \Big( w_{l}(2),u_{n}(2)\Big) \Big( u_{k}(1),w_{m}(1)\Big) \bigg |^2
\end{displaymath}
\begin{displaymath}\hsp
\mbox{~~~~~~~~~~~~~~~~~~~~} + { C^{2}\over 2} \bigg |  a^*_{mn} \pm  \sum_{kl}a^*_{lk} \Big( w_l(1),u_n(1) \Big) \Big( u_{k}(2),w_{m}(2)\Big) \bigg |^2. 
\end{displaymath}
\noindent As the scalar products in (D4) are zero the formula reduces to
\begin{displaymath}\hspp
P_{1}(u_{n},w_{m}) = C^{2}| a_{mn}| ^{2} .  
\end{displaymath}
The scalar products are zero because  the  final  wave  function $u_{l}$  of  the
particle in apparatus B and the final wave function $w_{l}$  of  the  particle  in
apparatus A are well separated from  each  other  and  do  not  overlap.  The
interaction of a  wavepacket  from  the  entangled  system  (D1)  with  that
apparatus that operated first, had led to reduction and to  disentanglement  of  the
system.

Second we consider the  probability  of B  observing  that  his  particle
assumes the state $u_{n}$ irrespective of the  state  of  A's  particle.  This  is
obtained by summing the probability $P_{1}(u_{n},w_{m})$ over all states of A's particle
\begin{displaymath}  (\mbox{D}5)  \hsp
P_{2}(u_{n}) = C^{2} \sum^{}_{m}| a_{mn}| ^{2} ,  
\end{displaymath}
\noindent and we want to show  that  this  probability  is  unchanged  when  A  uses  a
different apparatus. Let the eigenfunctions of the new operator corresponding
to the new apparatus be $w^{\prime}_m (x)$. They are related to the  eigenfunctions $w_{m}(x)$
of the original operator by 
\begin{displaymath}   (\mbox{D}6)  \hsp
w_{m}(x) = \sum_{k} U_{mk} w^{\prime}_k (x) .  
\end{displaymath}
\noindent $U_{mk}$ is a unitary matrix $(\sum _{m} U_{jm}^* U_{km}=\delta _{jk})$, and the  index $k$  may  even  be
continuous and the sum an integral. Actually, the transformation  (D6)  need
not even be unitary \cite{Jor83}, \cite{Bus87}, but we will not pursue this here. By  inserting
(D6) into (D3) we can write $\Psi _{\rm SA}(1,2)$ in the form
\begin{displaymath}\hsp
\Psi _{\rm SA}(1,2) = C \sum_{mn}a_{mn} \Big[ u_{n}(2) \sum_{k}U_{mk}w^{\prime}_k (1) \pm  u_{n}(1) \sum_{k} U_{mk}w^{\prime}_k (2) \Big]
\end{displaymath}
\begin{displaymath}\hspace{108pt}
= C \sum_{lk} \, \underbrace{
\sum_j a_{jk} U_{jl}}_{b_{lk}} \,  \Big[ w^{\prime}_{l}(1) u_k(2) \pm w^{\prime}_l (2) u_{k}(1) \Big] . 
\end{displaymath}
\noindent The probability that this changes into either $w^{\prime}_m (1)u_{n} (2)$  or $w^{\prime}_m (2) u_{n}(1)$  is
$C^{2} \allowbreak | b_{mn}| ^{2}$, by analogy with (D3) and (D5). Hence
\begin{displaymath}\hsp
P_{2}(u_{n}) = C^{2} \sum_{m}| b_{mn}| ^{2} = C^{2} \sum_{m}b^*_{mn} b_{mn} = C^{2} \sum_{mjk} a^*_{jn} U^*_{jm}a_{kn}U_{km}
\end{displaymath}
\begin{displaymath}\hspace{95pt}
= C^{2} \sum_{jk} a^*_{jn} a_{kn} \sum_{m}U^*_{jm}U_{km} = C^{2} \sum_{k}| a_{kn}| ^{2} , 
\end{displaymath}
\noindent and this coincides with (D5), concluding the proof of our assertion.

Finally we want to show that even if A chooses not to  use  his  apparatus
and to do nothing this will make  no  difference.  In  this  case,  when  B's
particle assumes the state $u_{n}$, A's  particle  will  assume  some  correlated
state $\zeta _{n}$. The probability of B's  particle  assuming  the  state $u_{n}$  is  the
probability of the transition where the state $\Psi _{\rm SA}(1,2)$  changes  either  into
the state $\zeta _{n}(1)u_{n}(2)$ or into the state $\zeta _{n}(2)u_{n}(1)$. When $\zeta _{n}$  is  expressed  in
terms of the $w_{m}$, according to formula (D2), the two states become
\begin{displaymath}\hspp
\sum_{m}a_{mn}w_{m}(1)u_{n}(2)\quad \mbox{and} \quad \sum_{m}a_{mn}w_{m}(2)u_{n}(1) ,
\end{displaymath}
\noindent respectively, and the transition probability becomes
\begin{displaymath}   (\mbox{D}7)       \hspace{30pt}
P^{\prime}_2 (u_{n}) =
\mbox{$\frac{1}{2}$}  \bigg |  \Big( \Psi _{\rm SA}(1,2), \sum_{m}a_{mn} w_{m}(1)u_{n}(2)\Big)  \bigg | ^{2} \times \bigg |   \sum_{m}a_{mn} w_{m}(1)u_{n}(2) \bigg |^{-2}
\end{displaymath}
\begin{displaymath}\hsp
 \hspace{20pt}+\quad  \mbox{$\frac{1}{2}$}\bigg | \Big( \Psi _{\rm SA}(1,2), \sum_{m}a_{mn} w_{m}(2)u_{n}(1)\Big)  \bigg | ^{2} \times \bigg |  \sum_{m}a_{mn} w_{m}(2)u_{n}(1) \bigg |^{-2} . 
\end{displaymath}
\noindent The denominators are different from 1 because the $a_{mn}$ come in via  the $\zeta _{n}(x)$
in formula (D2), and the $\zeta$'s are not normalized. With (D3) the  first  term
of expression (D7) becomes
\begin{displaymath}\hspace{3pt}
T_1= \bigg | \sum_{klm}a^*_{lk} a_{mn} \Big[\Big(  w_l(1), w_m(1) \Big)\Big( u_k(2), u_n(2)\Big) \pm \Big( w_l(2), u_n(2) \Big)\Big( u_k(1), w_m(1)\Big) \Big] \bigg |^2 
\end{displaymath}
\begin{displaymath}\hsp
 \times \bigg | \sum_{mk} a^*_{mn} a_{kn} \Big( w_{m}(1)u_{n}(2), w_{k}(1)u_{n}(2)\Big) \bigg | ^{-1}\times \frac{C^2}{2}
\end{displaymath}
\begin{displaymath}\hsp
= {C^{2}\over 2} \bigg | \sum_{m}a^*_{mn}a_{mn} \bigg |^2  \times \bigg | \sum_{m}a^*_{mn} a_{mn} \bigg |^{-1} = {C^{2}\over 2} \sum_{m}| a_{mn}| ^{2} .\mbox{~~~~~~~~~~}  
\end{displaymath}
\noindent The second term of (D7) leads to the same expression, so
\begin{displaymath}\hspp
P^{\prime}_2 (u_n) = C^2 \sum_{m} | a_{mn}| ^{2} = P_{2}(u_{n}) ,  
\end{displaymath}
\noindent which is what we wanted to show. 

\newpage
\noindent 
{\bf Notes and References}

\renewcommand{\labelenumi}{[\arabic{[enumi]}]}

\renewcommand{\section}[2]{}

\smallskip
\hspace{5cm}
--------------------------------------------
\end{document}